\input{epsf}

\documentstyle[prd,eqsecnum,aps,twocolumn]{revtex}

\catcode`\@=11

\def\maketitle2{\par 
\begingroup
\let\cite\@bylinecite
\def\thefootnote{\fnsymbol{footnote}}%
\twocolumn[\@maketitle2\vskip2pc]%
\thispagestyle{plain}\@thanks
\endgroup
\def\thefootnote{\arabic{footnote}}%
\setcounter{footnote}{0}%
\let\maketitle2\relax \let\@maketitle2\relax
\let\@thanks\relax \let\@authoraddress\relax \let\@title\relax
\let\@date\relax \let\thanks\relax \let\@abstract\relax 
\let\@pacs\relax}

\def\abstract#1{\gdef\@abstract{{\par 
\bgroup
\ifdim\prevdepth=-1000pt \prevdepth0pt\fi
\hsize\columnwidth
\dimen0=-\prevdepth \advance\dimen0 by17.5pt \nointerlineskip
\small\vrule width 0pt height\dimen0 \relax}{~~}#1\egroup}}

\def\pacs#1{\gdef\@pacs{{\par 
\bgroup
\hsize\columnwidth \parindent0pt
\ifdim\prevdepth=-1000pt \prevdepth0pt\fi
\dimen0=-\prevdepth \advance\dimen0 by20pt\nointerlineskip
\egroup} PACS numbers:~#1}}

\def\@maketitle2{
\@preprint
\@title
\ifdim\prevdepth=-1000pt \prevdepth0pt\fi
\@authoraddress
\@date
\begin{list}{}{\leftmargin=0.10753\textwidth \rightmargin=\leftmargin
\itemsep=1pc\partopsep=-1pc}
\item\@abstract
\item\@pacs
\end{list}
}

\catcode`\@=12

\begin{document}
\draft
\preprint{LA-UR-96-3336}
\title{Nonequilibrium Dynamics of Symmetry Breaking in $\lambda
\Phi^4$ Field Theory}
\author{Fred Cooper,$^1$ Salman Habib,$^1$ Yuval Kluger,$^2$ and
Emil Mottola$^1$} 
\address{$^1$Theoretical Division, T-8, MS B285, Los Alamos National
Laboratory, Los Alamos, NM 87545, USA} 
\address{$^2$Nuclear Science Division, Lawrence Berkeley National
Laboratory, MS 70A-3307, Berkeley, CA 94720, USA} 
\date{\today}

\newcommand{\sq}{\lower.25ex\hbox{\large$\Box$}}

\abstract
{The time evolution of $O(N)$ symmetric $\lambda \Phi^4$ scalar field
theory is studied in the large $N$ limit. In this limit the
$\langle\Phi\rangle$ mean field and two-point correlation function
$\langle\Phi\Phi\rangle$ evolve together as a self-consistent closed
Hamiltonian system, characterized by a Gaussian density matrix. The
static part of the effective Hamiltonian defines the True Effective
Potential $U_{eff}$ for configurations far from thermal
equilibrium. Numerically solving the time evolution equations for
energy densities corresponding to a quench in the unstable spinodal
region, we find results quite different from what might be inferred
from the equilibrium free energy ``effective'' potential $F$. Typical
time evolutions show effectively irreversible energy flow from the
coherent mean fields to the quantum fluctuating modes, due to the
creation of massless Goldstone bosons near threshold. The plasma
frequency and collisionless damping rate of the mean fields are
calculated in terms of the particle number density by a linear
response analysis and compared with the numerical results. Dephasing
of the fluctuations leads also to the growth of an effective entropy
and the transition from quantum to classical behavior of the
ensemble. In addition to casting some light on fundamental issues of
nonequilibrium quantum statistical mechanics, the general framework
presented in this work may be applied to a study of the dynamics of
second order phase transitions in a wide variety of Landau-Ginsburg
systems described by a scalar order parameter.}


\maketitle2
\narrowtext

\section{Introduction}
\label{sec:level1}

Spontaneous symmetry breaking by a scalar order parameter occurs in
many different physical systems, as diverse as liquid $^4$He at
temperatures of order $2\,K$, to the standard model of electroweak
interactions at temperatures of order $250$ GeV $= 3 \times
10^{15}\,K$. The prototype renormalizable quantum field theory
describing this symmetry breaking is a scalar field with a $\lambda
\Phi^4$ self-interaction. The behavior of the finite temperature
effective potential or Landau-Ginsburg-Helmholtz free energy in this
theory is well-known and forms the usual basis for discussion of the
symmetry restoration at high temperature. On the other hand,
surprisingly little effort has been devoted to the nonequilibrium or
time dependent aspects of the symmetry breaking phase transition. With
the development of practical general techniques for studying time
dependent problems in quantum field theory, as well as the advent of
high speed supercomputers it has become possible to address these
dynamical issues systematically for the first time. It is clear that a
detailed description of the time dependent dynamics will be necessary
to calculate nonequilibrium properties of the phase transition, such
as the formation and evolution of defects in the $^4$He system after a
rapid quench, or the efficiency of baryogenesis in the electroweak
phase transition. Other examples requiring the detailed time evolution
of scalar fields are the chiral phase transition of the strong
interactions such as may soon be probed in relativistic heavy-ion
colliders and the problem of reheating the very early universe after
it has passed through an epoch of rapid expansion and cooling.

All of these problems require a consistent treatment of time-dependent
mean fields such as the scalar expectation value $\langle
{\bf\Phi}\rangle$ in interaction with its own quantum and/or thermal
fluctuations. The technique we rely upon in this paper consists of
replicating the scalar field ${\bf\Phi} \rightarrow {\bf\Phi}_i,\,
i=1,\dots, N$. Then it becomes possible to compute the quantum
effective action in a systematic power series in the parameter $1/N$
\cite{largeN}.  Variation of this effective action yields equations of
motion for the mean fields coupled to Green's functions of the theory
which are suitable for implementation on a computer. The leading order
in large $N$ corresponds to a self-consistent mean field
approximation, {\em i.e.} a truncation of the infinite hierarchy of
Schwinger-Dyson equations for the $n$-point correlation functions to a
closed Hamiltonian system of just the one-point function, $\langle
{\bf\Phi} (x) \rangle$ and two-point function, $\langle
{\bf\Phi}(x){\bf\Phi}(x')\rangle$. Because no irreducible correlators
higher than these appear in the leading order of the large $N$
expansion, it is equivalent to a Gaussian approximation to the
time-dependent density matrix of the system. As will become apparent,
the approximation allows automatically for a mixed-state Gaussian
density matrix, ${\bf\rho}$, and is therefore more general than a
Gaussian ansatz for the pure state wave function in the Schr\"odinger
picture. Hence, the mixed-state Gaussian width of ${\bf\rho}$ can
describe at once and on the same footing both the quantum and
classical statistical fluctuations of the $\bf\Phi$ field about its
mean value far from thermal equilibrium.

An important property of the evolution equations in the large $N$
limit is that they are Hamilton's equations for an effective {\em
classical} Hamiltonian $H_{eff}$ (in which $\hbar$ appears as a
parameter). This effective Hamiltonian turns out to be nothing else
than the expectation value of the full quantum Hamiltonian $\bf H$ in
the general mixed state described by the time-dependent Gaussian
density matrix, $\bf \rho$, {\em i.e.},
\begin{equation}
H_{eff} = {\rm Tr} \left({\bf \rho H}\right)\,.
\end{equation}
The canonical variables of $H_{eff}$ are in one-to-one correspondence
with the parameters needed to specify the general Gaussian density
matrix in the Schr\"odinger picture, or the one- and two-point
functions of the Schwinger-Dyson hierarchy in the Heisenberg
picture. In order to highlight the Hamiltonian structure of the mean
field equations, our first purpose in this paper will be to establish
these various correspondences in detail for the $O(N)$ symmetric
$\lambda \Phi^4$ theory. The existence of an effective Hamiltonian
makes it clear from the outset that the leading order large $N$
approximation is self-consistent and energy conserving, and hence does
not introduce any time irreversibility or dissipation ``by hand'' into
the system.

A corollary of the identification of the effective Hamiltonian
$H_{eff}$ in the large $N$ limit is that its static piece $U_{eff}$
(obtained by setting all the canonical momenta to zero) is the {\em
True} Effective Potential which governs the nonequilibrium evolution
of the system. This True Effective Potential (TEP) is real, and
completely well-defined for states far from thermal equilibrium. In
the special case of thermal equilibrium it becomes the {\em internal}
energy $U$ of the closed Hamiltonian system described by $H_{eff}$. In
contrast, the Helmholtz free energy $F$ which is sometimes called the
``effective'' potential is not defined away from thermal equilibrium
and becomes complex in the unstable spinodal co-existence region
between the two spontaneously broken vacua in simple approximation
schemes.  This is easily understood in the more general nonequilibrium
context of this paper: it means simply that there is no stable
equilibrium state in the co-existence region with a fixed constant
value of $\langle {\bf\Phi}\rangle$. In other words, if we start at
$t=0$ with initial conditions corresponding to this putative thermal
equilibrium state, the system immediately begins to evolve in time
away from it, and this is the physical meaning of the imaginary part
of $F$ \cite{WW}.

We shall see that the free energy function $F$ (or simply its real
part) is a very poor guide to the time evolution of the system far
from thermal equilibrium. In particular, the oscillations of the time
dependent expectation value $\langle{\bf\Phi}\rangle (t)$ about the
spontaneously broken minima are characterized by a frequency (the
plasma frequency) which is {\em not} the second derivative of $F$ at
its minimum (which turns out to be zero), and moreover, even the {\em
location} of the minima of $F$ is in general different from the
stationary points of the mean field evolution. Explicitly solving for
the actual nonequilibrium mean field dynamics and demonstrating that
it is quite different from what might be inferred from an uncritical
use of the equilibrium free energy function $F$ is the second major
emphasis of this work. In this we corroborate the similar conclusions
reached in Ref. \cite{Boyan}.

The spontaneous breakdown of the global $O(N)$ symmetry in the model
leads to the existence of $N-1$ massless Goldstone bosons (in $d>1$
spatial dimensions), which dominate the dynamics in the large $N$
limit.  In $d \le 1$ spatial dimension there is no symmetry breaking
and the system is inevitably driven into the symmetric phase, no
matter what the initial state or energy density. The one dimensional
case is of more than passing interest in showing how the $O(N)$
symmetry is restored dynamically and the Mermin-Wagner-Coleman theorem
\cite{mwc} is satisfied in real time. For $d>1$, the absence of a finite
mass threshold means that an arbitrarily small amount of energy in the
mean field can create massless Goldstone boson pairs nearly at
rest. This open channel provides an efficient mechanism for the mean
field to continuously transfer its kinetic energy to the massless
particle modes over time. The presence of a symmetry which requires
massless particles is a feature which the $O(N)$ model shares with
other physically interesting theories such as non-abelian gauge
theories and gravity. The $O(N)$ scalar theory provides an instructive
example of dissipation by means of massless particle creation, which
should be applicable in other quite diverse contexts, such as gluon
production in relativistic heavy-ion collisions or graviton creation
in early universe phase transitions. Developing techniques and gaining
some valuable intuition for these more challenging problems is our
third reason for presenting a study of massless $\lambda\Phi^4$ theory
in some detail.

It is remarkable that despite the explicitly Hamiltonian structure of
the mean field equations, we observe quasi-dissipative features in the
evolution, in the sense that energy flows from the mean field $\langle
{\bf\Phi} \rangle$ into the fluctuating particle modes without
returning over times of physical interest. In other words, although
the underlying equations are fully time-reversal invariant, typical
evolutions beginning with energy concentrated in the mean fields are
{\em effectively irreversible}, at least over very long intervals of
time. This apparent irreversibility is quantifiable first, in terms of
the increase in particle entropy obtained by averaging over the
rapidly varying phases of the fluctuating modes, and second, by the
effective damping rate of the collective motion which we calculate by
a standard linear response analysis. Since the mean field Gaussian
approximation contains no collision terms, the particles interacting
with each other only through the mean fields, this effective
relaxation to a quasi-stationary (but non-thermal) state is a form of
collisionless damping, similar to Landau damping in non-relativistic
electromagnetic plasmas. We call this collisionless damping due to
effective loss of phase information in the fluctuating quantum modes
{\em dephasing}. Dephasing of the fluctuations is both an extremely
general and efficient mechanism for introducing effective dissipation
into the reversible Hamiltonian dynamics of mean field
evolution. Hence, even in this relatively simple collisionless
approximation to a quantum many-body theory one can begin to see how
irreversibility and the second law of thermodynamics emerge from a
consistent treatment of fluctuations in a closed Hamiltonian
system. Collisions which first appear at one order beyond the mean
field limit in the large $N$ expansion would be expected to make the
dephasing and dissipation found at lowest order still more efficient.

In addition to the effective dissipation of energy from the collective
plasmon mode to the fluctuations, dephasing is also responsible for
quantum decoherence, in the sense of suppression with time of the
off-diagonal elements of the Gaussian density matrix. The point is
that by going to the appropriate time-dependent number basis the
diagonal matrix elements of ${\bf\rho}$ are adiabatic invariants of
$H_{eff}$ and therefore slowly varying functions of time, while the
off-diagonal elements are very rapidly varying. These rapid phase
variations in the off-diagonal interference terms cancel out very
efficiently when the sum over the mode momentum is performed, or if
the phases in a given mode are averaged in time.  In either case, the
particle creation and interactions have the effect of bringing the
quantum system into what {\em effectively} looks more and more nearly
like a classical mixture in which the phase information in the
off-diagonal components may be discarded for most practical purposes
at late times. The symmetry breaking behavior of the density matrix
and effective disappearance of quantum interference between the two
classically allowed outcomes is the final result. An important
consequence of decoherence is the appearance of a diagonal effective
density matrix which may be sampled to generate smooth classical field
configurations. These configurations are free from spurious cut-off
dependences and can be used to address issues such as the generation
of topological defects in a nonequilibrium phase transition and the
modeling of individual events in heavy-ion collisions. 

The time-dependent scalar theory is an excellent theoretical laboratory 
for the study of general nonequilibrium phenomena such as decoherence 
and the quantum to classical transition quite aside from specific 
potential applications to phase transitions in many systems of physical 
interest. The detailed study of effective dissipation and decoherence 
in an explicit field theoretic example is the fourth major focus of the 
present work. Some of our results on the Hamiltonian nature of the 
evolution and on dephasing and decoherence have been reported earlier 
in condensed form \cite{ourprl}.
 
The paper is organized as follows. In the next Section we begin by
reviewing the large $N$ expansion of $\lambda{\bf\Phi}^4$ theory in
the real time effective action formulation. Then the Hamiltonian
structure of the equations of motion is exhibited and the effective
Hamiltonian and Gaussian density matrix are identified in Section
III. The static part of this effective Hamiltonian is the
nonequilibrium True Effective Potential (TEP), $U_{eff}$ which we
define and relate to the thermodynamic free energy $F$ in Section
IV. Numerical evolution of the actual equations of motion in both one
and three dimensions show clearly the difference with what might have
been inferred from $F$. In Section V
we identify the adiabatic particle number basis in which dissipation
through the increase of the effective particle entropy and decoherence
are described. We also present numerical evidence for the efficient
dephasing of the quantum modes in the time-dependent mean fields, and
show how it leads to typical classical configurations in the mixture
in which quantum interference effects have been washed out. In Section
VI we perform a linear response analysis of small perturbations away
from thermal equilibrium, as well as away from the non-thermal
stationary states found in the previous sections, compute the plasmon
damping rate, and compare it with the numerical results. There
is excellent agreement in the thermal case for both the plasmon
frequency and damping rate, whereas the non-thermal situation is more
complicated. We conclude in Section VII with a
discussion of our results and of their possible application to diverse
problems of interest in the real time dynamics of second order phase
transitions with a scalar order parameter. There are three Appendices,
the first on the renormalization of the energy and pressure of the
theory, the second cataloguing some mathematical properties of the
Gaussian density matrix used in the text, and the third containing the
details of the numerical methods used in solving the equations.

\section{The Large $N$ Effective Action}
\label{sec:level2}

The most direct method of deriving the equations of motion in the
large $N$ approximation is the method of the effective action, which
also has the advantages of exhibiting the covariance properties of the
theory explicitly and providing a general framework for the systematic
expansion in powers of $1/N$ to any desired order beyond the mean
field approximation.  Here we will present a short derivation of the 
effective action and refer the interested reader to the earlier work
\cite{largeN} for details of the derivation, which is quite
standard. The underlying $O(N)$ symmetric scalar field theory with
which we begin is described in $d$ space dimensions by the classical
action, 
\begin{eqnarray}
S_{cl} [\Phi, \chi] &=& \int dt d^dx\, L_{cl}[\Phi, \chi]\nonumber\\
&=& \int dtd^dx\,\left\{-\frac{1}{2} \Phi_i G^{-1}[\chi]\Phi_i +
{N\over \lambda} \chi\left({\chi \over 2}  +  \mu^2
\right)\right\}\nonumber\\ 
\label{lag}
\end{eqnarray}
where $i = 1, \ldots , N$ and
\begin{equation}
G^{-1}[\chi] \equiv -\sq + \chi\ ,
\label{Ginv}
\end{equation}
with the metric signature $(-,+,+,+)$. This form of the action is
equivalent to the more familiar Lagrangian density, 
\begin{eqnarray}
\tilde L_{cl}[\Phi] &=& -\frac{1}{2} (\partial_{\mu}
\Phi_i)(\partial^{\mu}\Phi_i) - V_{cl}\nonumber\\
&=& -\frac{1}{2} (\partial_{\mu}\Phi_i)(\partial^{\mu} \Phi_i)
-{\lambda\over 8N}\left(\Phi_i\Phi_i - {2N\mu^2\over\lambda}\right)^2
\nonumber\\   
\label{lag1} 
\end{eqnarray}
with the definition of the auxiliary field $\chi$ by
\begin{equation} 
\chi  = -\mu^2 + {\lambda\over 2N} \Phi_{i} \Phi_{i}~,  
\label{chi}
\end{equation}
since the two Lagrangians $L_{cl}$ and $\tilde L_{cl}$ are then equal
up to a surface term. The quartic coupling in the Lagrangian has been
taken to be $\lambda /N$ from the outset, rather than rescaling it
later by $1/N$ as is sometimes done \cite{bcg}.

If the parameter $\mu^2 > 0$, then this classical Lagrangian describes
spontaneous symmetry breaking since the minimum of the classical
potential $V_{cl}$ occurs at
\begin{equation}
\Phi_{i\,vac} = \delta_{iN}\, v_0 \,\sqrt {N}\quad ; \qquad v_0 = 
\sqrt{{2\over \lambda}}  \ \mu
\label{ssb}
\end{equation}
rather than at zero. The second derivative of the classical potential
at its minimum is
\begin{eqnarray}
{\partial^2 V_{cl}\over \partial\Phi_i\partial\Phi_j}&=& {\lambda
\over 2N} \left(\Phi_k\Phi_k - {2N\mu^2\over\lambda}\right)\delta_{ij}
+ {\lambda \over N}\Phi_i\Phi_j \nonumber\\
&=& \left( \chi\delta_{ij} + {\lambda \over
N}\Phi_i\Phi_j\right)\nonumber\\ 
&=& \left\{ \begin{array}{ll}0,\ & i \ {\rm or}\ j= 1,\dots, N-1\\
\lambda v_0^2,\ & i=j=N\ . 
\end{array} \right. 
\label{Vcl}
\end{eqnarray} 
At this minimum the $O(N)$ symmetry is spontaneously broken, $\chi =0$
and there are $N-1$ massless modes. Small oscillations in the
remaining $i=N$ direction describe a massive mode with bare mass equal
to $\sqrt 2 \mu = \sqrt\lambda v_0$. In this standard way of
describing the symmetry breaking, one direction is singled out and its
vacuum expectation value is scaled with $\sqrt N$ as in
(\ref{ssb}). It is clear that in the large $N$ limit the effects of
this single degree of freedom are down by $1/N$ compared to the $N-1$
massless modes which dominate the dynamics. As we shall see in Section
VII the effects of the single massive degree of freedom are recovered in
the behavior of the $\chi$ correlator at late times.

Passing now to the quantum theory, the large $N$ mean field
approximation is obtained by retaining only the leading terms in a
systematic expansion of Feynman diagrams in $1/N$. To lowest order,
the quantum effective action is 
\begin{equation}
{\cal S}_{eff} [\phi, \chi] = N S_{cl}[\phi, \chi] + N
{\frac{i\hbar}{2}}  \ {\rm Tr} \ln G^{-1}[\chi]\ . 
\label{Seff}
\end{equation}
We use units in which the speed of light and Boltzmann's constant are
unity, $c=k_B=1$, but we retain $\hbar$ in order to exhibit the
semiclassical nature of the large $N$ limit. The effective action,
${\cal S}_{eff}$ generates the proper vertices of the $O(N)$ symmetric
$\lambda\Phi^4$ theory, correct to leading order in $1/N$. It is just
$N$ times the classical action for a {\em single} component $\phi$
field, plus the self-consistent (or resummed) one-loop correction
represented by the last Tr $\ln G^{-1}[\chi]$ term, where $\chi$ is
the mass squared of the propagator $G$ appearing in the loops.  The
quantity $\phi \equiv \langle {\bf \Phi} \rangle$ is the mean value of
the quantum field $\bf \Phi$. Thus, in this description we effectively
return to a single component $\Phi^4$ theory. The role of the large
$N$ limit is to justify the neglect of the single massive degree of
freedom relative to the $N-1$ massless Goldstone bosons in the
propagator $G[\chi]$ of the ${\bf\Phi}$ field. However, the massive
degree of freedom determines the $\chi$ propagator which controls the
plasmon oscillations discussed in Section VI.
 
Before proceeding it is worth pausing at this point to compare and
contrast the large $N$ effective action (\ref{Seff}) with two
different but related approximations.  The simplest is the standard
one-loop approximation on the single component theory, which is
obtained by expanding the ${\bf\Phi}$ field about its mean value
$\phi$. The quantum effective action in this approximation is
\begin{equation}
{\cal S}_{1-loop}[\phi] = \tilde{\cal S}_{cl}[\phi] + 
{\frac{i\hbar}{2}}  \ {\rm Tr} \ln
G^{-1}\left[V_{cl}^{\prime\prime}\right] 
\label{onel}
\end{equation}
where the effective mass appearing in the two-point function is 
\begin{equation}
V_{cl}^{\prime\prime} = {\partial^2 V_{cl}\over \partial
{\bf\Phi}^2}\bigg\vert_{_{\Phi =\phi}} = -\mu^2 + {3\lambda\over 2}
\phi^2 \ .  
\end{equation}
The simple one-loop approximation includes only the one-loop
self-energy diagram of Fig. 1, without any further resummation of
diagrams.
 
\vspace{.4cm}
\epsfxsize=4cm
\epsfysize=2cm
\centerline{\epsfbox{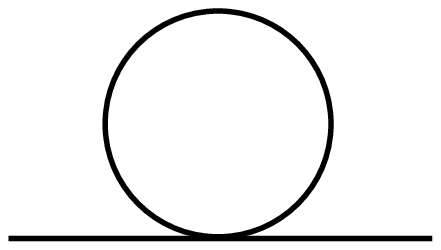}}
\vspace{.35cm}
{FIG. 1. {\small{The one-loop self-energy diagram which
contributes to the simple one-loop effective action of Eqn. 
(\ref{onel}).}}}\\ 
 
In contrast, the approximation usually called the Hartree or Gaussian
approximation in the literature is closely related to the large $N$
approximation in that the second derivative of the classical potential
$V_{cl}^{\prime\prime}$ in (\ref{onel}) is replaced by its Gaussian
mean value, {\em viz.} 
\begin{equation}
{\cal S}_{Hartree}[\phi, M^2] = 
{\cal S}_{cl} + {\frac{i\hbar}{2}}  \ {\rm Tr} \ln G^{-1}[M^2] 
\label{meanf}
\end{equation}
with
\begin{equation}
M^2_{ij}\equiv \langle {\partial^2 V_{cl}\over \partial {\bf \Phi_i}
\partial{\bf \Phi_j}}\rangle = \langle -\mu^2 +
{\lambda\over 2N} {\bf\Phi}_{k} {\bf\Phi}_{k}\rangle\delta_{ij} 
 + {\lambda\over N}\langle{\bf\Phi}_i {\bf\Phi}_j\rangle\,.
\label{secd}
\end{equation}
In the Hartree approximation the classical action should be expressed
in terms of the set of variational parameters, $\phi$ and $M^2_{ij}$.
In both the Hartree and large $N$ approximations the expectation
values of bilinears obey the factorization condition,
\begin{equation}
\langle {\bf\Phi}_i(x) {\bf\Phi}_j(y) \rangle =
\langle{\bf\Phi}_i(x)\rangle\,\langle{\bf\Phi}_j(y)\rangle
-i\delta_{ij} G(x,y)\,, 
\label{factor}
\end{equation}
where which two-point function $G$ appears here depends on the
application. In any case, this factorization implies that the last
term in (\ref{secd}) above is suppressed by a factor of $1/N$ compared
to the diagonal $\delta_{ij}$ term.  The difference between the
Hartree and large $N$ effective action is just this last term in
(\ref{secd}). In the large $N$ approximation it is discarded at
leading order as in (\ref{Vcl}) and (\ref{Seff}), whereas in the
Hartree approximation of Eqns. (\ref{meanf}) and (\ref{secd}) it is
retained. An important consequence of this difference between the two
approximation schemes is that unlike the Hartree approximation, the
large $N$ effective action is the leading term in a series in a
well-defined expansion in powers of $1/N$. Thus, it is possible to
improve on the large $N$ limit in a systematic way by retaining higher
order terms in this series. In contrast, the Hartree approximation is
simply a variational ansatz.

In addition to providing a practical expansion technique the large $N$
expansion has the significant conceptual advantage over the Hartree
approximation of placing mean field theory in its proper context with
respect to other systematic methods in nonequilibrium statistical
mechanics. In fact, the systematic truncation of the Schwinger-Dyson
hierarchy of connected $2n$-point functions by the large $N$ expansion
in quantum field theory is the precise analog of the truncation of the
BBGKY hierarchy of $n$-particle distribution functions in
nonequilibrium classical statistical mechanics. The existence of an
energy-conserving expansion parameter in $1/N$ which preserves all the
relevant symmetries of the underlying field theory is a powerful
technical tool for development of the quantum theory of nonequilibrium
processes from first principles in a systematic way.
 
The effective mass which appears in the large $N$ effective action is 
\begin{eqnarray}
\chi (x) &=& \langle -\mu^2 + {\lambda \over 2N} {\bf \Phi}_i(x){\bf
\Phi}_i(x)\rangle 
\nonumber\\
&=& -\mu^2 + {\lambda \over 2}\phi^2(x) - {i\lambda\over 2}
G[\chi](x,x) 
\label{ceq}
\end{eqnarray}
which will be recognized as just the expectation value of the
(operator) definition of the auxiliary field in (\ref{chi}) upon using
the factorization condition (\ref{factor}). Since $G[\chi]$ itself
depends on $\chi$ through the definition,
\begin{eqnarray}
G^{-1}[\chi]\circ G[\chi] &=& 1 \qquad {\rm or} \nonumber\\
\left( -\sq + \chi\right) G[\chi] (x,y) &=& \delta^{d+1} (x-y)
\label{invG}
\end{eqnarray}
in position space, (\ref{ceq}) is a nonlinear integral equation for
the auxiliary field $\chi$. The Feynman diagrams that contribute in
both the large $N$ limit and the Hartree approximation are the sum
over all daisy and superdaisy diagrams represented in Fig. 2, with the
only difference that the lines correspond to the propagator $G[\chi]$
in one case (large $N$) and $G[M^2]$ in the other case (Hartree). This
is just the resummation of self-energy diagrams required by a
renormalization group analysis, and in fact, $\chi$ is a
renormalization group invariant. In the large $N$ limit the
fundamental excitations are the $N-1$ Goldstone modes, whose
masslessness is fixed by the $O(N)$ symmetry which the $1/N$ expansion
respects order by order. In contrast, since the Hartree approximation
is not a systematic expansion in any small parameter it is not clear
{\em a priori} how to renormalize $M^2$, nor whether it has any
renormalization group invariant meaning.  The result is that although
the two approximations are very similar in some respects and may be
handled by the same techniques, the physics they describe is really
quite different. Which approximation is more reliable depends very
much on the application, and in particular whether or not the massless
Goldstone modes of the large $N$ limit actually play the lead role in
the physics we wish to describe.

\vspace{.4cm}
\epsfxsize=3cm
\epsfysize=2cm
\centerline{\epsfbox{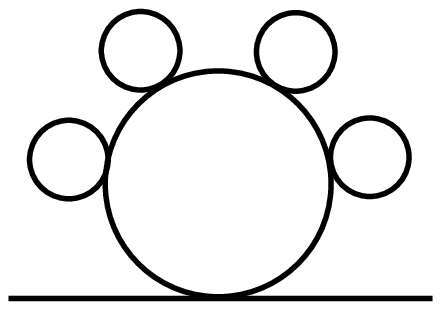}}
\vspace{.35cm}
\epsfxsize=3cm
\epsfysize=2cm
\centerline{\epsfbox{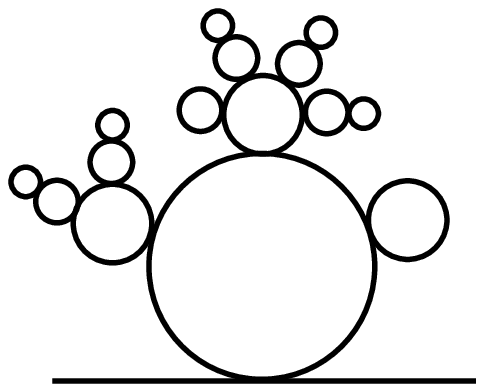}}
\vspace{.35cm}
{FIG. 2. {\small{Typical daisy and superdaisy diagrams
included in the Hartree or leading order large $N$ approximation
to the effective action.}}}\\ 

In the real time formulation, the effective action of (\ref{Seff})
becomes the starting point for all further analysis of the large $N$
dynamics. From the construction of $\cal S$ in (\ref{Seff}) as a
Legendre transform in the presence of external sources, it follows
that its first variation with respect to the independent mean fields
$\phi$ or $\chi$ is proportional to the sources for these fields. In
the absence of external sources, these variations yield the equations
of motion for the mean fields,
\begin{equation}
{1\over N}{\delta {\cal S}_{eff} [\phi, \chi]\over \delta \phi} =
G^{-1}[\chi]\phi(x)= \left(-\sq +\chi (x)\right)\phi (x) = 0\,,  
\label{feq}
\end{equation}
and
\begin{eqnarray}
{1\over N}{\delta {\cal S}_{eff} [\phi, \chi]\over \delta \chi} &=&
{\left(\chi(x) + \mu^2\right)\over \lambda} - {1\over 2} \phi^2(x)
+{i\hbar\over 2} G[\chi](x,x)\nonumber\\
&=& 0~,  
\label{chieq}
\end{eqnarray}
which just recovers (\ref{ceq}) to this order. Unlike the equation
(\ref{feq}) for $\phi$, (\ref{chieq}) involves no derivatives and is
therefore an equation of constraint (or gap equation), rather than an
equation of motion for an independent propagating degree of
freedom. That $\chi$ is nevertheless a useful indicator of symmetry
breaking should be clear from the spacetime independent form of
Eqn. (\ref{feq}),
\begin{equation}
\chi \phi  = 0\ , \qquad (\partial_{\mu}\phi = 0) 
\label{static}
\end{equation}
which tells us that either $\phi$ or $\chi$ (or both) must vanish in a
uniform, stationary state. The case, $\phi =0$ is the $O(N)$ symmetric
state with generally positive $\chi$, while the case of non-vanishing
$\phi$ is the spontaneously broken state in which $\chi=0$ is the
vanishing Goldstone boson mass. This is an explicit realization of
Goldstone's theorem, which is respected by the large $N$
approximation.

In the following the explicit functional dependence of $G[\chi]$ and
its inverse on the mean field $\chi$ will be suppressed, and we adopt
the simpler notation $G(x,y)$ or still more briefly $G$ hereafter.

The important difference of the large $N$ equations from the purely
classical ones is the last term in (\ref{ceq}) involving $\hbar
G$. Whereas the spontaneous symmetry breaking solution $\chi=0$ can be
achieved only when $\phi = \pm v_0 \equiv \pm \sqrt{{2/\lambda}}
\mu$ classically, in the large $N$ approximation the two-point
function $G$ contributes at the same order and can even dominate the
mean field $\phi$ in Eqn. (\ref{ceq}). Since $G$ itself depends on
$\chi$ this additional term also has the effect that the $\chi$ field
can undergo nonlinear collective plasmon-like oscillations as we shall
see explicitly in Section VII. Notice also that there is nothing to
prevent $\chi$ from being negative at some times which allows us to
explore the dynamics of the unstable spinodal region of Fig. 3. The
spinodal at a given temperature is the region where the second
derivative of the potential becomes negative. In the large $N$
approximation, this corresponds to $\chi < 0$.

\vspace{1cm}
\epsfxsize=6.5cm
\epsfysize=4.5cm
\centerline{\epsfbox{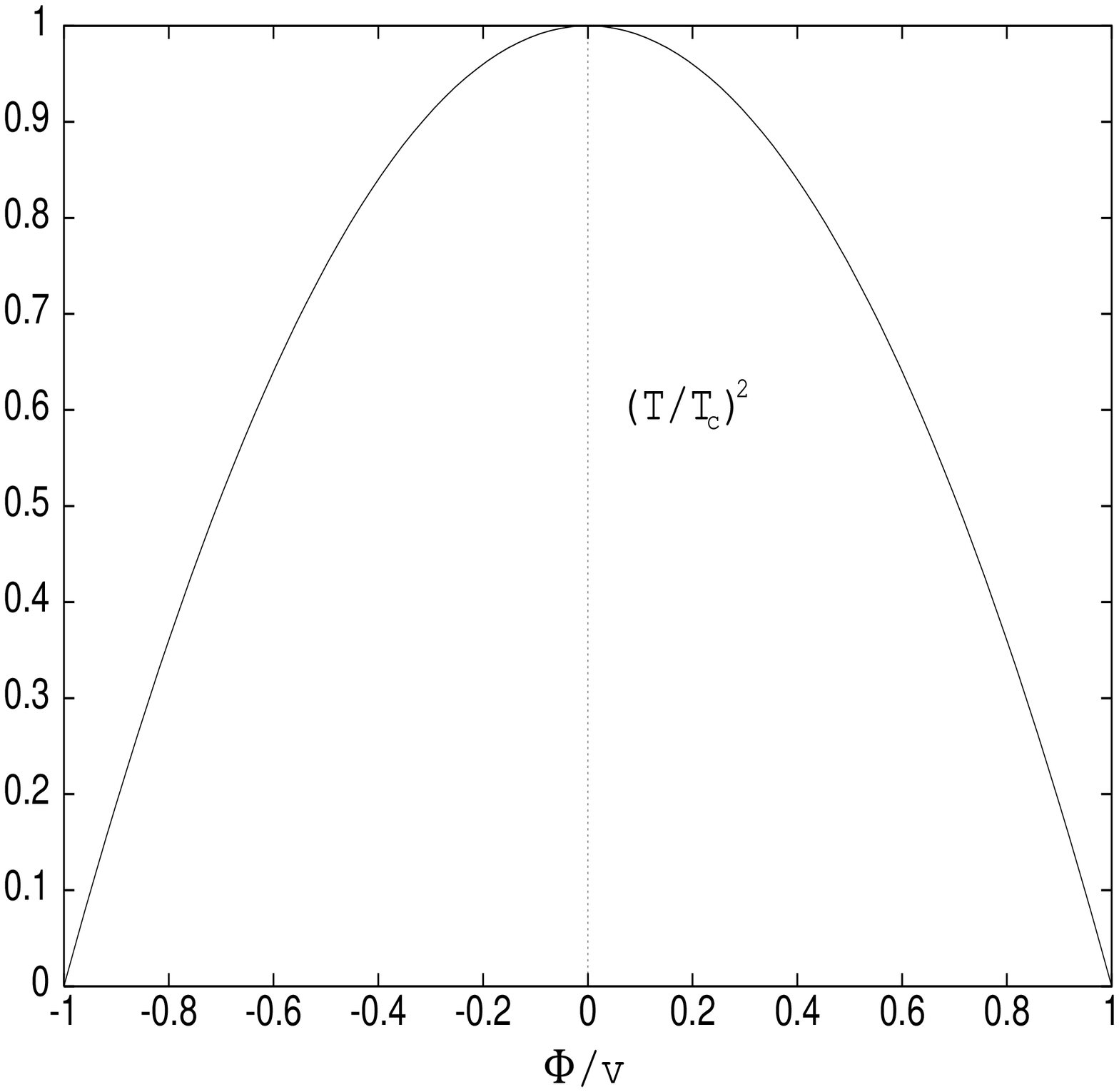}}
{FIG. 3. {\small{The spinodal value of $\phi$, $\phi_T$ is the value
of $\phi$ where $\chi$ first becomes negative. It satisfies the
equation $(\phi_T/ v)^2 = 1-({T/T_c})^2$ shown above (see
Eqn. (\ref{phiT}) below).}}}\\  

In this paper we shall be concerned only with spatially homogeneous
mean fields $\phi = \phi (t)$ and $\chi = \chi (t)$. It is not
difficult to treat spatially inhomogeneous mean fields by the same
methods, but as the homogeneous case is simpler and already contains
much of the essential physics, we restrict ourselves to that case in
this paper. When $\phi$ and $\chi$ are functions only of $t$, then the
two-point Green's function $G$ is a function only of the spatial
difference ${\bf x} -{\bf x'}$ and it is useful to introduce the
Fourier transform of the corresponding Wightman function $G_>$ (or
$G_<$),
\begin{eqnarray}
G_> (t, {\bf x} ; t' ,{\bf x}^{\prime}) &=& - G_<^*(t, {\bf x}; t',
{\bf x}^{\prime})\nonumber\\ 
&=& \int [d {\bf k}] e^{i {\bf k} \cdot ({\bf x} - {\bf x}^{\prime})} 
G_> (t, t' ; {\bf k}) 
\label{four}
\end{eqnarray}
with
\begin{equation}
[d {\bf k} ]   \equiv  {d^d {\bf k} \over (2 \pi)^d}\,.
\label{meas}
\end{equation}
In fact, $G_>$ is a function only of $k \equiv |{\bf k}|$ by
rotational symmetry of the spatially homogeneous state. The Wightman
functions are solutions of the source-free wave equation $G^{-1}\circ
G_> = G^{-1}\circ G_< = 0$. In terms of complex mode functions
$f_k(t)$ obeying 
\begin{equation}
\left[{d^2 \over dt^2} +  k^2 + \chi (t)\right] f_k (t) =  0~,
\label{modefn}
\end{equation}
it is convenient to express the Wightman functions as
\begin{equation}
G_> (t, t' ; k) = if_k (t) f_k^*(t') (N(k) + 1) + if_k^* (t) f_k(t')
N(k)\,, 
\label{wigh}
\end{equation}
where $N(k)$ (not to be confused with the $N$ of large $N$) is an
arbitrary time-independent function of $k$. It carries the
interpretation of particle number density in the general spatially
homogeneous initial state in the basis specified by the mode functions
$f_k$, and is arbitrary except for the requirement of being finitely
integrable with respect to the integration measure (\ref{meas}).
 
Explicitly, if the original quantum field ${\bf\Phi}$ is expanded
about its mean value in terms of the quantum modes $f_k$ and the
corresponding plane wave creation and destruction operators $a_{\bf
k}^{\dagger}$ and $a_{\bf k}$,
\begin{eqnarray}
{\bf\Phi} (t, {\bf x}) &=& \langle{\bf\Phi}\rangle + \delta{\bf\Phi}
(t,{\bf x}) \nonumber\\ 
&=& \phi (t) + L^{-{d\over 2}} \sum_{\bf k}\left( e^{i {\bf k} \cdot
{\bf x}} f_k (t)a_{\bf k} +  e^{-i {\bf k} \cdot {\bf x}} f_k^*
(t)a_{\bf k}^{\dagger}\right)\nonumber\\
\label{pexp}
\end{eqnarray}
in a cubical box of finite length $L$, then 
\begin{equation}
N(k=|{\bf k}|) = \langle a_{\bf k}^{\dagger} a_{\bf k}\rangle
\end{equation}
is just the expectation value of the particle number density in this
basis.  It is also a function only of $k$ by isotropy of the spatially
homogeneous state. The Wightman functions are given then by the usual
expressions,
\begin{eqnarray}
\hbar G_> (t, {\bf x}; t^{\prime}, {\bf x}^{\prime}) &=& i \langle 
\delta{\bf\Phi} (t,{\bf x}) \delta{\bf\Phi} (t^{\prime}, {\bf
x}^{\prime})\rangle\,, 
\nonumber\\
\hbar G_< (t, {\bf x}; t^{\prime}, {\bf x}^{\prime}) &=& i \langle 
\delta{\bf\Phi} (t^{\prime}, {\bf x}^{\prime})\delta{\bf\Phi} (t,{\bf
x})\rangle\,,  
\end{eqnarray}
and the condition $G_> = -G^*_<$ follows immediately from the
hermiticity of the underlying quantum field $\delta{\bf\Phi}$ in
(\ref{pexp}). The commutation relation
\begin{equation}
[a_{\bf k}, a_{\bf k^{\prime}}^{\dagger}]=\delta_{{\bf k},{\bf
k}^{\prime}} 
\end{equation}
implies that the complex mode functions should be chosen to satisfy
the Wronskian condition, 
\begin{equation}
f_k {df_k^*\over dt} - f_k^{*} {df_k\over dt} = i \hbar \,,
\label{wron}
\end{equation}
in order for the quantum field operator ${\bf\Phi}$ to obey the usual
canonical commutation relation,
\begin{equation}
\langle [{\bf\Phi} (t, {\bf x}), {\partial {\bf\Phi}\over \partial
t}(t,{\bf x'})] 
\rangle = i\hbar \delta^d ({\bf x} - {\bf x}^{\prime})\,.
\label{ccm}
\end{equation}
The normalization of the time-independent Wronskian condition
(\ref{wron}) is the only place where the constant $\hbar$ enters the
mean field equations, which otherwise are quite classical in their
time evolution dynamics.
 
We have not written the {\em a priori} possible bilinear terms
$f_kf_k$ or $f_k^*f_k^*$ in (\ref{wigh}), since they can always be
absorbed into a redefinition of $f_k$ and $N(k)$ under the
transformation,
\begin{equation}
f_k \rightarrow \cosh\gamma_k\ e^{i\theta_k + i\psi_k} f_k 
+ \sinh \gamma_k\ e^{i\theta_k-i\psi_k} f_k^*
\label{Bogl}
\end{equation}
without affecting the mode equation (\ref{modefn}) or Wronskian
condition (\ref{wron}). This is equivalent to making a Bogoliubov
transformation which sets to zero the expectation values of the pair
densities $\langle a_{\bf k} a_{\bf k}\rangle = \langle
a^{\dagger}_{\bf k} a^{\dagger}_{\bf k}\rangle = 0$. Hence there is a
natural $SU(1,1)$ Heisenberg group structure (for each $\bf k$)
inherent in the leading order large $N$ equations.
 
In the gap equation (\ref{ceq}) the coincidence limit of the Green's
function $G(x,x)$ appears. The coincidence limit of either $G_>, G_<$
or the Feynman propagator $G$ are all identical, so we obtain from
Eqns. (\ref{wigh}) and (\ref{four}) the coincidence limit of any of
the Green's functions in the form of an integral over $\bf k$ of the
Fourier mode functions,
\begin{eqnarray}
\hbar G_>(t, {\bf x}; t, {\bf x}) &=& \hbar G_<(t,{\bf x}; t,{\bf x})
\nonumber\\ 
&=& i\int [d {\bf k} ]\, \vert f_k(t)\vert^2 \ (2 N(k) + 1 )\,.
\label{cooi}
\end{eqnarray}
Since the mode functions obeying (\ref{modefn}) and (\ref{wron})
behave as  
\begin{equation}
f_k(t) \rightarrow  {\sqrt{\hbar\over 2 \omega_k (t)}}
\exp \left(-i\int^t dt'\omega_k (t')\right)
\label{zeroth}
\end{equation}
for large $k$, where
\begin{equation} 
\omega_k (t) \equiv \left(k^2 +\chi(t)\right)^{1\over 2}\,,
\end{equation} 
the integral in (\ref{cooi}) is quadratically divergent in $d=3$
spatial dimensions. Introducing an explicit momentum cut-off $\Lambda$
and performing the angular $\bf k$ integrations, the gap equation
(\ref{ceq}) may be rewritten in the form,
\begin{equation} 
\chi(t) = -\mu_{\Lambda}^2 + {\lambda_{\Lambda} \over 2} \phi^2(t) +
{\lambda_{\Lambda}\over 4\pi^2} \int_0^{\Lambda}\, k^2dk\ 
\vert f_k(t)\vert^2 \ \sigma_k 
\label{gap}
\end{equation}
for $d=3$, where we have defined the notation,
\begin{equation}
\sigma_k \equiv 2 N(k) + 1\,.
\label{sigdef}
\end{equation}
The fact that the bare parameters $\mu_{\Lambda}$ and
$\lambda_{\Lambda}$ must depend on the cut-off in order to render the
equations independent of $\Lambda$ in the end has been exhibited
explicitly as well.  The quadratic divergence in $d=3$ is the
divergence of the one-loop self energy diagram in Fig. 1 and is
absorbed into the bare mass parameter $\mu_{\Lambda}$. One convenient
way to effect this mass renormalization is to evaluate (\ref{gap}) in
the time-independent spontaneously broken vacuum, $\chi = 0, \phi =
v$, thereby absorbing the quadratic divergence into the relation
between the bare and physical expectation value of the field. In this
way we eliminate $\mu_{\Lambda}^2$ and obtain
\begin{eqnarray}
\chi(t) &=& {\lambda_{\Lambda} \over 2} \left(\phi^2(t) - v^2\right)
\nonumber\\ 
&& + {\lambda_{\Lambda}\over 4\pi^2} \int_0^{\Lambda}\, k^2dk\
\left\{\vert f_k(t)\vert^2 \ \sigma_k - {\hbar\over 2k}\right\} \,, 
\label{cren}
\end{eqnarray}
which is free of quadratic divergences. 

The remaining logarithmic divergence in the mode integral of
(\ref{cren}) is removed by the logarithmic coupling constant
renormalization in the usual way, {\em i.e.}
\begin{equation}
\lambda_{\Lambda} = Z^{-1}_{\lambda}(\Lambda , m)\, \lambda_R (m^2)
\label{lren1}
\end{equation}
with 
\begin{eqnarray}
Z_{\lambda}(\Lambda , m) &=& 1 - {\hbar\over 32\pi^2} \lambda_R (m^2)
\ln \left({\Lambda^2\over m^2}\right)\nonumber\\
&=& \left[1 + {\hbar\over 32\pi^2} \lambda_{\Lambda}\ln
\left({\Lambda^2\over m^2}\right)\right]^{-1} 
\label{lren2}
\end{eqnarray}
and $\lambda_R (m^2)$ the renormalized ${\bf\Phi}^4$ coupling defined
at some finite mass scale $m^2$. By dividing both sides of
(\ref{cren}) by $\lambda_{\Lambda}$ and using Eqns. (\ref{lren1}) and
(\ref{lren2}), it is straightforward to verify from the large $k$
behavior of the integrand that the logarithmic dependence on $\Lambda$
of the integral in (\ref{cren}) is cancelled by the logarithm in
(\ref{lren2}). Thus, the resulting equation for $\chi (t)$ is
independent of $\Lambda$ for $\Lambda$ large, and $\chi$ is in fact a
renormalization group invariant physical mass squared of the
theory. The condition that $Z_{\lambda}>0$ prevents us from taking the
cut-off strictly to infinity with $\lambda_R > 0$ fixed, for otherwise
$Z_{\lambda}$ from (\ref{lren2}) would eventually become negative and
the theory would become unstable. This is just a reflection of the
Landau ghost instability of scalar ${\bf\Phi}^4$ field theory, and
means that the theory can be sensible and non-trivial only as an
effective field theory equipped with a large but finite cut-off,
$\Lambda$. This presents no problem in practice as long as $\Lambda$
is large enough that the physical time evolution, plasma oscillations,
damping, {\em etc.} occur on time scales much greater than
$\Lambda^{-1}$. In that case, the evolution is
numerically quite insensitive to the value of $\Lambda$ over a very
wide range, provided $\lambda_R$ is not too large \cite{largeN}.
 
There is no wave function or $\phi$ renormalization at lowest order in
large $N$ so no further renormalization is required in $d=3$ to this
order, and Eqns. (\ref{feq}), (\ref{modefn}), and (\ref{cren})
together with the constraint on the initial data (\ref{wron}) specify
a well-defined closed system of evolution equations for the mean
fields $\phi$ and $\chi$ in interaction with the fluctuations $f_k$.
Let us emphasize again that these equations differ from the purely
classical tree level approximation or the simple one-loop
approximation in that the last term of (\ref{gap}) or (\ref{cren})
couples the fluctuations self-consistently and nonlinearly back on
the time-dependent mass gap function $\chi (t)$, corresponding to the
full sum of daisy and superdaisy diagrams in Fig. 2.

In $d=1$ space dimensions the integral in (\ref{ceq}) is only
logarithmically ultraviolet divergent and (\ref{cren}) is already
well-defined and finite, without any $\lambda$ renormalization.
However, in lower spatial dimensions the small $k$ or {\it infrared}
behavior of the integral becomes more delicate, and must be treated
carefully.

\section{The Effective Hamiltonian and Density Matrix}
\label{sec:level3}

The presentation of the large $N$ equations of motion of the previous
section was based on the functional method, in which the extremization
of the effective action ${\cal S}_{eff}$ in (\ref{feq}) and
(\ref{chieq}) takes the place of the usual Euler-Lagrange variational
principle of classical mechanics. The equations so obtained are for
the mean values of the field operators and their two-point Green's
functions which describe fluctuations about the mean fields. This
immediately raises a question: Is there a corresponding Hamilton form
of the variational principle for an effective large $N$ Hamiltonian
involving both the mean fields and their fluctuations? In addition one
would like to know: To what distribution of field amplitudes in the
Schr\"odinger wave function (or density matrix) do the large $N$
equations for the one and two-point functions correspond? It is to
these questions that we turn in this section. Answering them will lead
directly to the true effective potential of nonequilibrium large $N$
mean field theory.

Let us begin by consideration of the case of $d=0$ spatial dimensions,
{\it i.e.} quantum mechanics. The generalization to higher $d$ will
turn out to be straightforward. For $d=0$ the Lagrangian $L_{cl}$ or
$\tilde L_{cl}$ of (\ref{lag}) or (\ref{lag1}) is that of an $N$
component anharmonic oscillator. If $N=1$ it reduces to the usual
anharmonic double well oscillator. In the $d=0$ case there is no $\bf
k$ index to be integrated and the equations derived in the last
section become simply
\begin{eqnarray}
&&\left({d^2\over dt^2} + \chi(t)\right)\phi (t)= 0\ ,\nonumber\\
&&\chi(t) = {\lambda\over 2} \left(\phi^2 (t) + \xi^2 (t) -
v_0^2\right) , 
\qquad (d=0) \label{zerod}
\end{eqnarray}
where we have introduced the notation,
\begin{equation}
\xi^2 (t) \equiv \sigma\,  \vert f(t)\vert^2 \equiv (2N + 1)\,\vert
f(t)\vert^2 
\label{xidef}
\end{equation}
in terms of the expectation of the number operator, $N=\langle
a^{\dagger} a \rangle$. We recognize that
\begin{equation}
\xi^2 (t) = \langle {\bf\Phi} (t) {\bf\Phi} (t)\rangle - \phi^2 (t)  =
\langle \left({\bf\Phi}(t) - \phi(t)\right)^2\rangle  
\end{equation}
is just the quadratic variance of the quantum variable ${\bf\Phi}(t)$
from its mean value $\phi (t)$. By differentiating this relation and
defining $\eta \equiv \dot\xi$ we find
\begin{eqnarray}
2\xi(t)\eta (t) &=& \langle \left(\dot{{\bf\Phi}} (t) {\bf\Phi} (t) + 
{\bf\Phi} (t) \dot{{\bf\Phi}} (t) - 2\phi(t)
\dot{\phi}(t)\right)\rangle\nonumber\\ 
&=& 2\sigma {\rm Re}(\dot{f}f^*)\,.
\label{etadef}
\end{eqnarray}
The advantage of defining these quantities will become apparent from
the role they play in the physical interpretation of the equations of
motion (\ref{zerod}) in the Schr\"odinger picture.  Indeed, by
differentiating (\ref{etadef}) again and using both the equation of
motion (\ref{modefn}) (with $k=0$ in $d=0$ spatial dimensions) and the
Wronskian condition (\ref{wron}) for $f(t)$, we find
\begin{equation}
\dot{\eta} = \ddot{\xi} = -\chi\,\xi - {\dot\xi^2\over \xi} + {\sigma
\vert \dot f\vert^2 \over \xi} = - \chi\,\xi + {\hbar^2\sigma^2\over 4
\xi^3} 
\label{etaeqn}
\end{equation}
in the new notation. This equation of motion together with
(\ref{zerod}) are just Hamilton's equations,
\begin{eqnarray}
\dot p &=& -{\partial H_{eff} \over \partial \phi}\,,\nonumber\\
\dot\eta &=& -{\partial H_{eff} \over \partial \xi}
\end{eqnarray}
for the effective {\em two}-dimensional classical Hamiltonian,
\begin{eqnarray}
H_{eff}(p, \phi ;\eta, \xi; \sigma) &=& {1\over 2}(p^2 + \eta^2)
\nonumber\\
&&+ {\lambda \over 8}(\phi^2 + \xi^2 - v_0^2)^2 + {\hbar^2\sigma^2 
\over 8\xi^2} 
\label{Heff}
\end{eqnarray}
with
\begin{eqnarray}
p &\equiv& \dot{\phi}= {\partial H_{eff} \over \partial p}~,\nonumber
\\  
\eta &\equiv &\dot{\xi} = {\partial H_{eff} \over \partial \eta}
\end{eqnarray}
the canonical momenta conjugate to the two generalized coordinates
$\phi$ and $\xi$. A different but equivalent set of canonical
variables was discussed in Ref. \cite{semi}.

Hence by the simple change of notation in (\ref{xidef}) and
(\ref{etadef}) we have recognized that the large $N$ equations for the
quantum anharmonic oscillator with classical potential
\begin{equation}
V_{cl}(\Phi) = {\lambda \over 8N}\left(\sum_{i=1}^N\Phi_i\Phi_i -
v_0^2\right)^2\ , 
\end{equation}
derived by the effective action technique of the last section are
precisely equivalent to Hamilton's equations for the effective
Hamiltonian (\ref{Heff}). This answers the first question we posed at
the beginning of this section in the affirmative, at least for the
case of $d=0$.

An immediate corollary of the Hamiltonian structure in the extended
phase space $(p, \phi ;\eta, \xi)$ is that the large $N$ evolution
equations are energy conserving with $H_{eff}$ the value of the
conserved energy. It is also interesting to note that $\hbar\sigma$ is
a constant of the motion which enters the effective potential
\begin{equation}
U_{eff} (\phi, \xi; \sigma) = {\lambda \over 8}(\phi^2 + \xi^2 -
v_0^2)^2 + {\hbar^2\sigma^2 \over 8\xi^2} 
\end{equation}
together as a single ``centrifugal barrier'' term, whose effect is to
repel the variance $\xi$ away from zero. The physical and mathematical
analogy to an angular momentum barrier is made even stronger by the
fact that the three symmetric bilinears $aa$, $a^{\dagger}a^{\dagger}$
and $aa^{\dagger} + a^{\dagger}a$ generate the Lie algebra of
$su(1,1)$ or $so(2,1)$ which is the non-compact version of the
ordinary angular momentum algebra, $su(2)$ or $so(3)$, and moreover,
the Casimir invariant of this rank one Lie algebra is exactly
$\hbar^2\sigma^2/4$, in the standard normalization. The corresponding
Lie group is just the three parameter group of homogeneous linear
Bogoliubov transformations (\ref{Bogl}).

To answer the second question posed at the beginning of this section
let us recall that both the time-dependent Hartree (TDH) and large $N$
equations have been studied in the Schr\"odinger representation,
and they are known to correspond to a Gaussian trial wave function
ansatz \cite{Gauss}. Indeed, it is straightforward to verify that the
Gaussian ansatz for the normalized pure state Schr\"odinger wave
function,
\begin{eqnarray}
\Psi (x ; t)&\equiv& \langle x \vert \Psi (t)\rangle\nonumber\\
&=& (2\pi\xi^2(t))^{-{1\over 4}} \exp \left\{i {p(t)x\over
\hbar}\right. 
\nonumber\\
&& -\left.\left({1\over 4 \xi^2(t)} + i {\eta (t) \over 2 \hbar \xi
(t)} \right)\left(x - {\phi}(t)\right)^2\right\}
\label{wavefn}
\end{eqnarray}
obeys the expectation value of the Schr\"odinger equation,
\begin{eqnarray}
&&\langle \Psi (t)\vert\,\left\{ -{\hbar^2\over 2}\sum_{i=1}^N
{\partial^2 \over \partial {\bf\Phi}_i\partial{\bf\Phi}_i} +
V({\bf\Phi})\right\}\,\vert \Psi (t)\rangle \nonumber\\
&&= i\hbar\langle \Psi (t)\vert\, {\partial\over\partial t}\,\vert
\Psi (t)\rangle   
\end{eqnarray}
in the coordinate representation where, with ${\bf P}$ as the
canonical momentum,
\begin{eqnarray}
&&\langle \Psi (t)\vert {\cal O}\left( {\bf\Phi} , {\bf P}
\right)\vert \Psi (t)\rangle \nonumber\\
&&= \int_{-\infty}^{\infty} dx\, \Psi^* (x ; t) {\cal O}\left( x ,
-i\hbar {\partial \over \partial x }\right) \Psi (x ; t)\ , 
\end{eqnarray}
provided the large $N$ limit is taken and the equations of motion
(\ref{zerod}) and (\ref{etaeqn}) are satisfied for $\sigma = 1$.
Thus, the Gaussian ansatz (\ref{wavefn}) is a special case of the
general large $N$ equations of motion, where $\xi$ and $\eta$ are
related to the real and imaginary parts respectively of the Gaussian
covariance.

In earlier work \cite{init} it had been recognized that the Gaussian
ansatz for the Schr\"odinger wave function(al) imposed one constraint
on the {\em ab initio} three independent symmetrized variances,
\begin{eqnarray}
\langle \Psi (t)\vert({\bf\Phi}- \phi)^2\vert \Psi (t) \rangle &=&
\xi^2~,\nonumber\\ 
\langle\Psi (t)\vert({\bf P}{\bf\Phi} + {\bf\Phi} {\bf P} - 
2 \phi p)\vert \Psi (t)\rangle &=&  2\xi \eta~, \label{var}\\  
\langle\Psi (t)\vert({\bf P}- p)^2 \vert \Psi (t)\rangle\ &=& 
\eta^2 + {\hbar^2 \over 4 \xi^2}~~{\rm (pure\ state)},\nonumber
\end{eqnarray} 
expressing all three in terms of only the two variables $\xi$ and
$\eta$, in the present notation. The one anti-symmetrized variance is
fixed by the commutation relation, $[{\bf\Phi}, {\bf P}] = i\hbar$. To
what does the restriction to $\sigma = 1$ correspond and
how can it be relaxed? The answer to this question is suggested by the
form of (\ref{var}), and definition of $\sigma$ in (\ref{sigdef}) or
(\ref{xidef}), which shows that $\sigma = 1$ corresponds to zero
expectation of the number operator $a^{\dagger}a$, {\it i.e.} to the
pure state vacuum annihilated by $a$. However, the mean field
equations of the Section II allow for the more general possibility that
this expectation value may take on any constant value $N$. For
example, we might consider the finite temperature Bose-Einstein
distribution,
\begin{eqnarray}
N_T &=& \left[\exp \left({\hbar\omega_0\over k T}\right) -
1\right]^{-1}~,    
\nonumber\\
\sigma_T &=& 1 + 2 N_T = \coth \left( {\hbar\omega_0\over 2 T}\right)
> 1 
\label{BE}
\end{eqnarray}
for some $\omega_0$ and temperature $T$. Since such a thermal state
corresponds not to a pure state Schr\"odinger wave function, but
rather to a {\em mixed} state density matrix, it is not surprising
that a pure state Gaussian wave function ansatz cannot describe this
case. However, the general structure of the mean field equations
involves only the one and two-point functions of the quantum variable,
so we should expect them to correspond to a Gaussian ansatz but for a
{\em mixed}-state density matrix instead of a pure state wave
function. The most general form for this mixed-state normalized 
Gaussian density matrix ${\bf\rho}$ is
\begin{eqnarray}
&&\langle x'|{\bf\rho} (p,\phi; \eta, \xi; \sigma) |x\rangle = (2\pi
\xi^2)^{-{1\over 2}} \exp \biggl\{ i\,{ p\over\hbar} (x'-x)
\nonumber\\
&&-{\sigma^2 + 1\over 8 \xi^2}\left[ (x'- \phi)^2 + 
(x- \phi)^2\right] \nonumber \\ 
&&+i{\eta \over 2 \hbar\xi} \left[ (x'- \phi)^2 - (x-
\phi)^2\right] + {\sigma^2 - 1\over 4 \xi^2} (x'-  \phi)(x-  \phi)
\biggr\},\nonumber\\
\label{gauss} 
\end{eqnarray}
in the coordinate representation. In the special case that $\sigma =
1$ the last (mixed) term in the exponent vanishes and ${\bf\rho}$
reduces to the pure state product,
\begin{equation}
{\bf\rho} (t) \Big\vert_{\sigma = 1} = \vert \Psi (t)\rangle\langle
\Psi (t)\vert\,, 
\end{equation}
with $\vert\Psi (t)\rangle$ given by (\ref{wavefn}).  For $\sigma > 1$
the general Gaussian ${\bf\rho}$ does not decompose into a product,
and
\begin{eqnarray}
{\rm Tr}~{\bf\rho}^2 (t) &\equiv& \int_{-\infty}^{\infty} dx
\int_{-\infty}^{\infty} dx' \langle x|{\bf\rho} (t)|x'\rangle \langle
x'|{\bf\rho} (t)|x\rangle \nonumber\\
&=& \sigma^{-1} < 1 
\end{eqnarray}
which is characteristic of a mixed state density matrix. 

That (\ref{gauss}) is indeed the correct generalization of the pure
state Gaussian wave function (\ref{wavefn}) is easily verified by checking
that ${\bf\rho} (t)$ satisfies the expectation value of the quantum
Liouville equation,
\begin{equation}
{\rm Tr} \left( i\hbar {\partial \over \partial t}{\bf\rho}\right) =
{\rm Tr}~[{\bf H},{\bf\rho}]\ ,
\end{equation}
provided the large $N$ limit as before is taken and the 
equations of motion (\ref{zerod}) and (\ref{etaeqn}) are satisfied for 
{\it arbitrary} $\sigma$. Taking the large $N$ limit is
equivalent here to the replacement of the full anharmonic Hamiltonian
by a time-dependent harmonic oscillator Hamiltonian, 
\begin{equation}
{\bf H} \rightarrow {\bf H}_{osc} = {1 \over 2}\left({\bf P}^2 +
\omega^2(t)\, {\bf\Phi}^2\right) 
\end{equation}
where 
\begin{equation}
\omega^2(t) = \left\langle{\partial^2 V\over
\partial{\bf\Phi}_i\partial{\bf\Phi}_j}\right\rangle\bigg\vert_{i=j}
\rightarrow {\lambda \over 2} (\phi^2 (t) + \xi^2(t) - v_0^2) = \chi
(t) 
\end{equation}
is the self-consistently determined frequency of the oscillator in the
large $N$ limit. With this replacement it is straightforward to verify
that the Gaussian form is preserved by the time evolution under ${\bf
H}_{osc}$ \cite{classham}\cite{eboli}\cite{rj},
\begin{equation} 
i\hbar {\partial \over \partial t}{\bf\rho} = [{\bf
H}_{osc},{\bf\rho}]\,. 
\end{equation}
In fact, subsitution of the Gaussian form (\ref{gauss}) into this
Liouville equation and equating coefficients of $x$, $x'$, $x^2$,
$x'^2$ and $xx'$ gives five evolution equations for the five
parameters specifying the Gaussian which are none other than the
Eqns. (\ref{zerod}) and (\ref{etaeqn}) together with $\dot\sigma =
0$. 

The effective classical Hamiltonian for the large $N$ equations,
(\ref{Heff}) is just the expectation value of the quantum Hamiltonian
in this general mixed state Gaussian density matrix, {\it i.e.}
\begin{equation}
H_{eff}({\phi}, {p} ; \xi , \eta ;\sigma) = {\rm Tr}~({\bf\rho} {\bf
H}) 
 = {\rm Tr}~({\bf\rho} {\bf H}_{osc}) = \varepsilon L^d\ , 
\end{equation}
in the large $N$ limit where $\varepsilon$ is the energy density
defined in (\ref{enerden}).  The three symmetrized variances are
indeed now all independent with
\begin{eqnarray}
{\rm Tr}\left({\bf\rho}\, ({\bf\Phi}- \phi)^2 \right) &=&
\xi^2~,\nonumber\\
{\rm Tr} \left({\bf\rho}\, ({\bf P}{\bf\Phi} + {\bf\Phi}{\bf P} - 2
\phi  p)\right) &=&  2\xi \eta~,\nonumber \\   
{\rm Tr}\left({\bf\rho}\,({\bf P}- p)^2\right)  &=& \eta^2 + {\hbar^2
\sigma^2 \over 4 \xi^2} 
\label{vargen}
\end{eqnarray}
replacing (\ref{var}) of the pure state case. The mean values,
\begin{equation}
\phi = \langle{\bf\Phi}\rangle = {\rm Tr}~({\bf\Phi} {\bf\rho})\ ;
\qquad {\rm and}\qquad  p = \langle \dot{\bf\Phi}\rangle = {\rm
Tr}~({\bf P}{\bf\rho})  
\end{equation}
remain valid for both the pure and mixed state cases. 

The physical interpretation of the five parameters of the general
large $N$ equations $(p,\phi,\eta, \xi ; \sigma)$ in terms of the
general time-dependent mixed-state Gaussian density matrix of the
Schr\"odinger picture is now explicit in $d=0$ quantum
mechanics. Since $\sigma$ is the constant parameter which determines
the degree of mixing and $\hbar$ and $\sigma$ appear only in the
combination $\hbar\sigma$ it is clear that the large $N$ equations
allow for a smooth interpolation between the quantum pure state case
in which $\hbar \sigma = \hbar$ to the high temperature or classical
limit where
\begin{equation}
\hbar\sigma_T \rightarrow {2 T\over \omega_0}\qquad {\rm as}\qquad
T\rightarrow\infty\qquad {\rm or}\qquad \hbar \rightarrow 0\ , 
\end{equation}
in which $\hbar$ drops out entirely. Thus, quantum and classical
thermal fluctuations are treated on the same footing in the large $N$
limit, with the value of the constant parameter $\hbar\sigma$
determining whether the fluctuations described by $\xi$ and $\eta$ are
to be regarded as predominantly quantal or thermal, or intermediate
between the two.  Moreover, we see that large $N$ Gaussian dynamics is
really classical dynamics of a Gaussian distribution function, except
that $\hbar\sigma$ which measures the second moment of the classical
distribution cannot be taken to zero as it could be classically, but
instead is bounded from below by $\hbar$. This is made explicit by the
form of the Wigner function corresponding to the density matrix
(\ref{gauss}), {\em viz.}
\begin{eqnarray}
&&f_W(x, p_x) \equiv {1\over 2\pi \hbar}\int_{-\infty}^{\infty} dy 
e^{-i p_x y /\hbar} \langle x + {y\over 2}\,\big\vert \rho 
\big\vert\, x - {y\over 2} 
\rangle\nonumber\\
&&= {1\over \pi\hbar\sigma} \exp\left\{-{(x-\phi)^2\over 2\xi^2} - 
{2\xi^2\over \hbar^2\sigma^2}\left[p_x-\dot\phi-{\eta\over
\xi}(x-\phi)\right]^2\right\}.\nonumber\\
\label{wigner}
\end{eqnarray} 

Before leaving our $d=0$ example it is instructive to examine the
static solutions of the effective Hamiltonian, {\it viz.} the
simultaneous vanishing of 
\begin{eqnarray}
\dot p &=& -{\partial H_{eff}\over \partial  \phi} =
-{\lambda \over 2}( \phi^2 + \xi^2 - v_0^2) \phi = 
\chi\,\phi = 0\ ,\qquad {\rm and}\nonumber\\
\dot\eta &=& - {\partial H_{eff}\over \partial \xi} = -\chi\, \xi + 
{\hbar^2\sigma^2 \over 4 \xi^3} = 0 \ .
\end{eqnarray}
If we look for a spontaneosly broken solution $\phi \ne 0$, then
$\chi$ must vanish from the first of these conditions. But then we
cannot satisfy the second condition for finite $\hbar\sigma$ and
$\xi$. This is just a rederivation of the fact that there can be no
spontaneous symmetry breaking in $d=0$ quantum mechanics (or in fact
for any $d\le 1$). We are forced instead to the symmetry restored
situation for which $\phi = 0$ and $\chi = \chi_0 > 0$ is determined
from the real positive root of the cubic equation
\begin{equation}
\chi_0 = {\lambda \over 2}(\xi_0^2 - v_0^2) = {\hbar^2\sigma^2 \over 4
\xi_0^4} 
\end{equation}
for $\xi_0^2(\sigma)$. The Gaussian density matrix centered at $\phi
=0$ with variance $\xi_0^2(\sigma)$ is the solution of the
time-independent Liouville equation for the anharmonic double well
oscillator with a trial variational density matrix of the form
(\ref{gauss}). In the limit that the height of the energy barrier
between the two wells, $E_b=\lambda v_0^4/8$ is much greater than the
fluctuation energy in either well, $E_f=\hbar\sigma \sqrt\lambda
v_0/2$, the width of the Gaussian $\xi^2_0\rightarrow v_0^2$, and its
energy $E_0 \rightarrow \hbar^2\sigma^2/8v_0^2 = E_f^2/16E_b\ll E_f$,
corresponding to a probability density spread over the entire region
$(-v_0, v_0)$ in all $N$ components of $\Phi_i$.

The entire development of the Hamiltonian equations and Gaussian
density matrix is quite easy to generalize to any number of spatial
dimensions $d$, at least for the case of spatially homogeneous mean
fields. Since in Fourier space the mode equations are just
replicated at every spatial momentum $\bf k$ we have simply to
introduce the subscript $\bf k$ on all of the relevant definitions in
this section. For example, the density matrix for $d>0$ can be written
as a product of Gaussians in Fourier space, {\it viz.}
\begin{eqnarray}
&&\langle \{\varphi_{\bf k}\}'|{\bf\rho} |\{\varphi_{\bf k}\}\rangle =
\prod_{\bf k }\langle\{\varphi_{\bf k}\}^{\prime}\vert{\bf\rho}
(p_{\bf k},\phi_{\bf k}; \eta_{\bf k}, \xi_{\bf k}; \sigma_{\bf
k}))\vert\{\varphi_{\bf k}\}\rangle 
\nonumber\\
&&=\prod_{\bf k }(2\pi \xi_{\bf k}^2)^{-{1\over 2}}
\exp \biggl\{ i\,{ p_{\bf k}\over\hbar} (\varphi_{\bf k}'-\varphi_{\bf k}) 
\nonumber\\
&&~~-{\sigma_{\bf k}^2 + 1\over 8 \xi_{\bf k}^2}\left[ (\varphi_{\bf
k}'- \phi_{\bf k})^2 + (\varphi_{\bf k}- \phi_{\bf k})^2\right]  
\nonumber \\ 
&&~~+i\,{\eta_{\bf k} \over 2 \hbar\xi_{\bf k}} \left[ (\varphi_{\bf
k}'- \phi_{\bf k})^2 -  
(\varphi_{\bf k}-  \phi_{\bf k})^2\right]  
\nonumber\\
&&~~+ {\sigma_{\bf k}^2 - 1\over 4 \xi_{\bf k}^2} (\varphi_{\bf k}'-
\phi_{\bf k})(\varphi_{\bf k}-  \phi_{\bf k}) 
\biggr\}~,
\label{gaussd} 
\end{eqnarray}
where $\varphi_{\bf k}$ is the generalized coordinate of the field
amplitude in Fourier space and $p_{\bf k}$ the corresponding canonical
momentum. The mean fields and their canonical momenta
\begin{equation}
\phi_{\bf k} = \delta_{{\bf k}0} \phi (t),~~~p_{\bf k}=\delta_{{\bf k}
0} p(t)
\label{meanfk}
\end{equation}
all vanish except for ${\bf k} = 0$ in the spatially homogeneous
case. The definitions
\begin{equation}
\xi_k^2 (t) \equiv \sigma_k  \vert f_k(t)\vert^2 \equiv 
(2N(k) + 1)\vert f_k(t)\vert^2 \ , \qquad \eta_k \equiv \dot\xi_k
\label{xikdef}
\end{equation}
and
\begin{eqnarray}
\chi(t) &=& {\lambda\over 2} \left(\phi^2 (t) + \int [d{\bf
k}]\,\xi_k^2 (t) - v_0^2\right) \nonumber\\
&=& {\lambda\over 2} \left[\phi^2 (t) + \int [d{\bf k}]\,\left(\xi_k^2
(t) - {\hbar\over 2k}\right)- v^2\right] 
\label{ddim}
\end{eqnarray}
have been introduced in obvious analogy to the $d=0$ case.  The
effective Hamiltonian density which gives rise to these equations is
\begin{equation}
{H_{eff}\over L^d} = \varepsilon ={1\over 2} p^2 + {1\over
2\lambda}\chi^2 + {1\over 2} \int [d{\bf k}]\left(\eta_k^2 + k^2
\xi^2_k + {\hbar^2\sigma_k^2 \over 8\xi_k^2}\right)
\label{Heffd}
\end{equation}
with $\chi$ regarded as a dependent variable of $\phi$ and the $\xi_k$
through the gap equation (\ref{ddim}) above. The $k^2$ term arising
from the spatial gradient of the field in $d>0$ dimensions is the most
significant difference from the $d=0$ case considered previously.  The
Hamiltonian equations of motion
\begin{eqnarray}
\dot\phi = p\ , \qquad\dot{ p} &=& -{\partial H_{eff}\over \partial
\phi} = -\chi\,\phi \ ,\nonumber\\
\dot\xi_k = \eta_k\ , \qquad\dot\eta_k 
&=& - {\partial H_{eff}\over \partial \xi_k} = 
-(k^2 + \chi)\, \xi_k + {\hbar^2\sigma_k^2 \over 4 \xi_k^3}\nonumber\\
\label{Hameom} 
\end{eqnarray}
are completely equivalent to the equations of motion for the mode
functions (\ref{modefn}) and mean fields (\ref{feq}) and (\ref{ceq})
in the spatially homogeneous case in any number of dimensions.

This completes our demonstration that the large $N$ mean field
equations of the previous section are Hamilton's equations with
$H_{eff}$ given explicitly by (\ref{Heffd}) above, and the
identification of the time-dependent Gaussian density matrix
(\ref{gaussd}) to which the homogeneous mean field equations
correspond. The case of spatially inhomogeneous mean fields involves
only the straightforward generalization of (\ref{gaussd}) to Gaussian
covariances which are off-diagonal in the momentum index $\bf k$, and
a corresponding coupling between the different momentum modes in the
effective Hamiltonian (\ref{Heffd}). Since we have no need of these
expressions in the present work we do not give the explicit formulae
here, though they may be easily worked out within the present
framework.

\section{The Nonequilibrium True Effective Potential (TEP)}
\label{sec:level4}

Having obtained the effective Hamiltonian which describes the
evolution of the closed system of mean fields and fluctuations in the
large $N$ limit, it is natural to define the dynamical or true
effective potential to be the static part of the effective Hamiltonian
density (\ref{Heffd}), {\it i.e.}
\begin{eqnarray}
&&U_{eff}(\phi , \{\xi_k\}\ ;\{\sigma_k\}) \nonumber\\
&&\equiv{1\over 2\lambda}\chi^2 + {1\over 2} \int [d{\bf k}]\left(k^2
\xi^2_k + {\hbar^2\sigma_k^2 \over 8\xi_k^2}\right)\label{ueff}\\
&&= {\chi\over 2}\left(\phi^2 - v_0^2 - {\chi\over\lambda}\right)
+  {1\over 2} \int [d{\bf k}]\left( (k^2 + \chi) \xi^2_k +
{\hbar^2\sigma_k^2 \over 4\xi_k^2}\right)~.\nonumber
\end{eqnarray}
This $U_{eff}$ is the energy density of an initial state with a
Gaussian density matrix (\ref{gaussd}) centered around $\phi$ with
instantaneously zero velocities $p = \eta_k = 0$ in both the mean
field and the fluctuation variables. We call $U_{eff}$ the ``True
Effective Potential'' (TEP) because it determines the true
out-of-equilibrium time evolution dynamics of the system according to
Hamilton's Eqns. (\ref{Hameom}). It must be clearly distinguished from
the finite temperature free energy effective potential often discussed
in the literature. The relationship between the two we would like to
discuss next.

The standard method of calculating the free energy at temperature $T$
is to analytically continue the effective action ${\cal S}_{eff}$ to
imaginary time \cite{effV} and evaluate the Tr ln $G^{-1}$ term over 
fluctuations
which are periodic in imaginary time with period $\beta = \hbar/T$. In
this way one finds for the case at hand,
\begin{eqnarray}
&&{\chi\over 2}\left(\phi^2 - v_0^2 - {\chi\over\lambda}\right)
\nonumber\\
&&+ \int [d{\bf k}]\, \left\{{\hbar\omega_k\over 2} + T \ln \left
[1-\exp\left(-{\hbar\omega_k\over T}\right) \right]\right\}\ , 
\label{fep}
\end{eqnarray}
which is a real function of $\phi$ provided
\begin{equation}
\omega_k^2 = k^2 + \chi \ge 0\ .
\label{rfreq}
\end{equation}
The infinite zero point energy in $F$ is handled in the usual way by
subtracting the energy of the zero temperature vacuum at $\chi = 0$,
{\em i.e.} $\int [d{\bf k}] \hbar k/2$. With the subtraction of this
temperature independent constant and the previous definitions of $v$
and the logarithmic coupling renormalization by Eqns. (\ref{lren1}) and
(\ref{lren2}), the thermodynamic free energy becomes finite and is
given by \cite{BM}:
\begin{eqnarray}
F(\phi, T) &=& {\chi_{_T}\over 2}\left(\phi^2 - v^2 -
{\chi_{_T}\over\lambda}\right) \nonumber\\
&&+ {\hbar\over 4\pi^2}\int_0^{\infty} k^2\, dk\, \left\{\omega_k - k
- {\chi_{_T}\over 2k} \right.\nonumber\\
&&\left.+ 2\,T \ln \left [1-\exp\left(-{\hbar\omega_k\over
T}\right) \right]\right\}\ , 
\label{fepr}
\end{eqnarray}
with $\chi_{_T}$ the solution of the finite temperature gap equation,
\begin{eqnarray}
\chi_{_T}(\phi) &=& {\lambda \over 2}(\phi^2 - v^2) \nonumber\\
&& + {\hbar\lambda\over 8\pi^2} \int_0^{\infty} k^2\, dk\,
\left\{{1\over \omega_k}{\rm coth}\left({\hbar\omega_k\over 2T}\right)
- {1\over k}\right\}.  
\label{gapT}
\end{eqnarray}
As long as $\chi_{_T} \ge 0$ the free energy function (\ref{fepr}) is
real and well-defined.

The first derivative of the free energy function is given by
\begin{equation}
{d F\over d\phi} = \chi_{_T}\phi
\label{firder}
\end{equation}
upon using the gap equation (\ref{gapT}). In the spontaneously broken
state defined by $\chi_{_T} = 0$, we have 
\begin{eqnarray}
\phi^2 = \phi_T^2 &=& v^2 - {\hbar\over 2\pi^2}\int_0^{\infty} k\,
dk\, {1 \over \exp \left({\hbar k\over T}\right) - 1}\nonumber\\
&=& v^2 - {T^2 \over 12\hbar}\,.
\label{phiT}
\end{eqnarray}
The last relation informs us that the expectation value $\phi_T$
vanishes at the critical temperature,
\begin{equation}
T_c = 2\sqrt{3\hbar}\, v\,.
\end{equation}
At $T= T_c$, the second derivative of $F(\phi, T)$ vanishes at $\phi =
0$, and therefore the phase transition is second order in the large
$N$ approximation. We remark that this is {\em not} the case in the
simple one-loop or Hartree approximation where the transition is
weakly first order \cite{ACP}. Since the $\Phi^4$ field theory lies
in the same universality class as spin models which are known to have
second order phase transitions, the large $N$ mean field approximation
gets the correct order of the transition, and is another reason why it
is to be preferred over the other approximations \cite{bagr}. Indeed, 
the correlation length, which is given by the the inverse mass of the
radial excitation,
\begin{equation}
\ell (T) = \lambda_R^{-{1\over 2}}\phi_T^{-1} \rightarrow 
\left({6\over\lambda_R T_c}\right)^{1\over 2} (T_c-T)^{-{1\over2}}
\end{equation}
diverges as $T\rightarrow T_c$ with the $-{1\over 2}$ critical
exponent of mean field theory.

Actually the second derivative of the free energy function vanishes
at its minimum for {\em any} $T \le T_c$. This follows the fact that
the derivative of the gap equation (\ref{gapT}) evaluated at
$\chi_{_T} = 0$ involves an infrared linear divergence so that 
\begin{equation}
{d\chi_{_T}(\phi)\over d\phi}\bigg\vert_{\chi_{_T} = 0} = 0\,,
\end{equation} 
and
\begin{equation}
{d^2 F\over d\phi^2} = \chi_{_T} + {d\chi_{_T}(\phi)\over d\phi}\phi = 0
\label{secder}
\end{equation}
at $\chi_{_T} = 0$. This perhaps surprising feature of the large $N$
free energy is illustrated in Fig. 4. It is a direct consequence of
the massless Goldstone particles present in the spontaneously broken
phase.  As we shall see in Section VII, it also makes the equilibrium
free energy useless for describing the plasma oscillations about the
spontaneously broken thermal minimum.

\vspace{.4cm}
\epsfxsize=7cm
\epsfysize=5.5cm
\centerline{\epsfbox{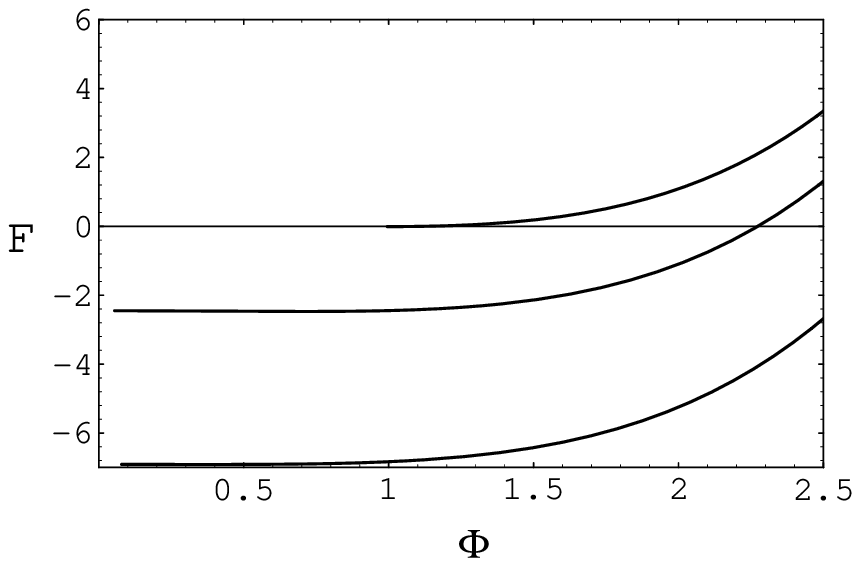}}
\vspace{.35cm}
{FIG. 4. {\small{The Helmholtz Free Energy $F$ as a function of $\phi$
for three temperatures $T=.1,~1,~1.5$ (in descending order from the
top) in units of $T_c$ in the leading order large $N$ approximation.
The curves terminate at the minimum of $F$ at $\phi= \phi_T$, $\chi_T
= 0$, given by Eqn. (\ref{phiT}), which is the boundary of the
spinodal region, {\em i.e.} for $\phi <\phi_T$ there is no stable
equilibrium state. The second derivative of $F$ at $\phi=\phi_T$
vanishes, showing in particular that the phase transition at $T=T_c$
is second order.}}}\\

For $T < T_c$ and $\phi < \phi_T$ the gap equation (\ref{gapT}) has no
positive real solutions. This is the unstable spinodal region of
Fig. 3. If $\chi <0$ then the frequency $\omega_k$ becomes imaginary
for $k < |\chi|^{1\over 2}$. This means that strictly speaking the
thermodynamic free energy $F$ is not well defined in this
region. Despite the fact that no thermal equilibrium uniform state can
exist in this spinodal region, what is sometimes done is to {\em
define} the free energy by an analytic continuation from $\chi >0$ to
$\chi = -|\chi| <0$. By this procedure $F$ acquires an imaginary part,
while its real part is no longer a convex function of $\phi$ (as
general theorems require). This has led to much discussion in the
literature \cite{confusion}\cite{Wbook}. However, we needn't
wonder at the meaning of this analytic continuation and imaginary part
of $F$ in the more general nonequilibrium context. The analytic
continuation from $\chi >0$ to $\chi = -|\chi| <0$ simply defines a
Gaussian density matrix centered at $\phi <\phi_T$ with
instantaneously zero velocities, $\dot\phi = \eta_k=0$ and parameters
$\xi_k$ and $\sigma_k = 2N_T(k) +1$ determined by the thermal
distribution at $|\chi_{_T}| = - \chi_{_T}$ where the free energy is
well defined. The density matrix will certainly evolve away from this
unstable configuration and the imaginary part is a measure of the rate
at which this evolution will occur initially \cite{WW}. However, shortly
afterwards the nonlinear effects of the evolution described by our
general nonequilibrium equations of motion will set in, and this
imaginary part can only be at best a simple order of magnitude
estimate for the evolution away from the initial state in the weak
coupling limit $\lambda_R \ll 1$.

The nonconvexity of the real part of the free energy is also
related to the same instability \cite{Wbook}. The point is that 
the construction of the equilibrium free energy by a loop expansion
or the large $N$ expansion tacitly
assumes the existence of a stable configuration as
the starting point for the expansion. It is precisely
this assumption which breaks down in the unstable
spinodal region. A careful definition of the 
``effective potential" $F$ as the minimum of the free energy
for fixed $\phi = \langle{\bf \Phi}\rangle$ independent of space and time leads
to a flat constant function between the two minima at
$\pm v$ which is convex in accordance with the general theorems.
However this minimization cannot be achieved
with a Gaussian density matrix of the form (\ref{gaussd}) and
the flat convex form of $F$ defined in this manner
tells us nothing about the dynamical evolution
of an initial Gaussian centered at a value of $\phi$ in the
spinodal region. For this reason we propose to drop all
attempts to refine the definition of a free energy
``effective potential" and focus instead on the potential
which actually governs the nonequilibrium dynamics in a given
approximation scheme. In the leading order large $N$ scheme
this dynamical potential is the True Effective Potential $U_{eff}$. 

In contrast to the free energy $F$, the TEP $U_{eff}$ is a
function(al) of all of the generalized fluctuation coordinates $\xi_k$
as well as the mean field $\phi$. It also depends on the constant
parameters $\sigma_k$, which need not be that in the thermal
distribution (\ref{BE}). Hence it is defined much more generally than
$F$ and is manifestly real and positive. If we wish to consider a
function only of the mean field $\phi$ it is possible to try first
minimizing $U_{eff}$ with respect to the Gaussian parameters $\xi_k$:
\begin{equation}
{\partial U_{eff} \over \partial \xi_k} = (k^2 + \chi)\, \xi_k  
- {\hbar^2\sigma_k^2 \over 4 \xi_k^3} = 0
\label{minU}
\end{equation}
which has the real solution,
\begin{equation}
\bar\xi_k^2 = {\hbar\sigma_k \over 2 \omega_k}
\label{extxi}
\end{equation}
provided again that (\ref{rfreq}) holds.  

If we restrict $U_{eff}$ further by requiring $\sigma_k$ to be
the Bose-Einstein thermal distribution, $\sigma_{T\, k} = 1 + 2N_T(k)$
of (\ref{BE}) we find
\begin{eqnarray}
U (\phi , T) &\equiv& U_{eff}(\phi , \{\xi_k\} = \{\bar\xi_k\}
\ ;\sigma_k = \sigma_{T\, k})\nonumber\\
&=& {\chi\over 2}\left(\phi^2 - v_0^2 - {\chi\over\lambda}\right)
\nonumber\\ 
&&+ {\hbar\over 2} \int [d{\bf k}]\, \omega_k\, \left (1 +
2N_T(k)\right)\,. 
\label{iep}
\end{eqnarray}
The first term is the classical potential energy density $V_{cl}$ of
the mean field $\phi$ and the second term is the quantum plus thermal
energy of the fluctuations at the extremum (\ref{extxi}).  We
recognize the expression (\ref{iep}) as the {\it internal} energy of
the Gaussian configuration specified by (\ref{extxi}) and
(\ref{BE}), which differs from the Helmholtz free energy $F$ by the
standard thermodynamic relation,
\begin{equation} 
F = U - T\, S\ ,
\end{equation}
with
\begin{eqnarray}
&&S= - {\rm Tr}~{\bf\rho}_T\,\ln{\bf\rho}_T \nonumber\\
&&=\int [d{\bf k}] \left[\left(N_T(k) + 1\right)\ln\left(N_T(k) +
1\right) - N_T(k)\ln N_T(k)\right]\nonumber\\
\label{vnentropy} 
\end{eqnarray}
the von Neumann entropy of a Gaussian thermal density matrix
${\bf\rho}_T$.

\vspace{.4cm}
\epsfxsize=7cm
\epsfysize=5.5cm
\centerline{\epsfbox{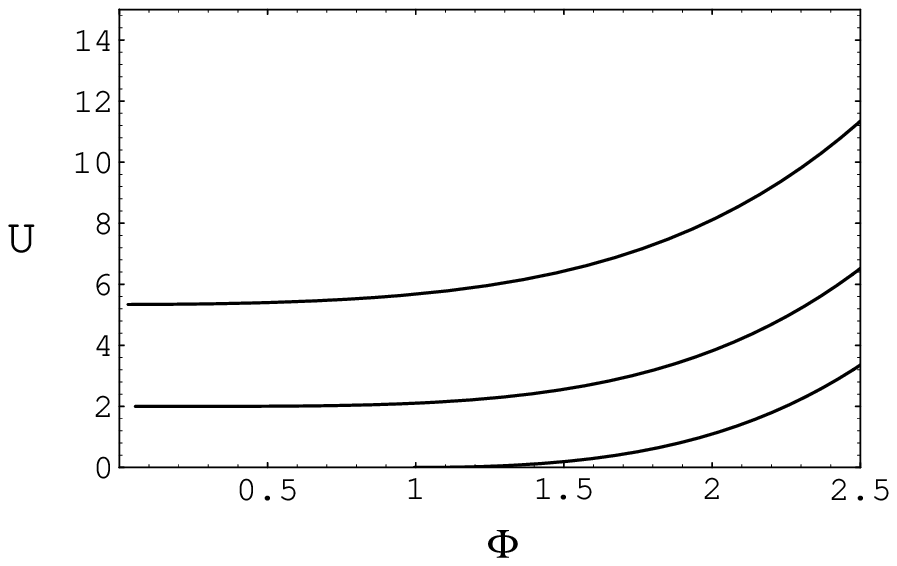}}
\vspace{.35cm}
{FIG. 5. {\small{The True Effective Potential $U_{eff}$ as a function
of $\phi$ and evaluated in a thermal equilibrium state for the same
three temperatures as in the previous figure (note here that the
lowest temperature is now the bottom curve). The Gaussian width
parameter $\xi_k$ has been set equal to its minimum value by
Eqn. (\ref{minU}). Although qualitatively similar to the free energy
$F$ in Fig. 4, $U_{eff}$ becomes equal to the internal energy $U$ in
thermal equilibrium.}}}\\
   
Although $F$ and $U$ have qualitatively similar behaviors in the
region $\chi \ge 0$ where they are both defined unambiguously, as shown
in Figs. 4 and 5, clearly $F$ and $U$ have quite different physical
meanings 
and applicability. The first is the negative of the pressure of the
gas of scalar particles against which work must be done in compressing
the system at fixed temperature, {\it i.e.}  if the compression is
performed while in contact with an external heat reservoir at
temperature $T$. The second is the energy density of the {\it closed}
system which is conserved if it is allowed to evolve, isolated from
all external sources or sinks of energy. The free energy $F$ can be
defined only in thermal equilibrium. The internal energy $U$ is a
special case of the more general true effective potential $U_{eff}$
which is defined for any Gaussian density matrix, equilibrium or not,
and which determines the evolution of the system away from the
instantaneous stationary state where $\dot\phi = \dot{\xi}_k = 0$
according to the Hamiltonian equations, (\ref{Hameom}). Moreover, in
the spinodal region where $T <T_c$ and $\phi < \phi_T$ it is clear
that the minimization condition (\ref{minU}) cannot be satisfied for
real $\xi^2_k$, which informs us that the time derivative of the
canonical momentum, $\dot\eta_k >0$ for $k^2 < |\chi|$, {\em i.e.} the
configuration is unstable against the fluctuation widths $\xi_k$
growing in time for these low $k$.  Note that by the definition
(\ref{xikdef}), in terms of the mode function $f_k$, $\xi_k$ is
necessarily real.  The physical interpretation is thus quite
straightforward in the more general nonequilibrium framework we have
laid out: there simply is no stationary Gaussian density matrix for
$\chi < 0$, and any such initial configuration will necessarily evolve
in time by growth of the small $k$ Fourier components of the Gaussian
parameters $\xi_k$. Indeed, this is obvious from the mode equations
(\ref{modefn}) which show that all the long wavelength modes with
$k^2 \le |\chi|$ will grow exponentially rather than oscillate when
$\chi < 0$ \cite{WW}.  Notice that the non-existence of a solution to
the minimization condition (\ref{minU}) implies that the exponentially
growing instability lies in the fluctuations $\xi_k$ for small $k$.
Only the nonlinear back reaction of this exponential growth of the
modes in the time-dependent gap equation (\ref{ceq}) can eventually
bring $\chi$ to non-negative values and turn off the instability. It
is clear that this time-dependent process involves many low $k$
Fourier components of the field $\Phi$ and cannot be described
adequately by a single function of one variable such as the
``effective'' potential $F$, much less a function of one variable
defined only in thermal equilibrium in the spinodal region where no
such thermal equilibrium exists.

That the time evolution as determined by $U_{eff}$ is quite different
from what might be inferred from the free energy function $F$ is
illustrated by our numerical solution of the evolution equations,
presented in Fig. 6.  The oscillation frequency and even the
final point of the evolution of $\phi$ as $t \rightarrow \infty$ in
general bear no simple relation to the minimum $\phi_T$ of the free
energy ``effective'' potential. This is easily understood from the
stationary points of the TEP, for a true stationary state does exist
in the spontaneously broken phase if we also require
\begin{equation}
{\partial U_{eff} \over \partial \phi} = \chi\phi = 0 \,
\end{equation}
since this is satisfied by
\begin{equation}
\chi = 0 \ ; \qquad {\rm and} \qquad \xi_k^2 = {\hbar \sigma_k\over
2k}\ . 
\label{ssbd}
\end{equation}
Notice that there is such a static solution for {\it any} $\phi$ and
$\sigma_k$. The finite temperature case, (\ref{BE}) is only one of
infinitely many possibilities for a static solution of the mean field
equations. Correspondingly there are also an infinite number of
stationary density matrices satisfying $[{\bf \rho}, {\bf H}_{osc}]
=0$, one for each choice of $\sigma_k$.  If $\sigma_k = 1 + 2N(k)$ is
given, then $\phi$ is determined by the static condition $\chi =0$,
{\em i.e.}
\begin{equation}
\phi^2\big\vert_{\chi =0} = v^2 - \hbar\int [d{\bf k}]\ {N(k) \over k}
= v^2 - {\hbar\over 2\pi^2}\int_0^{\infty} kdk\,N(k)\,.
\label{sumrul}
\end{equation}
For arbitrary $N(k)$ the relation (\ref{sumrul}) may be viewed as a
kind of sum rule which allows us to distribute any fraction of the
$v^2$ in the coherent vacuum expectation value $\phi^2$ and the
remainder in the integral over particle fluctuation modes.  In
particular, one may even consider the case where the mean field $\phi
= 0$ identically, but the theory is still in the spontaneously broken
phase since $\chi$ vanishes and the integral in (\ref{sumrul})
saturates $v^2$. This case is a kind of marginal symmetry breaking
since any small perturbation of $\phi$ will take us away from zero
mean field.

With the help of the sum rule (\ref{sumrul}) one can easily understand
the perhaps initially surprising result, first found numerically in
Ref. \cite{Boyan2}, that the evolution of the $\phi$ mean field can
settle to a value different from, and even much smaller than the
thermal equilibrium value $\phi_T$. It simply corresponds to the fact
that the distribution of particles in the final state is not at all a
thermal one. Obviously such a situation cannot be described by the
thermodynamic free energy $F$ since its evaluation by continuation to
imaginary time assumes from the outset a thermal distribution of
particles. For this reason one could not expect the true evolution
according to the mean field equations of motion to bear much relation
to what one might infer from an uncritical use of $F$, particularly in
the region $\phi < \phi _T$ where no thermodynamic uniform equilibrium
state exists and the Helmholtz free energy is not even strictly
defined.

\vspace{.4cm}
\epsfxsize=7cm
\epsfysize=8cm
\centerline{\epsfbox{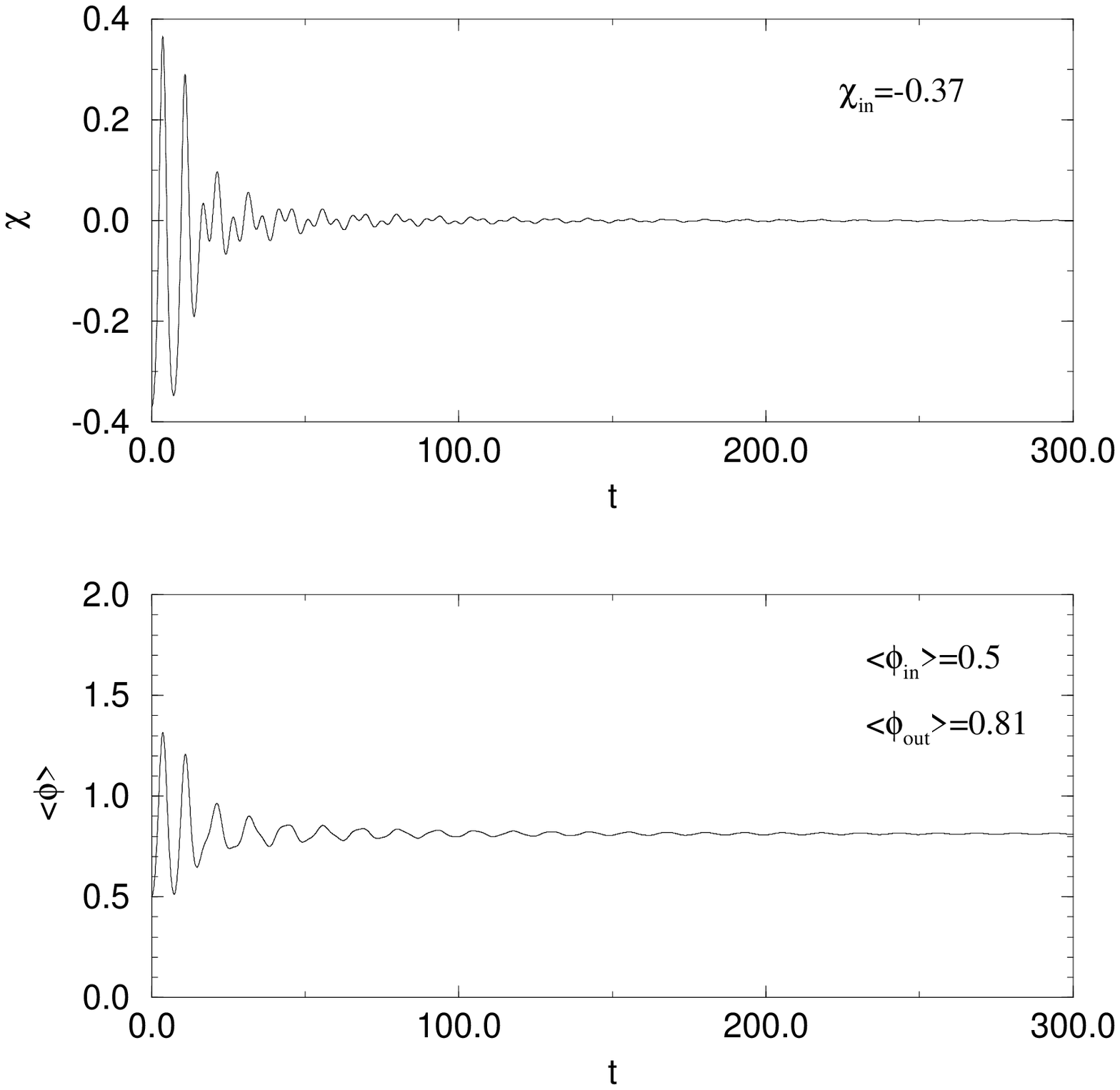}}
\vspace{.35cm}
{FIG. 6. {\small{A typical evolution of the mean field $\phi$ and
the mass squared $\chi$ in the spontaneously broken phase, starting
from an unstable initial state with $\chi(0) = -0.37 < 0$. All time
and lengths are measured in units of $v=1$. Note that if we use the
initial energy density $\varepsilon = 0.0697$ to infer an effective
temperature of massless bosons, then the expected minimum of $F$
occurs at $\phi_{out} = 0.98$ which is considerably different than
the observed $\phi_{out} = 0.81$. The oscillation frequency about
$\phi=\phi_{out}$ also bears no relation to the second derivative of
$F$ which vanishes at its minimum.}}}\\          

The inclusion of collisions in the next order of the expansion in 
$1/N$ makes it more likely to bring the distribution of particles
closer to a thermal one. However, since the evolution is unitary, 
this can happen only in some effective sense. This aspect is closely
connected to the notion of dephasing which will be discussed in 
detail in Section VI.

Finally we point out that the stationary spontaneously broken solution
$\chi = 0$ is disallowed in $d=1$ spatial dimension (as it was in
$d=0$ quantum mechanics), since the integration in the definition of
$\chi$ in (\ref{ddim}) diverges logarithmically as $k\rightarrow 0$
which is inconsistent with $\chi$ vanishing in (\ref{ssbd}). It is
this infrared divergence which prevents spontaneous symmetry breaking
of the global $O(N)$ symmetry in one space dimension, consistent with
the Mermin-Wagner-Coleman theorem \cite{mwc}. In two or higher space 
dimensions there is no such divergence and the spontaneous symmetry 
breaking static solution (\ref{ssbd}) is allowed.

\vspace{.4cm}
\epsfxsize=7cm
\epsfysize=9cm
\centerline{\epsfbox{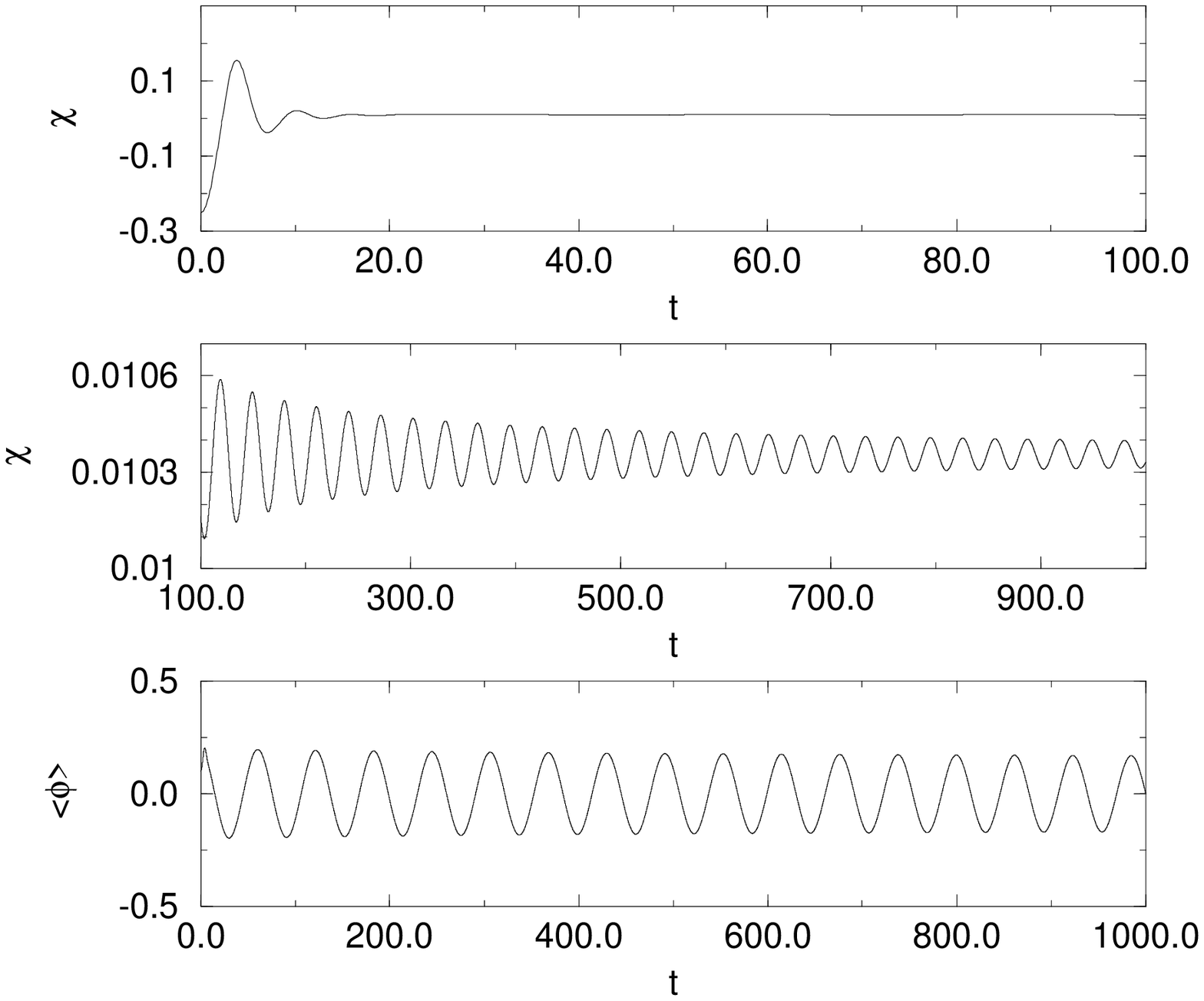}}
\vspace{.35cm}
{FIG. 7. {\small{Evolution of the mean field $\phi$ and
the mass squared $\chi$ in one space dimension, starting from an
unstable initial state. The middle figure shows that $\chi$ does not
go to zero asymptotically, and the oscillations do 
not damp as effectively as in three dimensions. Both these behaviors
are consequences of the Mermin-Wagner-Coleman theorem which prevents
spontaneous symmetry breaking in one space dimension.}}}\\

The consequences of the Mermin-Wagner-Coleman theorem for nonequilibrium 
dynamics in one dimension are illustrated in Fig. 7 where the numerical
evolution of $\chi$ and $\phi$ is shown beginning from an unstable
initial state with negative $\chi$. We see that $\chi$ does not go 
to zero at late times (no symmetry breaking) and that both the $\chi$ 
and $\phi$ oscillations damp very slowly as compared to the behavior in
three dimensions.

\section{Dephasing, Dissipation and Decoherence}
\label{sec:level5}

Although we have shown by explicit construction of the effective
Hamiltonian (\ref{Heffd}) that the large $N$ mean field equations are
completely time reversible, nevertheless one observes in typical
evolutions as in Fig. 6 an {\em effective irreversibility}, in the
sense that energy flows from the mean fields $\phi$ and $\chi$ to the
fluctuating modes $f_k$ without returning over times of physical
interest. It is our purpose in this section to explain this apparent
irreversibility in terms of {\em dephasing, i.e.} the dynamical
averaging to zero in the sums over $k$ of the rapidly varying phases
of the fluctuations at a given time. To the extent that this phase
averaging is exact and the information in the phases cannot be
recovered, the time evolution is irreversible. This is the sense in
which the fluctuations $f_k$ act as a ``heat bath'' or ``environment''
for the mean field evolution of $\phi$ and $\chi$. Of course, since
information is never truly ``lost'' in a closed Hamiltonian system
evolved with arbitrarily high accuracy, the information in the phases
can be recovered in principle and we should expect Poincar\'{e}
recurrences in the mean field evolution after very long times. As the
number of modes $f_k$ (and particularly as the number of relevant
infrared modes) approaches infinity, we would expect the recurrence
time to go to infinity. In typical evolutions we followed several tens
of thousands of modes, and recurrences were never observed in
practice. The precise dependence of the recurrence time on the number
of modes is an interesting question which may be studied
quantitatively within our mean field framework by numerical
methods. However, we have not undertaken such a systematic study here,
and leave the question of the recurrence time for future research. In
any case, we cannot expect the leading order large $N$ collisionless
approximation to continue to be valid for times longer than the
collisional relaxation time in the full theory.
The study of recurrence times in the leading order approximation may
be interesting nevertheless as a model for how such recurrences and
rephasing can occur in more realistic situations.

To understand precisely what is meant by dephasing let us return to
the Gaussian density matrix of only one degree of freedom
(\ref{gauss}).  The coordinate representation is only one of many
possible representations of the density matrix. The number basis,
defined by the integer eigenvalues of
\begin{equation}
a^{\dagger}a \vert n \rangle = n \,\vert n \rangle 
\end{equation}
defines a time-independent basis associated with the Fock
decomposition of the Heisenberg operators (for $d=0$),
\begin{eqnarray}
{\bf\Phi} (t) &=& \phi (t) + a f(t) + a^{\dagger} f^*(t)\,,\nonumber\\
\dot{\bf\Phi} (t) &=& {\bf P}(t) = \dot\phi (t) + a \dot f(t) +
a^{\dagger} \dot f^*(t)\,, 
\label{fock}
\end{eqnarray}
in terms of its mean value and quantized fluctuations. By using the
Wronskian condition (\ref{wron}) we may solve these relations for $a$
and $a^{\dagger}$,
\begin{eqnarray}
({\bf\Phi} - \phi)\dot f^* - ({\bf P} - \dot \phi)f^* &=& i\hbar
a\,,\nonumber\\ 
({\bf\Phi} - \phi)\dot f - ({\bf P}- \dot \phi)f &=& -i\hbar
a^{\dagger}\,. 
\end{eqnarray}
In this time independent basis, the Hamiltonian is given by
\begin{equation}
H=\hbar \omega(t=0)\left(a^{\dagger}a+{1\over 2}\right)~.
\label{hzero}
\end{equation}
This transformation is just a (complex) canonical transformation which
on the quantum level is implemented by a unitary transformation of
bases, with the transformation matrix, $\langle x\vert n\rangle$ in
Dirac's notation. To find this transformation matrix explicitly we use
the definitions (\ref{xidef}) and (\ref{etadef}), the equation of
motion (\ref{etaeqn}), and the Wronskian condition again to secure
\begin{eqnarray} 
\hbar^2 \sigma\left(a^{\dagger} a + {1\over 2}\right) &=& ({\bf\Phi} -
\phi)^2 \left( {\eta^2\over \sigma} +
{\hbar^2\sigma^2\over\xi^2}\right) + \xi^2({\bf P}-p)^2 \nonumber\\
&-& \xi\eta\left( ({\bf\Phi} - \phi)({\bf P} - p) +  
({\bf P}-p)({\bf\Phi} - \phi)\right)\nonumber 
\end{eqnarray}
in terms of $\xi$, $\eta$, and $\sigma$.  Then by acting with this
operator identity on the matrix $\langle x\vert n\rangle$ and
replacing ${\bf \Phi} \rightarrow x$ and ${\bf P} \rightarrow -i\hbar
{d\over dx}$ we obtain the differential equation,
\begin{eqnarray}
&&\hbar^2\sigma (n + {1\over 2})\langle x\vert n\rangle = \nonumber\\
&&\left\{(x - \phi)^2 \left( \eta^2 +
{\hbar^2\sigma^2\over\xi^2}\right) +  
\xi^2\left(-i\hbar {d\over dx}-p\right)^2\right.\nonumber\\
&&\left.- 2\xi\eta\left[-i\hbar (x-\phi) {d\over dx} -{i\hbar\over 2}
-(x-\phi)p \right]\right\}\langle x\vert n\rangle
\end{eqnarray}
which is easily solved in terms of the ordinary harmonic oscillator
wavefunctions,
\begin{equation}
\psi_n (x; \omega) = {1\over (2^n n!)^{1 \over 2}}
\left({\omega\over\pi\hbar}\right)^{1 \over 4}H_n \left ({\sqrt
{\omega\over \hbar}}x\right) \exp \left(-{\omega\over
2\hbar}x^2\right) 
\end{equation}
by 
\begin{equation}
\langle x\vert n\rangle = \exp\left({ip(x-\phi)\over\hbar}
+{i\eta(x-\phi)^2 
\over 2\hbar\xi}\right) \psi_n \left(x-\phi; {\hbar\sigma\over
2\xi^2}\right).
\label{xn}  
\end{equation}

With the transformation matrix element determined it is a
straightforward exercise in Gaussian integration and the properties of
Hermite polynomials to obtain the form of the density matrix
(\ref{gauss}) in the time-independent number representation,
\begin{eqnarray}
\langle n'|{\bf\rho}|n\rangle &=& \int_{-\infty}^{\infty} dx\, 
\int_{-\infty}^{\infty} dx^{\prime}\, 
\langle n'\vert x^{\prime}\rangle \langle x'|{\bf\rho}|x\rangle
\langle x\vert n\rangle \nonumber\\
&=& {2\delta_{n'n}\over \sigma +1}\left({\sigma
-1\over \sigma +1}\right)^n \,,
\label{diag}
\end{eqnarray} 
where $\sigma$ is defined by (\ref{sigdef}). The derivation of these
results and some further properties on the transformation matrices are
given in Appendix B.

The matrix elements of the Liouville equation in this basis are
\begin{eqnarray}
i\hbar\langle n|\dot{\rho}|n'\rangle&=&\langle n|[H,\rho]|n'\rangle \nonumber\\
&=&(E_n-E_{n'})\langle n|{\rho}|n'\rangle~.
\end{eqnarray}
Clearly, if the density matrix in this basis is initially diagonal, which 
can always be accomplished by a Bogoliubov transformation (Appendix B), then
it stays diagonal and time-independent. 
The time independence of the matrix elements of $\rho$
in the Heisenberg number basis $|n\rangle$ is simply a reflection
of the fact that the density matrix (like the state
vector $|\psi\rangle$) is time independent in the Heisenberg
picture where the field operator $\Phi$ depends on time
according to Eqn. (\ref{pexp}). All the
time-dependent dynamics resides in the transition matrix element
$\langle x\vert n\rangle$ which depends on the five variables
$(\phi(t),p(t);\,\xi(t),\eta(t);\,\sigma)$ while the matrix elements
(\ref{diag}) remain forever unchanging under the mean field
evolution. Indeed, this is just another reflection of the unitary
Hamiltonian nature of the evolution, since the von Neumann entropy of
the Gaussian density matrix
\begin{eqnarray}
S &=& -{\rm Tr}~{\bf\rho}\,\ln\,{\bf\rho} \nonumber\\
&=& \left({\sigma + 1\over2}\right) 
\ln\left({\sigma + 1\over 2}\right) - \left({\sigma - 1\over
2}\right)\ln\left({\sigma - 1\over 2}\right)\nonumber\\ 
&=& (N + 1) \ln (N + 1) - N \ln N 
\end{eqnarray}
is strictly constant under the time evolution. Hence, there is no
information lost in the evolution in any strict sense.

There is however another basis which is more appropriate for
discussing ``physical'' particle number. This is the time-dependent
Fock basis specified by replacing the mode function $f$ which
satisfies
\begin{equation}
\left({d^2\over dt^2} + \omega^2(t)\right)f(t) = \left({d^2\over dt^2}
+ \chi(t)\right)f(t) = 0 
\end{equation}
by the adiabatic mode function,
\begin{equation}
\tilde f (t) \equiv \sqrt {\hbar\over 2 \omega (t)}\, \exp \left(
-i\int_0^t \, dt' \omega(t')\,\right)\,,
\label{ftilde}
\end{equation}
and the Fock representation (\ref{fock}) by
\begin{eqnarray}
{\bf\Phi} (t) &=& \tilde a(t) \tilde f(t) + \tilde a^{\dagger}(t)
\tilde f^*(t)\,,\nonumber\\ 
\dot{\bf\Phi} (t) &=& {\bf P}(t) = -i\omega(t)\tilde a(t)\tilde f(t) +
i\omega(t)\tilde a^{\dagger}(t) \tilde f^*(t)\,,
\label{focka}
\end{eqnarray}
where the $\tilde a$ and $\tilde a^{\dagger}$ operators must now be
time-dependent. The corresponding time-dependent number basis is
defined by
\begin{equation}
\tilde a^{\dagger}\tilde a \vert \tilde n \rangle = \tilde n \,\vert 
\tilde n \rangle\,, 
\end{equation}
in which
\begin{equation}
{\bf H}_{osc} = {\hbar\omega\over 2} (\tilde a^{\dagger} \tilde a + 
\tilde a \tilde a^{\dagger})
\end{equation}
is diagonal for each time. In the $\tilde a^{\dagger} \tilde a$ number
basis, ${\bf\rho}$ is no longer diagonal, $\langle\tilde a \rangle$,
$\langle \tilde a\tilde a\rangle$, {\em etc.}, are non-vanishing, and
$\tilde N \equiv\langle\tilde a^{\dagger}\tilde a\rangle \neq N$ in
general, except in the static case of constant $\omega$.

In addition to diagonalizing the time-dependent harmonic oscillator
Hamiltonian ${\bf H}_{osc}$ which describes the evolution of the
Gaussian density matrix, the $\tilde n$ basis has another important
property, namely the existence of an adiabatic invariant $W$. This
adiabatic invariant may be constructed from the Hamilton-Jacobi
equation corresponding to the effective classical Hamiltonian
$H_{eff}$ in the usual way,
\begin{equation}
{1\over 2} \left({\partial W \over \partial x}\right)^2 +
{\omega^2\over 2} x^2 + {1\over 2} \left({\partial W \over \partial
\xi}\right)^2 + {\omega^2\over 2} \xi^2 + {\hbar^2\sigma^2\over
8\xi^2} = E\,. 
\end{equation}
This equation is separable and the ansatz $W(x,\xi) = W_1(x) +
W_2(\xi)$ along with $E=E_1+E_2$ yields, in the first variable, 
\begin{equation}
W_1 = \oint dx\, \sqrt{2E_1-\omega^2x^2} = 2\pi {E_1\over \omega}
\end{equation}
while in the second variable $\xi$ one obtains
\begin{equation}
W_2 = \oint d\xi \sqrt{2E_2 - \omega^2\xi^2 - {\hbar^2\sigma^2\over
4\xi^2}}~. 
\end{equation}
Under the substitution $r = \xi^2/2$ this turns out to be the same
integral which occurs in the Kepler problem, again pointing out the
formal similarity with the angular momentum barrier in a central
potential.  Using standard methods \cite{goldstein}, one finds
\begin{equation}
W_2 = 2\pi {E_2\over \omega} - \pi \hbar\sigma\,.
\end{equation}
Hence the full adiabatic invariant is
\begin{equation}
{W\over 2\pi \hbar} = {E\over \hbar \omega} - {\sigma\over 2} = \tilde
N - N\,. 
\label{adbinv}
\end{equation}
upon using the definition of $\tilde N$, the definition of $\sigma$
and 
\begin{equation}
E = \langle {\bf H}\rangle = \langle {\bf H}_{osc}\rangle = \hbar
\omega  
\left(\tilde N + {1\over 2}\right)\,.
\end{equation}
Since $N$ is strictly a constant of the motion, Eqn. (\ref{adbinv})
informs us that $\tilde N$ is an adiabatic invariant and therefore
slowly varying if $\omega^2 = \chi$ is a slowly varying function of
time. In the language of classical action-angle variables, $W$ is an
action variable which is slowly varying while the angle variable
conjugate to it varies rapidly in time (linearly with time in the
limit $\omega$ is a constant, as in the exponent in
(\ref{ftilde})). This means that although the density matrix
${\bf\rho}$ is certainly not diagonal in the $\tilde n$ number basis,
its off-diagonal elements which depend on the angle variable will be
very rapidly varying functions of time, whereas its diagonal elements
in this basis will be only slowly varying. This is clearly seen in the
explicit form for the matrix elements of this basis in
Eqn. (\ref{app:rhofd}) of Appendix B. Hence if we are interested only in
the motion of the mean fields which are slowly varying functions of
time we may average over the rapid phase variations in the
off-diagonal matrix elements of $\rho$ in this adiabatic number
basis. For only a single quantum degree of freedom this amounts to a
time-averaging, and can be implemented only by fiat.  However in field
theory there are many momentum modes, so that this averaging is
actually performed for us in the mean field evolution equations by the
integrations over $k$ at fixed $t$.

\vspace{.4cm}
\epsfxsize=7cm
\epsfysize=5.5cm
\centerline{\epsfbox{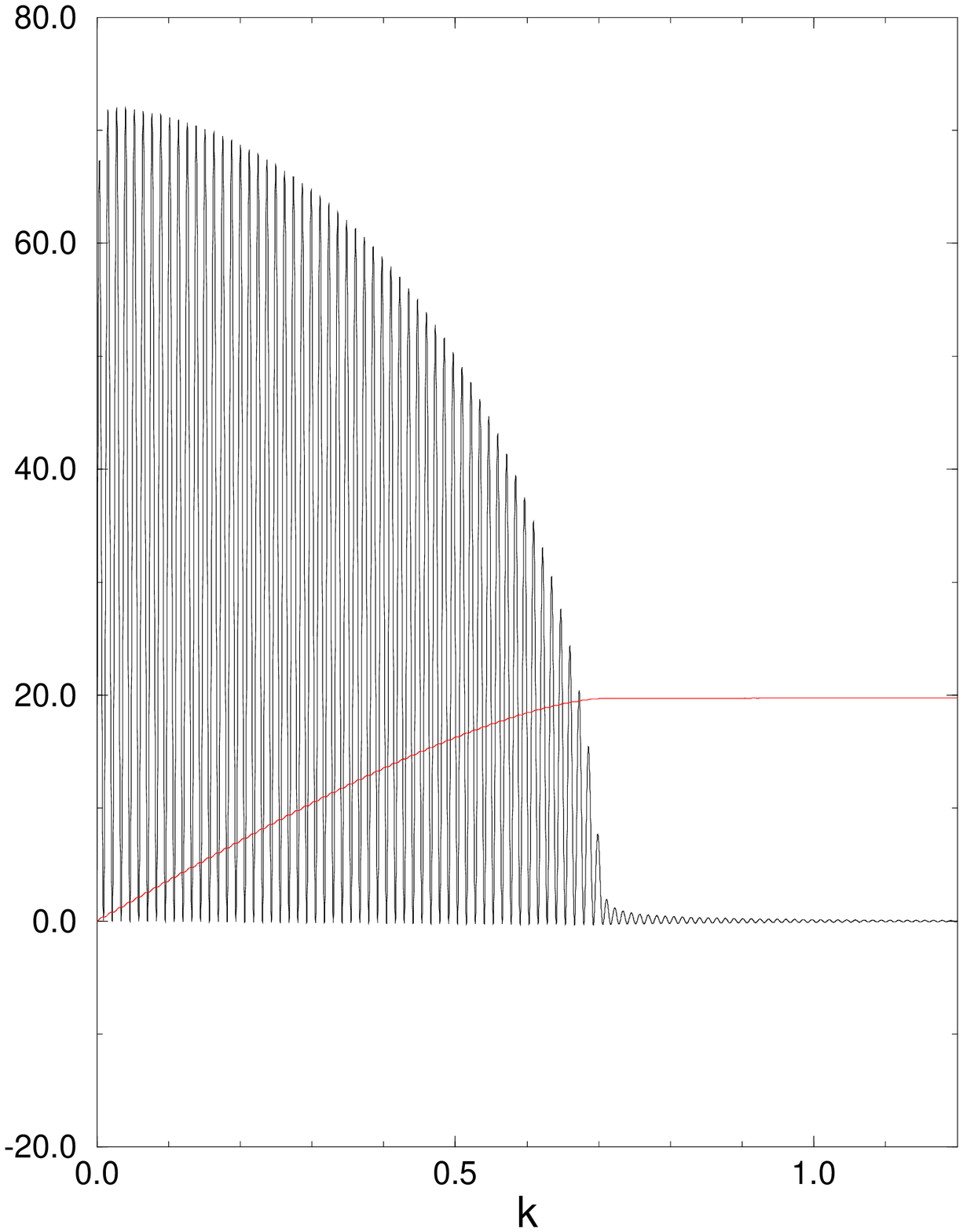}}
\vspace{.35cm}
{FIG. 8. {\small{Integrand and integral of the RHS of
Eqn. (\ref{gapint}) as a function of $k$ for fixed $t = 257$.}}}\\

\vspace{.2cm}
\epsfxsize=8cm
\epsfysize=5.5cm
\centerline{\epsfbox{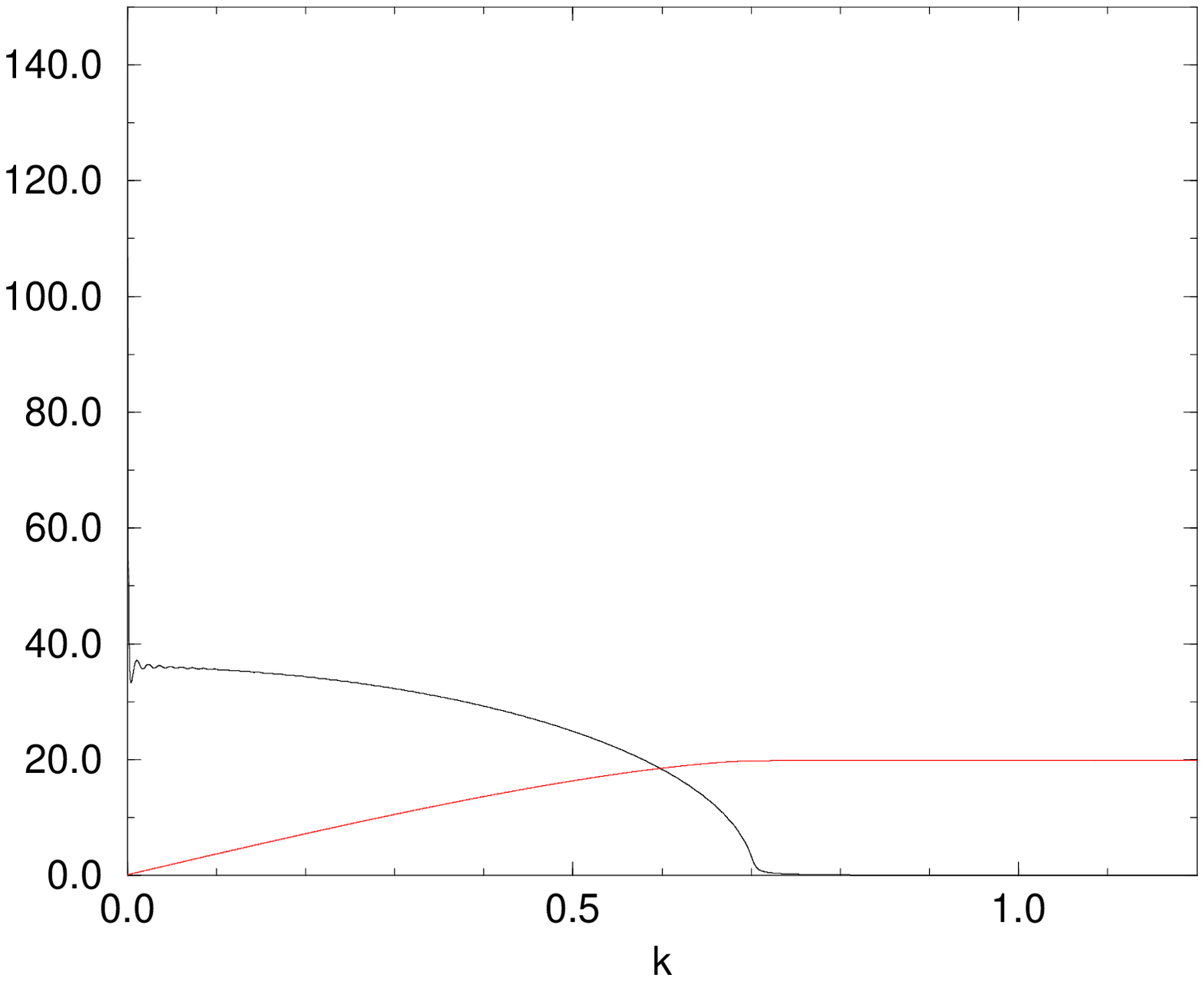}}
{FIG. 9. {\small{Integrand and integral of the phase averaged quantity
defined by the adiabatic particle number basis as a function of $k$
for the same fixed $t = 257$. The value of the integral using the
replacement (\ref{repl}) is the same as that of the previous
figure.}}}\\

The term {\em dephasing} has a precise meaning in this adiabatic
particle number representation of the density matrix.
To the extent that the phases in the off-diagonal matrix elements of
the density matrix in the adiabatic $\tilde n$ basis are rapidly
varying in time, they should have little or no effect on the
evolution of more slowly varying quantities such as the mean fields
$\phi$ and $\chi$ or the mean adiabatic particle number $\langle \tilde N
\rangle$ itself. Thus, a natural approximation to the full density
matrix $\bf \rho$ is immediately suggested by the existence of the
adiabatic number basis, namely to discard the off-diagonal elements
(see (\ref{app:rhofd}) below) of $\bf \rho$ in this basis, which is 
equivalent to
time-averaging the distribution of fluctuations in the exact Gaussian
density matrix (\ref{gaussd}) over times long compared to their rapid
variations. In field theory, where there are many Fourier modes, each
with its own rapid phase variation, the effect of dephasing may be
obtained by simply integrating over the momentum index $k$ at fixed
time. In either case, we expect this averaging procedure to scarcely
affect the actual evolution of the mean fields, and the extent to
which this expectation is realized is the extent to which dephasing of
the fluctuations is effective, and the evolution is irreversible. In
Figs. 8 and 9 we compare the actual $k$ dependence of the function
$G(k,t)$ appearing in the mean field evolution equation (\ref{ceq})
with the phase averaged quantity defined by the adiabatic particle
number $\tilde N(k)$, {\em i.e.} we make the replacement,
\begin{equation}
-i\hbar G(k,t) = |f_k(t)|^2 \rightarrow {1\over 2k} (2 \tilde N (k) + 1)\,.
\label{repl}
\end{equation}
The relevant integral in the renormalized gap 
equation (\ref{cren}) is
\begin{eqnarray}
&&G(x,x,\chi)-G(x,x,\chi=0)=\nonumber\\
&&{1\over 4\pi^2} \int_0^{\Lambda}\,
k^2dk\left\{\vert f_k(t) \vert^2\sigma_k - {\hbar\over 2k}\right\}.
\label{gapint}
\end{eqnarray}
Notice that although there are many more
oscillations in the integrand of the RHS of Eqn. (\ref{gapint}) than in 
the phase averaged quantity, the integrals over $k$ of the two quantities 
are indistinguishable. 

The adiabatic particle basis may also be used to define effective
Boltzmann and von Neumann entropies \cite{Kand} through the diagonal matrix
elements (\ref{app:rhofd}) below. Neglecting the rapidly varying
off-diagonal matrix elements gives
\begin{eqnarray}
S_{B} &\equiv &\sum_{\vec k} \left\{ \left( \tilde N(\vec k) 
+ 1 \right) \ln \left( \tilde N(\vec k) + 1\right) - \tilde N(\vec
k) \ln \tilde N(\vec k) \right\}\nonumber\\ 
S_{eff} &\equiv& - {\rm Tr}~\rho_{eff}\ln\,\rho_{eff} \nonumber\\
&=& -\sum_{\vec k}\sum_{\ell = 0}^{\infty} \rho_{2\ell}(\vec k) \ln\, 
\rho_{2\ell} (\vec k)
\label{seff}   
\end{eqnarray}
for the Boltzmann and effective von Neumann entropy of the truncated
density matrix, where $\rho_{2\ell}(\vec k)$ is given by
Eqn. (\ref{rhopair}) below. The evolution of these two quantities for
a typical solution of the mean field equations is shown in
Fig. 10. Both display general increase due to continuous creation of
massless Goldstone particles near threshold \cite{CalHu}. Neither
quantity is a strictly monotonic function of time and neither obeys a
strict Boltzmann $H$-theorem. Since the particle modes $f_{k}$
interact with the mean field $\chi$ but not directly with each other,
the effective damping observed is certainly {\em collisionless}, and
the dephasing here is similar to that responsible for Landau damping
of collective modes in classical electromagnetic plasmas. The entropy
$S_{eff}$ of the effective density matrix provides a precise measure
of the information lost by treating the phases as random. The
Boltzmann ``entropy'' would be expected to equal $S_{eff}$ only in
true thermodynamic equilibrium, which is not achieved in the
collisionless approximation of Eqns. (\ref{modefn}). Notice the
non-thermal distribution of particles in Fig. 9. In this non-thermal
distribution we see from Fig. 10 that the Boltzmann entropy $S_B$
generally overestimates the amount of information lost by phase
averaging.

\vspace{.4cm}
\epsfxsize=7cm
\epsfysize=5.5cm
\centerline{\epsfbox{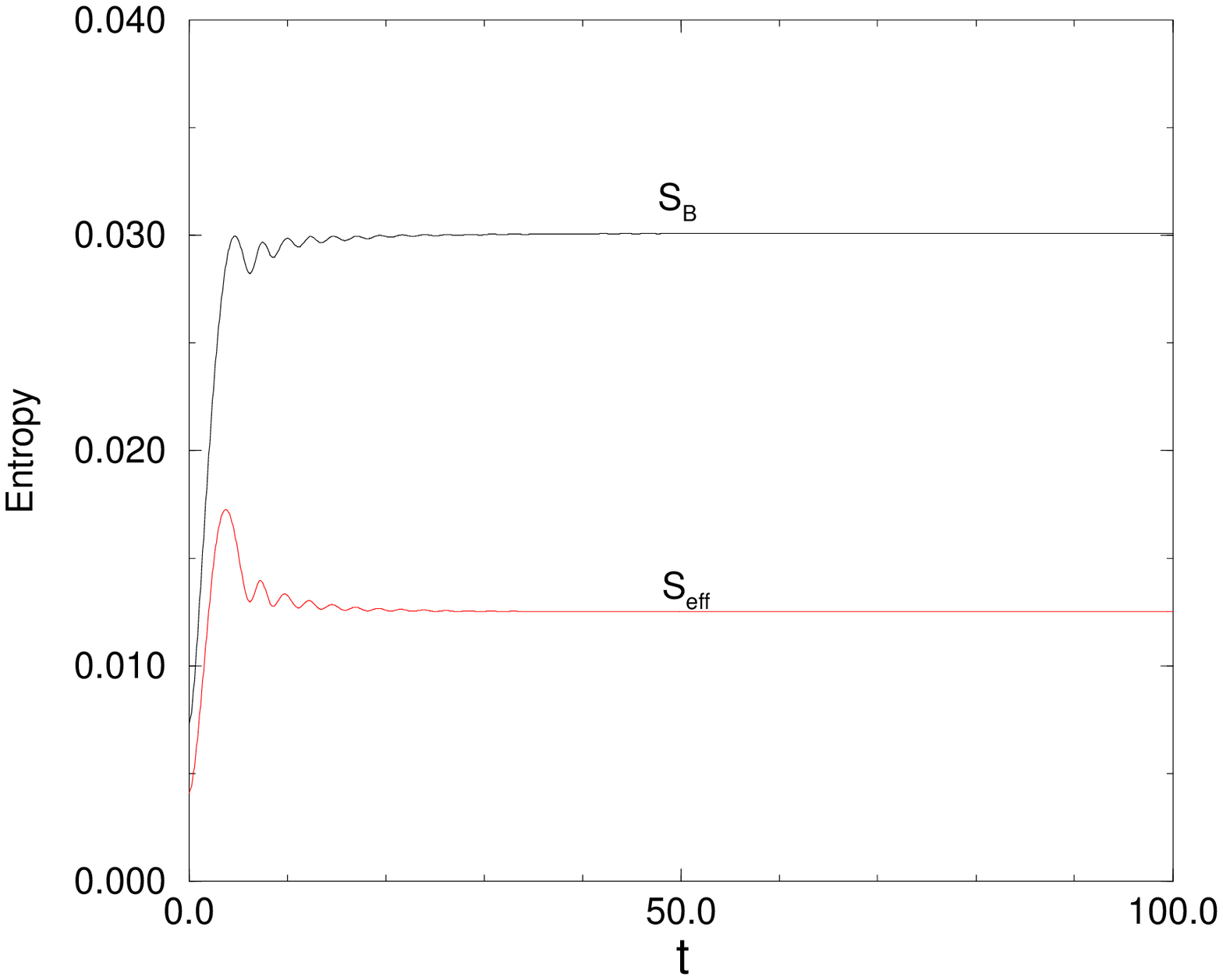}}
\vspace{.35cm} 
{FIG. 10. {\small{Evolution of the Boltzmann and effective von
Neumann entropies of the diagonal elements of the density matrix
in the adiabatic particle number basis.}}}\\   

To the extent that the phase information in the off-diagonal matrix
elements of $\rho$ is irretrievable the system has become effectively
classical, in the sense that the quantum interference effects present
in the original ensemble represented by $\rho$ are washed out as
well. In that case we might as well regard the ensemble represented by
the diagonal, truncated effective density matrix in the adiabatic
number basis as a {\em classical} probability distribution with the
diagonal elements of $\langle \tilde n = 2\ell|\rho|\tilde n=
2\ell\rangle$ giving the classical probabilities of observing $\ell$
pairs in the ensemble. This probability distribution is derived in
Appendix B and given by
\begin{eqnarray}
\rho_{2\ell}(\vec k) &=&
\langle \tilde n_{\vec k}=2\ell|\rho|\tilde n_{\vec k} = 2\ell\rangle
\Big\vert_{_{\stackrel{\sigma =1}{\phi =\dot\phi = 0}}}\nonumber\\ 
&=& {(2\ell -1)!!\over 2^{\ell}\, \ell !} {\rm sech} \gamma_{\vec k}
\,\tanh^{2\ell} \gamma_{\vec k}
\label{rhopair}  
\end{eqnarray}
where $\gamma$ is the magnitude of the Bogoliubov transformation
between the $a$ and $\tilde a$ bases, given explicitly in terms of the
mode functions by
\begin{equation}
\tilde N (\vec k) = \sinh^2 \gamma_{\vec k}  = {\vert \dot f_k + i
\omega_k f_k\vert^2 \over 2 \hbar\omega_k}\,, 
\label{bogg}   
\end{equation}
which depends only on the magnitude of $\vec k$ by spatial
homogeneity.  Sampling this distribution with random phases will yield
typical classical field amplitudes which make up the
distribution. Obtaining such typical classical fields in the ensemble
can give us explicit realizations of the symmetry breaking behavior of
the system, as well as providing the starting point for the study of
topological defects produced during the phase transition.

\vspace{.4cm}
\epsfxsize=7cm
\epsfysize=5.5cm
\centerline{\epsfbox{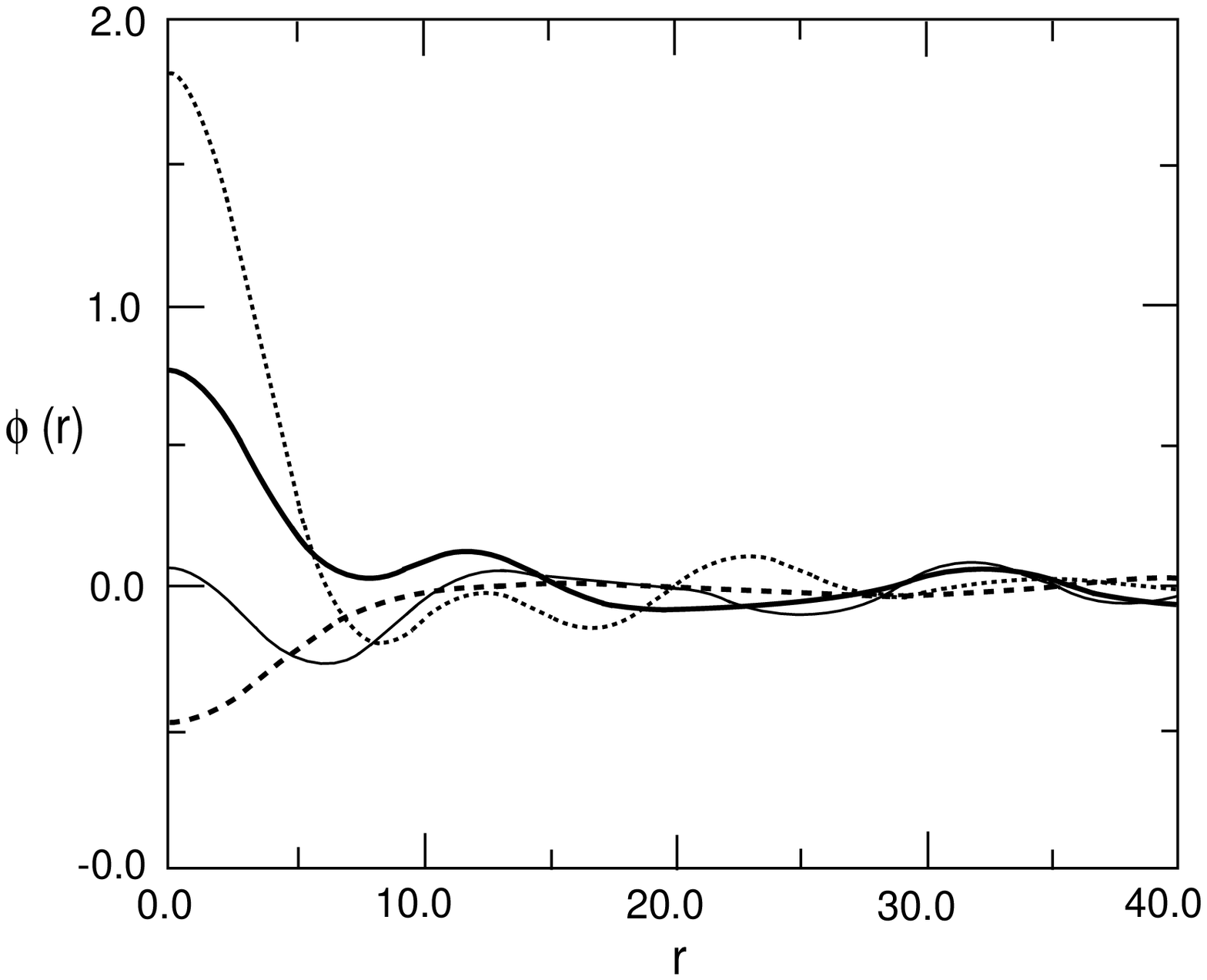}}
\vspace{.35cm}
{FIG. 11. {\small{Four typical field configurations drawn
from the same classical distribution of probabilities in the
adiabatic particle number basis, according to Eqn. (\ref{sample}),
for t=490 in the case the mean field $\phi = 0$. The units of $r$
are $v^{-1}$ and $\lambda_{\Lambda}=1$.}}}\\

The sampling of the field configurations proceeds in several steps.
For a fixed late time and Fourier wave number $k$ we calculate the
Bogoliubov transformation coefficient from the static $n$ to
time-dependent adiabatic particle number $\tilde n$ basis from
Eqn. (\ref{bogg}). This gives
a set of numbers $\rho_{2\ell} = \langle \tilde n=2\ell|\rho|\tilde
n = 2\ell\rangle$ normalized to unit total probability,
\begin{equation}
\sum_{\ell = 0}^{\infty} \rho_{2\ell} =1
\end{equation}
and which typically fall off very rapidly with $\ell$ so that only a
finite number of the $\rho_{2\ell}$ need be retained.  Then we sample
this distribution by drawing a random number $q$ in the unit interval
$[0, 1]$. Looking at the table of $\rho_{2\ell}$ and the partial sums,
\begin{equation}
Q_{\ell} = \sum_{\ell' = 0}^{\ell} \rho_{\ell'}
\end{equation}
we find $\ell$ such that
\begin{equation}
Q_{\ell -1} < q \le Q_{\ell}
\end{equation}
to determine $\tilde n = 2 \ell$ of this random drawing for the given
value of $k$. We then write 
\begin{eqnarray}
\phi (r, t) &=& {1\over \sqrt V}\sum_{\bf k} {1\over \sqrt {2
\omega_k}}\left( a_{\bf k} e^{i{\bf k \cdot x}} + a^*_{\bf k}
e^{-i{\bf k \cdot x}}\right)  
\nonumber\\
&\rightarrow &{\sqrt V \over \pi^2 r} \int_0^{\infty} \, dk\, k\, \sin
kr {\sqrt{\tilde n_k \over 2 \omega_k}} \cos \theta_k\,,
\label{sample}
\end{eqnarray}
where we have performed the angular integrations in $d=3$ dimensions,
and written
\begin{equation}
a_k = \sqrt{{\tilde n}_k}\, e^{i\theta_k}
\label{ranph}
\end{equation}
in terms of a random phase $\theta_k$. After performing the Fourier
transform in (\ref{sample}) the result is a typical field
configuration as a function of radial $r$ at the fixed time $t$. In
this way we obtained the four field configurations shown in
Fig. 11. We observe that although the mean field $\phi =0$ when
averaged over the entire ensemble, typical field configurations in the
ensemble are quite far from zero. In fact they sample values of $\phi$
between the two minima at $v$ and $-v$.  We observe as well that there
is a typical correlation length in the classical field configurations
which is of the order of the inverse momentum in which the power of
the two-point function is distributed, as in Fig. 9.

It is clearly possible by such sampling techniques to generate typical
field configurations with any number of components $\Phi_i$ in any
number of spatial dimensions $d$. Then by appropriately matching the
number of components to the number of dimensions we could search for
different types of defect-like structures such as vortices, strings,
and domain walls. This classical description of the quantum mean field
theory is possible only when a classical decoherent limit exists
through the diagonal density matrix in the adiabatic particle number
basis. Moreover, the classical fields generated by this procedure are
smooth and free of any cut-off dependent short distance effects. The
extraction of smooth classical field configurations from a quantum
mean field description is a direct consequence of dephasing and the
definition of defect number in these smooth configurations is free of
the difficulties encountered when classical definitions of defects are
applied uncritically to quantum field theories. It would be very
interesting to pursue these ideas further in more realistic models of
phase transitions where such topological defects are expected. This we
leave for a future investigation.

The dephasing of the density matrix justifies the replacement of the
exact Gaussian $\bf \rho$ by its diagonal elements only in the
adiabatic number basis leading to the effective density matrix ${\bf
\rho}_{eff}$. Now if this effective density matrix is transformed back
into the coordinate basis it corresponds to a density matrix of the
original Gaussian form but with zero mean fields, zero momentum
$\eta_k =0$ and $\tilde \sigma_k \equiv 1 + 2 \tilde N(k)$ replacing
$\sigma$, {\em i.e.} we obtain a product of Gaussians of the form,
\begin{eqnarray}
&&\langle \varphi'_k|{\bf\rho}_{eff} |\varphi_k\rangle = \nonumber\\
&& (2\pi \xi_{\bf k}^2)^{-{1\over 2}}
\exp \biggl\{-{\tilde\sigma_{\bf k}^2\over 8 \xi_{\bf k}^2}(\varphi_k'- 
\varphi_k)^2  
-{1\over 8 \xi_{\bf k}^2}(\varphi_k'+ \varphi_k)^2 \biggr\}~,\nonumber\\
\label{gaueff} 
\end{eqnarray}
in all the $k \neq 0$ modes where the mean field vanishes. This form
shows that the Gaussian distribution off the diagonal $\varphi_k'=
\varphi_k$ is strongly suppressed compared to the diagonal
distribution. Indeed in the numerical evolutions we have found that $k
\tilde N(k)$ behaves like a constant for small $k$. This means that
$\tilde\sigma\propto 1/k$ which is large for small $k$.  Hence whereas
the root mean square deviation from zero along the diagonal
$\varphi_k'= \varphi_k$ is $\xi_k \cong \sqrt{\hbar\tilde N(k)/k}$
which is large, the root mean square deviation from zero in the
orthogonal direction off the diagonal is $${\xi_k\over \sigma_k} \cong
\sqrt{\hbar\over 2k\tilde N(k)}$$ which is much smaller. The result is
that there is virtually no support in this distribution for
``Schr\"odinger cat'' states in which quantum interference effects
between the two classically allowed macroscopic states at $v$ and $-v$
can be observed.  Instead we see in a different way how dephasing
produces an ensemble which may be regarded as a classical probability
distribution over classically distinct outcomes ({\em i.e.} the
diagonal field amplitude distribution) but with essentially no
components in the off-diagonal quantum interference between these
classical configurations. This is illustrated in Fig. 12. Hence even
in this rather simple collisionless approximation the particle
creation effects in the time dependent mean field give rise to strong
suppression of quantum interference effects and mediate the quantum to
classical transition of the ensemble.

\vspace{.4cm}
\epsfxsize=7cm
\epsfysize=5cm
\centerline{\epsfbox{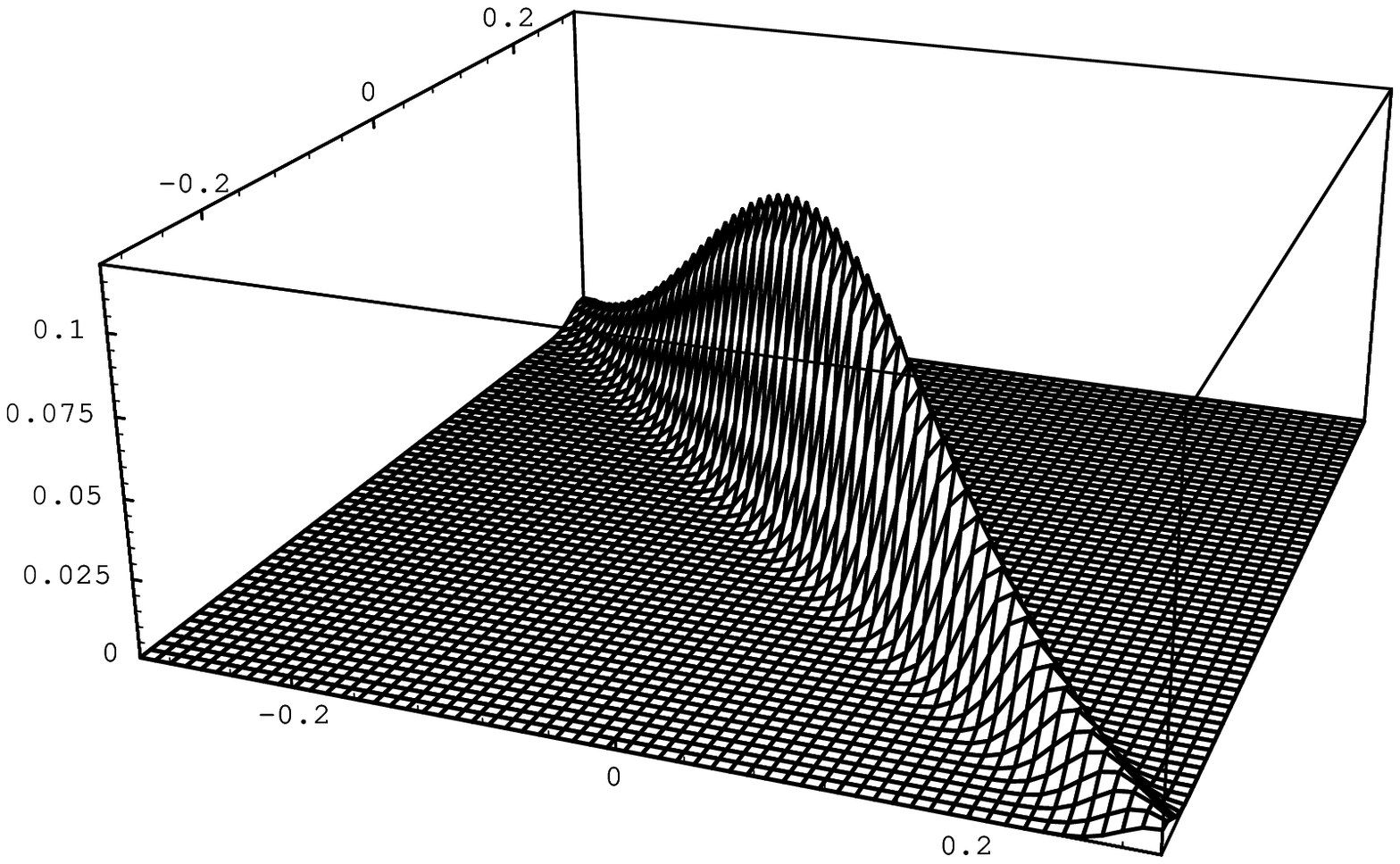}}
\vspace{.35cm}
{FIG. 12. {\small{The Gaussian $\rho_{eff}$ for $k=.4$ for the data
and parameter values of Fig. 9 illustrating the strong suppression of
off-diagonal components due to dephasing.}}}\\

When more realistic collisional interactions are included at the
next and higher orders beyond the leading mean field limit we can
expect this transition to decoherent classical behavior to be even
more pronounced. As the Gaussian assumption is relaxed we expect the
single peak at the origin to split into two peaks. Hence we can begin
to see how a quantum phase transition leads to an effective
classically broken symmetry in which large domains are in a definite
classical ground state or another but not in a quantum superposition
of ground states.
  
\section{Linear Response, Plasma Oscillations and Damping Rate}
\label{sec:level6}

The very efficient damping of the oscillations around the final state
which we have studied in the density matrix formalism of the last
section may also be understood as due to the continuous creation of
massless Goldstone particles near threshold. In this appearance of
strictly massless modes the spontaneously broken $\lambda\Phi^4$ model
is qualitatively different from our previous studies of QED in the
large $N$ approximation \cite{QED}. In that case the charged particles
are massive and there is finally a tunneling barrier which shuts off
particle creation effects after the mean electric field has decreased
below a certain critical value. Beyond this point the mean field
undergoes essentially undamped plasma oscillations since there is no
transfer of energy to created particles. In contrast, the present
model has no such critical threshold in the mean field $\chi$ which is
free to continue creating massless bosons for arbitrarily small
amplitude and $\chi \rightarrow 0$ asymptotically. This very efficient
asymptotic damping of the $\chi$ mean field towards its stationary
spontaneously broken solution at $\chi =0$ is very well illustrated by
the numerical results shown in Figs. 13 and 14 for
$\lambda_{\Lambda}=1$ and an initial $\chi=-.5$ 

\vspace{1cm}
\epsfxsize=7cm
\epsfysize=5cm
\centerline{\epsfbox{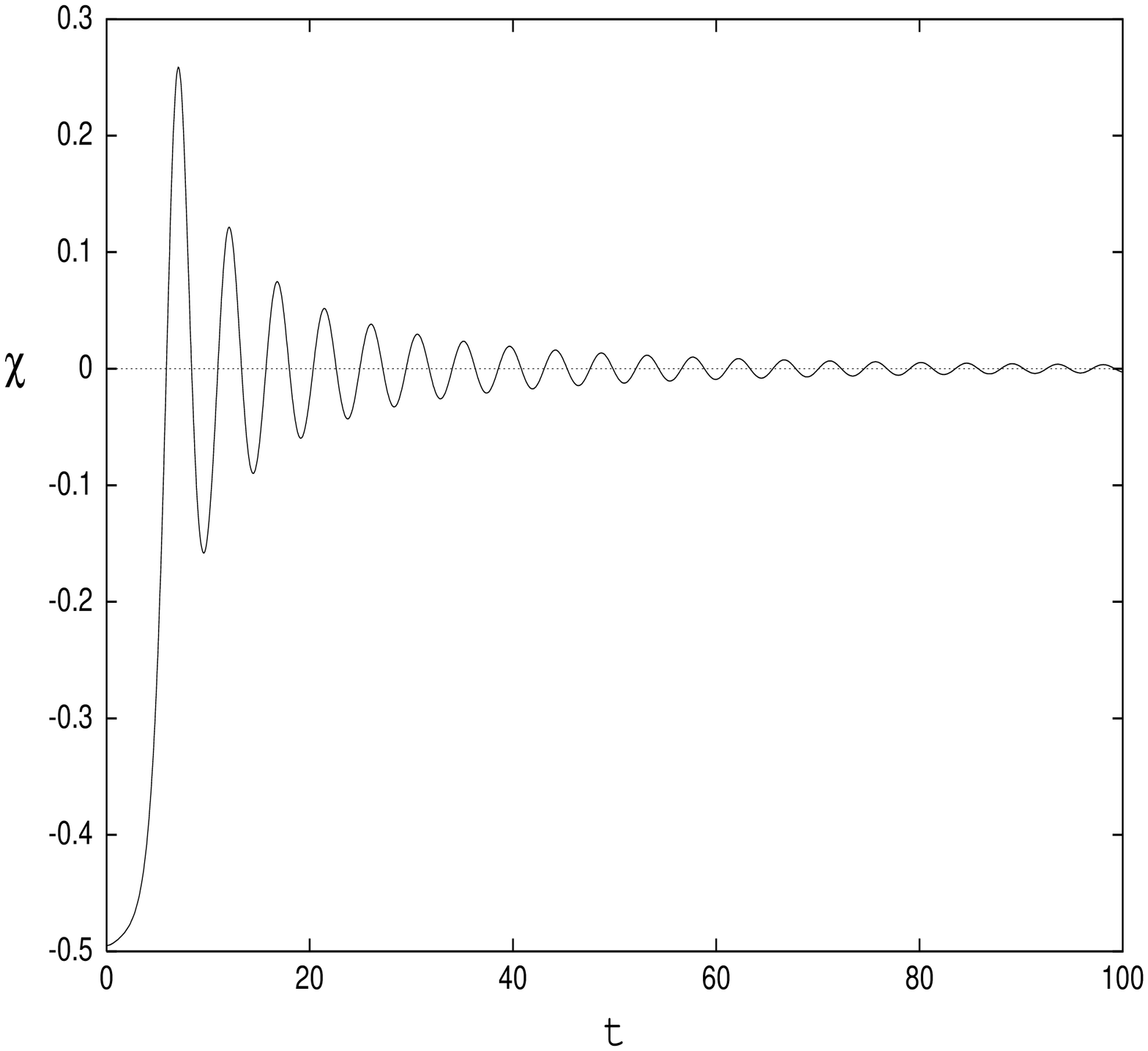}}
{FIG. 13. {\small{Time evolution and effective damping of
the $\chi$ mean field towards zero, its stationary value in the
spontaneously broken phase.}}}\\    

Since $\chi$ becomes very small at late times, it is possible to
analyze the approach to zero by linear response methods. In principle
this is possible for any initial condition but we restrict ourselves
to the thermal and vacuum cases. With thermal initial conditions, both
the frequency and the damping rate of the plasma oscillations can be
calculated and are in very good agreement with the simulations as will
be discussed further below. The long time behavior for initial
conditions with $N(k)=0$ is somewhat more complicated. While the
oscillation frequency is still very well determined by the linear
response analysis, the oscillation envelope does not damp
exponentially. Instead it falls off as an exponential multiplied by a
power law in time.

\vspace{1cm}
\epsfxsize=7cm
\epsfysize=5cm
\centerline{\epsfbox{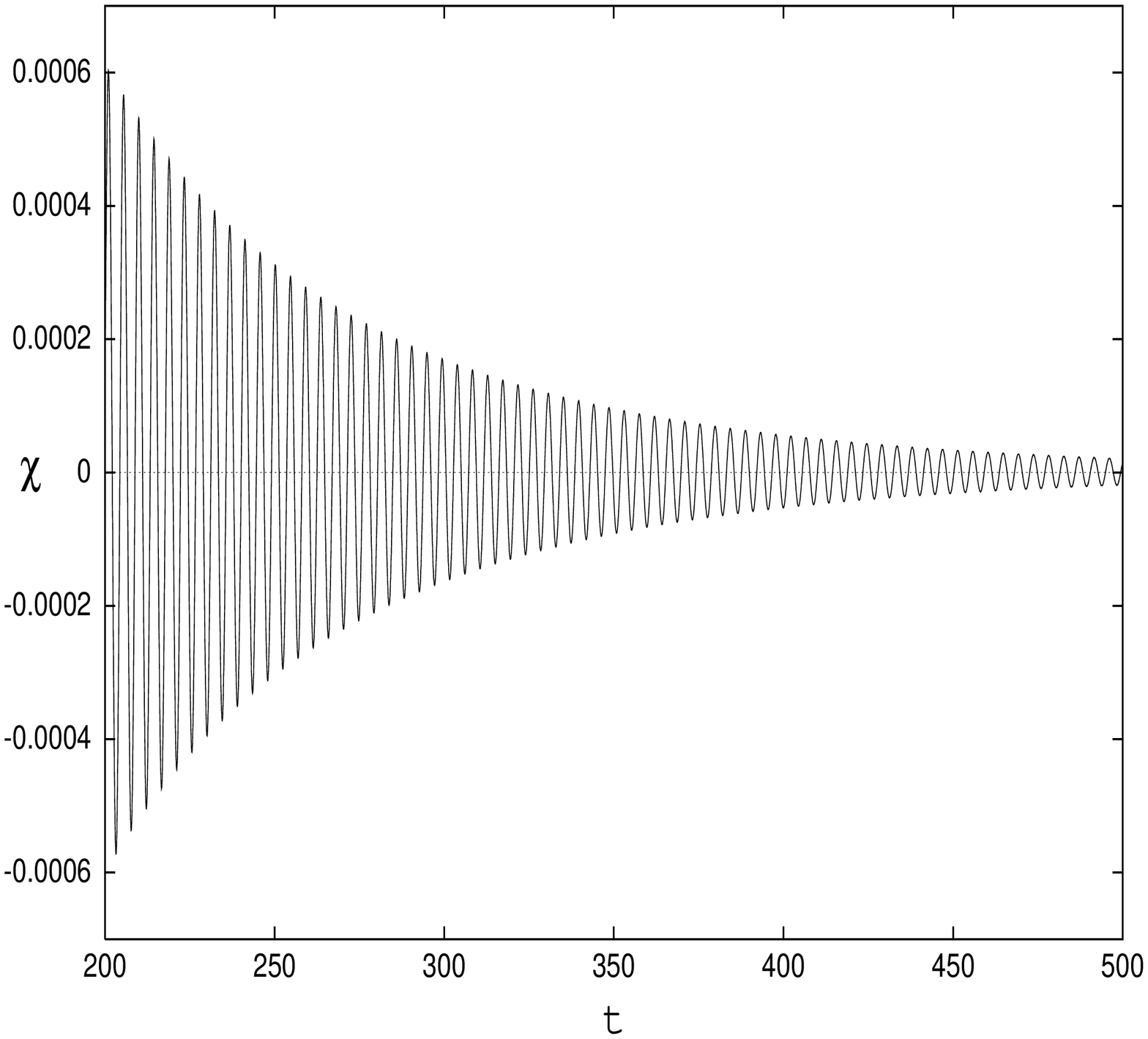}}
{FIG. 14. {\small{Same evolution as in Fig. 13 but followed to large
times and with an expanded scale for $\chi$ to show the long time
behavior of the time-dependent Goldstone mass squared.}}}\\

Given any static solution to the equations of motion with spontaneous
symmetry breaking, as in (\ref{ssbd}), we may consider the real time
linear response of the system away from this static solution.  This is
accomplished by linearizing the evolution equations in the deviations
from the static solution,
\begin{eqnarray}
\phi &\rightarrow& \phi + \delta\phi\, \nonumber\\
\chi &\rightarrow& 0 + \chi\, \nonumber\\  
f_k &\rightarrow& f_k(t) + \delta f_k(t)\ ,
\end{eqnarray}
with
\begin{equation}
f_k(t) = {\sqrt{\hbar\over 2 k}}\exp \left(-ikt\right)
\end{equation}
the mode functions corresponding to the static solution, $\chi = 0$.

The linearized mode equation,
\begin{equation}
\left[{d^2 \over dt^2} +  k^2 \right] \delta f_k (t) = 
- \chi (t)f_k(t) 
\end{equation}
may be solved by use of the retarded 
Green's function,
\begin{equation}
G_R(t-t';k) = {\sin k(t-t')\over k} \theta (t-t')
\end{equation}
in the form,
\begin{eqnarray}
\delta f_k(t) &=& A_kf_k(t) + B_k f_k^*(t)\nonumber\\
&& - \int_0^t\,dt'\, G_R(t-t';k)\chi(t')f_k(t')
\label{delf}
\end{eqnarray}
where $A_k$ and $B_k$ are coefficients of the solutions to the
homogeneous equation. Because the Wronskian condition (\ref{wron}) is
maintained under the linearization, the $A_k$ must have vanishing real
part,
\begin{equation}
{\rm Re}\, A_k = 0\ .
\label{reA}
\end{equation}
The linearized $\phi$ equation gives in the same way,
\begin{eqnarray}
\delta\phi(t) &=& t\,\delta\dot\phi(0) + \delta\phi(0)\nonumber\\
&&- \int_0^t\,dt'\, G_R(t-t';k=0)\chi(t')\phi(t')\,.
\label{delph}
\end{eqnarray}
At the same time the linearized gap equation reads
\begin{equation}
\chi  = \lambda \phi \,\delta\phi  + \lambda \int [d {\bf k}]\,
{\rm Re} (\delta f_k\, f_k^*) ( 1 + 2 N(k))\ .
\label{lingap}
\end{equation}
Upon substituting Eqns. (\ref{delf}-\ref{delph}) this becomes a 
linear integral equation for the perturbation, $\chi$,
\begin{equation}
\chi (t) = -\lambda \int_0^t\, dt'\, \Pi (t-t') \chi(t')
+ \lambda B(t)\ ,
\label{linchi}
\end{equation}
where
\begin{equation}
\Pi (t) = t\phi^2 + \int [d{\bf k}]{(1 + 2 N(k)) \over 2 k^2} \sin
kt\, \cos kt 
\end{equation}
is the polarization part in the static background and
\begin{eqnarray}
B(t) &\equiv& t\phi\, \delta\dot\phi (0)  + \phi\delta\phi(0) 
\nonumber\\
&&+ \int [d{\bf k}]{(1 + 2 N(k)) \over 2 k} {\rm Re} (B_k e^{2ikt})
\end{eqnarray}
depends only on the initial perturbation away from the static
solution. The linearized integral equation (\ref{linchi}) may be put
in the form
\begin{equation}
\int_0^t\, dt'\, D^{-1}(t-t')\chi (t') = -B(t)
\end{equation}
where $D^{-1}$ the $\chi$ inverse propagator function. 
 
The most direct method of solving such integral equations is to make
use of the Laplace transforms,
\begin{eqnarray}
\tilde{\Pi} (s) &\equiv& \int_0^{\infty} dt\, e^{-st} \Pi (t) \nonumber\\
&=&{\phi^2\over s^2} + \int [d{\bf k}]\ {(1 + 2 N(k)) \over 2 k (s^2 +
4k^2)}~,  \nonumber\\
\tilde B(s) &\equiv& \int_0^{\infty} dt\, e^{-st} B(t) \nonumber\\ 
&=&{\phi\,\delta\dot\phi(0)\over s^2} 
+ \int [d{\bf k}]\ {\left(1 + 2 N(k)\right) \over 2 k} {\rm
Re}\left({B_k\over s-2ik}\right)~,\nonumber\\ 
\tilde{\chi}(s)  &\equiv& \int_0^{\infty} dt\, e^{-st} \chi(t) = -\tilde
D(s) B(s)~, 
\end{eqnarray}
and $\tilde D(s)$, the simple algebraic reciprocal of the Laplace
transform of $D^{-1}(t)$,
\begin{equation}
-\tilde D^{-1}(s) = {1\over\lambda} + \tilde \Pi (s) = 
{1\over\lambda} + {\phi^2\over s^2} + \int [d{\bf k}]\ {(1 + 2 N(k))
\over 2 k (s^2 + 4k^2)}\ . 
\end{equation}  
Both the bare coupling $\lambda = \lambda_{\Lambda}$ and the integral
over $\bf k$ involving the $N(k)$ independent term are logarithmically
divergent. They combine to give the ultraviolet finite contribution,
\begin{eqnarray}
{1\over\lambda_R (s^2/4)} &=& {1\over\lambda_{\Lambda}} + {\hbar\over
32\pi^2} \ln \left({4\Lambda^2 \over s^2}\right)\nonumber\\
&=& {1\over\lambda_R(m^2)} +  {\hbar\over 32\pi^2} \ln \left({4m^2 
\over s^2}\right)\ .
\label{lams}
\end{eqnarray}
The renormalized sum rule (\ref{sumrul}) may be used in the $N(k)$
dependent term to secure as well
\begin{eqnarray}
-\tilde D^{-1}(s) &=& {1\over\lambda_R (s^2/4)} + {v^2 \over s^2}
\nonumber\\ 
&+&{\hbar \over 2\pi^2} \int_0^{\infty} k dk
N(k) \left({1\over s^2 + 4k^2} - {1\over s^2}\right)
\label{Dinvlt}
\end{eqnarray}
which is now independent of the static value of $\phi$.

\vspace{.4cm}
\epsfxsize=4.5cm
\epsfysize=2.5cm
\centerline{\epsfbox{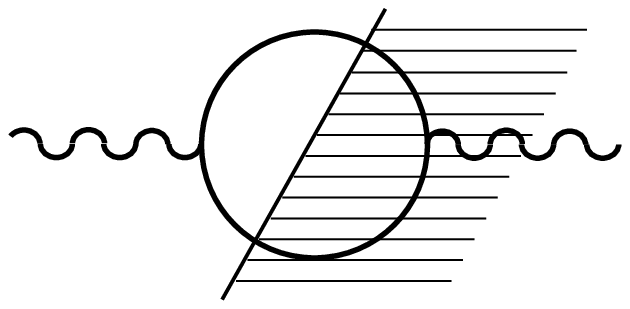}}
\vspace{.35cm}
{FIG. 15. {\small{One loop diagram which contributes to the
polarization $\Pi= -{i\over 2}G\,G$ in the leading order large $N$
approximation.}}}\\  

The function $\tilde D^{-1}(s)$ has a singularity at $s=0$ and
depending on the exact form of $N(k)$ near $k =0$ in general, a branch
cut starting from the origin in the complex $s$ plane which may be
taken along the negative real $s$ axis. Its physical origin is the
zero mass Goldstone bosons propagating in the internal loop of the
polarization $\Pi$ in Fig. 15. The discontinuity across this cut
arises from the non-zero probability for the time dependent $\chi$
field to create Goldstone pairs with arbitrarily small spatial
momentum $k$ above the massless threshold.  The function $\tilde
D^{-1}(s)$ also may have one or more zeroes in the left-half complex
$s$ plane. The zero at
\begin{equation}
s^2 \cong \pm 4m^2\exp \left({32\pi^2\over \hbar\lambda_R(m^2)}\right)
\rightarrow \pm \infty  
\end{equation}
for $\lambda_R \ll 1$ is the Landau ghost pole in the far ultraviolet
which lies outside the range of validity of the large $N$ expansion
and which in any case does not affect the long-time behavior of the
inverse Laplace transform. 

In the case of a strictly thermal distribution of massless bosons
(\ref{BE}), the integral in (\ref{Dinvlt}) may be performed in closed
form in terms of the digamma function $\psi$ and we find
\begin{eqnarray}
-\tilde D_T^{-1}(s) &=& {1\over\lambda_R (s^2/4)} + {\phi_T^2 \over
s^2} \nonumber\\ 
&&+ {\hbar \over 16\pi^2}\left[\ln\left(-{s\over 4\pi T}\right) + 
{2\pi T\over s} - \psi\left(-s\over 4\pi T\right)\right] \nonumber\\
&=& {1\over\lambda_R (4\pi^2 T^2)} + {\phi_T^2 \over s^2} \nonumber\\
&&+ {\hbar \over 16\pi^2}\left[{2\pi T\over s} - 
\psi\left(-{s\over 4\pi T}\right)\right]\,.
\label{DinvT}
\end{eqnarray}
A particular case is $T=0$ in which case the integral vanishes and
$\tilde D_0^{-1}(s)$ possesses a zero at complex $s_{\pm}$ in the
left-hand complex plane, 
\begin{equation}
s_{\pm} = \pm i \omega - \gamma
\label{zer}
\end{equation}
with
\begin{eqnarray}
\omega^2 &\cong &\lambda_{pl} v^2
\qquad {\rm and}\nonumber\\
\gamma &\cong & {\lambda_{pl} \over 64\pi}\,\omega =
{\lambda_{pl}^{3 \over 2}\over 64\pi} v
\label{omz}
\end{eqnarray}
where 
\begin{equation}
\lambda_{pl}=\lambda_R(s^2/4=-\omega_{pl}^2/4)~,
\label{ompl}
\end{equation}
for $\lambda_{pl} \ll 1$. In the frequency (\ref{omz}) the effective
mass of the ``radial'' mode, ignored in the direct quantization of the
$N-1$ massless modes in (\ref{pexp}) reappears in the large $N$ limit
as an oscillation in the real time linear response to perturbations
about the vacuum $\phi =v$, $ \chi=0$. It may be viewed either as this
radial degree of freedom of the $N$ component $\Phi_i$ field, with its
effective mass dressed by the polarization effects of $\Pi$ in the
presence of massless Goldstone modes, or alternately and just as
correctly, as a genuine collective excitation of the composite field
$\chi$. The two descriptions are equivalent since the oscillations of
$\chi$ and $\phi$ are constrained by the linearized equations
(\ref{delph}) and (\ref{lingap}), and there is only one degree of
freedom between them. The oscillation frequency (\ref{omz}) should be
compared with the second derivative of the free energy potential at
$T=0$ which vanishes from Eqn. (\ref{secder}). Thus the ``effective"
potential is completely ineffective at predicting the radial
oscillation frequency at zero temperature. It is clear that a zero
temperature the origin of the decay rate $\gamma$ is the imaginary
part of the two-particle cut in the polarization diagram illustrated
in Fig. 16.

At finite temperature the analytic structure of $\tilde D_T^{-1}(s)$
in the complex plane with Re $s < 0$ is as anticipated on general
physical grounds, {\em i.e.} there is a logarithmic branch cut
beginning at $s=0$ along the negative real $s$ axis with a
discontinuity that monotonically increases from zero to its asymptotic
value of $\pm i\hbar/32\pi$ from the logarithm in (\ref{lams}) as $s$
runs from $0$ to $-\infty$. Indeed, the digamma function of small
argument behaves like
\begin{equation}
\psi(z) \equiv {d\ln\Gamma(z)\over dz} \rightarrow -{1\over z} -
\gamma_E \qquad {\rm as} \qquad z\rightarrow 0
\label{dig}
\end{equation}
where $\gamma_E= 0.577215\dots$ is Euler's constant, so that we
see from the second form of (\ref{DinvT}) that the logarithms cancel
and the discontinuity vanishes, leaving a simple pole structure,
\begin{equation}
-\tilde D_T^{-1}(s) \rightarrow {1\over \lambda_{R, T}} +
{\phi_{_T}^2\over s^2} - {T\over 8\pi s}
\end{equation}
as $s \rightarrow 0$. Setting this to zero and solving the quadratic
equation informs us that $\tilde D_T^{-1}(s)$ has a zero at $s_{\pm}$
as in (\ref{zer}) with
\begin{eqnarray}
\omega_{pl\, T}^2 &= &\lambda_R (T'^2)\, \phi^2_T
\qquad {\rm and}\nonumber\\
\gamma_T &= & {\lambda_R (T'^2) \over 16\pi}\,T
\label{omT}
\end{eqnarray} 
for $|s_{\pm}| \ll T$ and $T'\equiv 2 \pi e^{-2\gamma_E}T$. 
This condition implies that Eqn. (\ref{omT}) is
only valid in the critical region,
\begin{equation}
\left( {T_c-T\over T_c}\right)^2 \ll {T^2\over \hbar\lambda_R T_c^2}
\end{equation}
where one would legitimately question the validity of the leading
order mean field approximation. The approach of $\omega_{pl\, T}^2$ to
zero as $T \rightarrow T_c$ is a manifestation of critical slowing
down. Notice also that this near critical temperature plasma
oscillation (\ref{omT}) does not go over smoothly to the zero
temperature case (\ref{omz}) since opposite limits of the digamma
function are involved, and $T \rightarrow 0$ is not permitted in
Eqn. (\ref{omT}) without violating the assumption $|s_{\pm}| \ll
T$. However, it seems likely that the simple poles found at zero
temperature migrate continuously from their location at $s_{\pm}$
given by (\ref{omz}) to the negative real axis as in (\ref{omT}) as the
temperature is raised from zero to $T_c$. It would be interesting to
map out the analytic structure of $\tilde D_T^{-1}(s)$ in the left
half complex $s$ plane as a function of $T$ to study the intermediate
temperature behavior of the linear response function from its low
temperature form to the critical region, without recourse to the
asymptotic expansion (\ref{dig}).  We reiterate that $\tilde
D_T^{-1}(s)$ describes the real time oscillations of the $\phi$ mean
field coupled to the collective plasma oscillations of the auxiliary
field $\chi$, and that the equilibrium free energy ``effective"
potential $F$ has no such real time information at either zero or
finite temperature, as conclusively demonstrated by (\ref{secder}).

The linear response predictions for the oscillation frequency and the
damping rate in the case of finite temperature were compared with a
numerical evolution with $\lambda_{\Lambda}=.01$ and $T=1$. The
theoretical predictions of $\omega_{plT}=.09574$ and
$\gamma_T=1.9894\times10^{-4}$ are in very good agreement with the
numerical results $\omega_{plT}=.09585$ and $\gamma_T=2\times
10^{-4}$. A similar test was carried out at a larger value of the
coupling $\lambda_{\Lambda}=.1$ again with excellent results:
predictions of $\omega_{plT}=0.3028$ and
$\gamma_T=1.9894\times10^{-3}$ compared with numerical values of
$\omega_{plT}=0.3031$ and $\gamma_T=1.9944\times10^{-3}$. The 
extraction of the behavior of the $\chi$ envelope from the numerical
data is described below.

Instead of plotting $\chi$ directly as a function of time, 
information may be extracted more conveniently by plotting 
\begin{equation}
y_t\equiv{\log(y(t+\epsilon))-\log(y(t))\over\log(t+\epsilon)-\log(t)}
\label{yy}
\end{equation}
as a function of time, where each time point $t$ is taken at either a 
peak or a trough of the oscillation and $\epsilon$ is the time 
separation between two such neighboring points. At late times, if the 
frequency stabilizes, $\epsilon$ goes to a constant. Using this fact, 
along with the late time condition $t\gg\epsilon$, one can show from 
(\ref{yy}) that if the $\chi$ oscillation envelope can be fit by an 
expression of the form
\begin{equation} 
At^{-\alpha}\exp{-\gamma t}~,
\label{envfit}
\end{equation}
then at late times,
\begin{equation}
y_t=-\gamma t-\alpha~,
\end{equation}
enabling the direct reading off of the power $\alpha$ and the exponent
$\gamma$ as the $y$-intercept and slope of a straight line
respectively. It is important to note that this method is also a test
of whether the frequency is stable as otherwise a straight line will
not be obtained. The behavior of $y_t$ against time is shown in Fig. 17
for the two values of the coupling constant mentioned above.

\vspace{.4cm}
\epsfxsize=7cm
\epsfysize=5cm
\centerline{\epsfbox{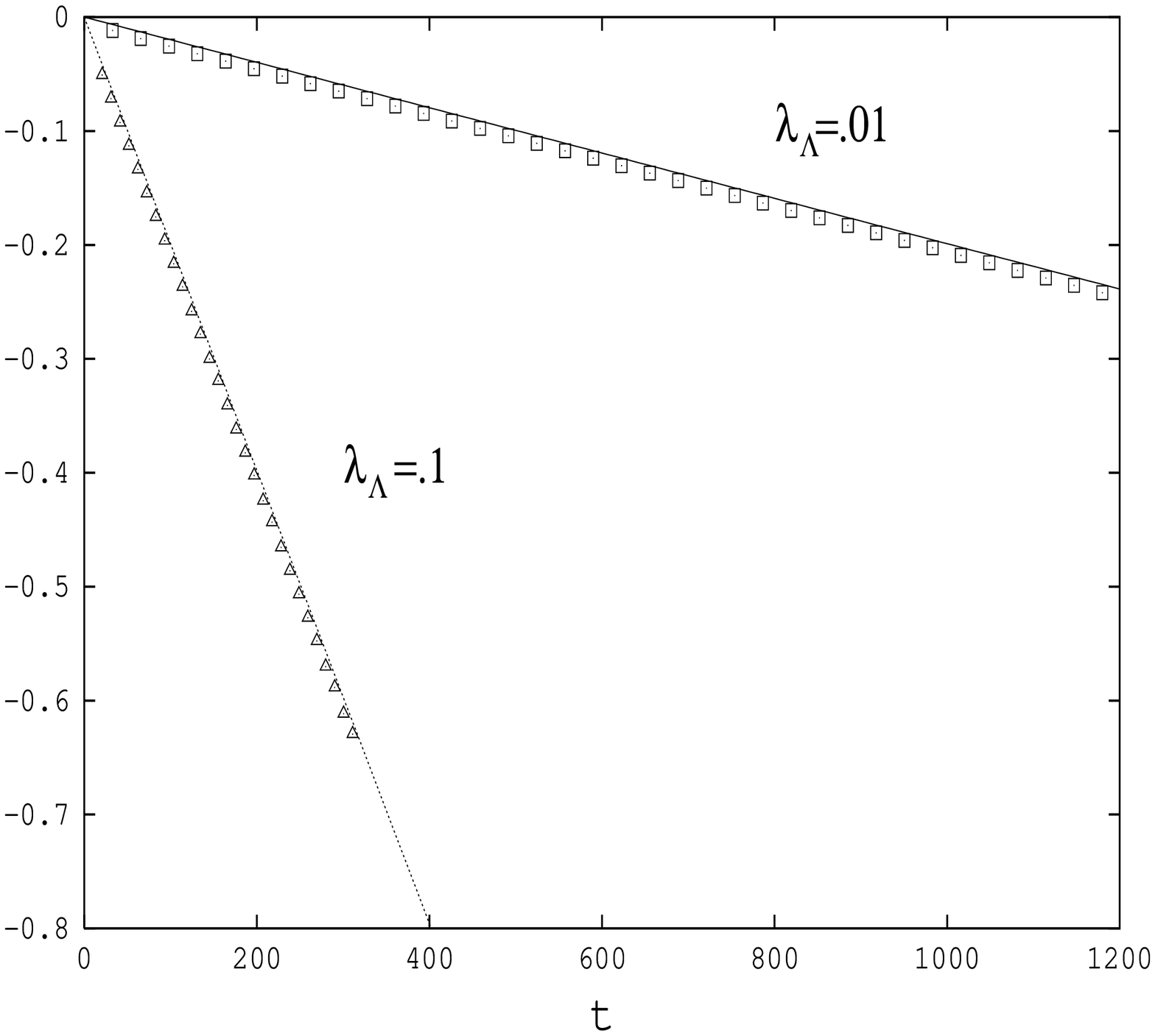}}
\vspace{.35cm}
{FIG. 16. {\small{The function $y_t$ defined in Eqn. (\ref{yy}) 
plotted for perturbed thermal initial conditions at two values of
$\lambda_{\Lambda}$. The solid lines are the corresponding theoretical
predictions.}}}\\

When the distribution $N(k)$ is non-thermal the behavior of $\tilde
D^{-1}(s)$ is more difficult to analyze, and there is no guarantee
that it possesses a zero in general. If $N(k)$ is peaked at small $k$
then a reasonable first approximation is to neglect the last term in
(\ref{Dinvlt}). In that case we find a zero of $\tilde D^{-1}(s)$ at
with 
\begin{eqnarray}
\omega_{pl}^2 &\approx & \lambda_{pl} v^2\  \qquad {\rm and}\nonumber\\  
\gamma &\approx & {\lambda_{pl} \over 64\pi} \omega_{pl} = 
{\lambda_{pl}^{3 \over 2} v\over 64\pi}\ . 
\label{plf}
\end{eqnarray}
If the running renormalized $\lambda_R$ is not very small and/or
$N(k)$ is not sharply peaked at $k=0$ then the position of this zero
of $\tilde D^{-1}(s)$ will be shifted from (\ref{plf}), and at some values it
may even cease to exist. Comparing with our numerical data for which
$k \tilde N (k)$ is shown in Fig. 18, we observe that the
approximations leading to the estimates (\ref{plf}) can only be order
of magnitudes at best. Indeed the numerical value of the plasma
frequency from the data at late times is $\omega_{pl} = 1.405$ whereas
(\ref{plf}) gives $\omega_{pl} =v$. The damping rate is estimated to
be $\gamma=0.005 v$ (or $0.007 v$ if the correct $\omega_{pl}$ is
used), however the envelope of the oscillations at late times is not
even strictly exponential as we shall see below. Hence we cannot
expect to obtain an accurate estimate of the damping rate by a simple
pole analysis of the Laplace transform $\tilde D(s)$. Nevertheless, if
one fits an exponential to the data at late times a value $\gamma =
0.009 v$ is obtained, agreeing only roughly with the estimate.
 
Despite the envelope of the oscillations being non-exponential,
$\omega_{pl}$ is very well determined by the regular oscillations
observed (to a few parts in $10^3$). To obtain a more accurate
approximation to the plasma frequency consider the following analytic
approximation to $kN(k)$,
\begin{equation}
k N(k)= 8 \pi \omega_0 ~ \theta\left(\omega_0^2 - 4 k^2\right)~ 
\sqrt{1- {4 k^2 \over \omega_0^2}}\,,
\label{Nfit}
\end{equation}
where $\omega_0$ is a free parameter which is close to the numerical
value of the plasma frequency. The accuracy of this analytic fit to
the data is shown in Fig. 18 for $\omega_0 = \omega_{pl}$. Notice that
because of dephasing, the $N(k)$ of the initial state in the linear
response analysis may be identified with the $\tilde N(k)$ of the
particles created earlier in the nonlinear evolution from the spinodal
region (except for very small $k < 1/t$ where the dephasing is as yet
ineffective).

\vspace{.4cm}
\epsfxsize=7cm
\epsfysize=4cm
\centerline{\epsfbox{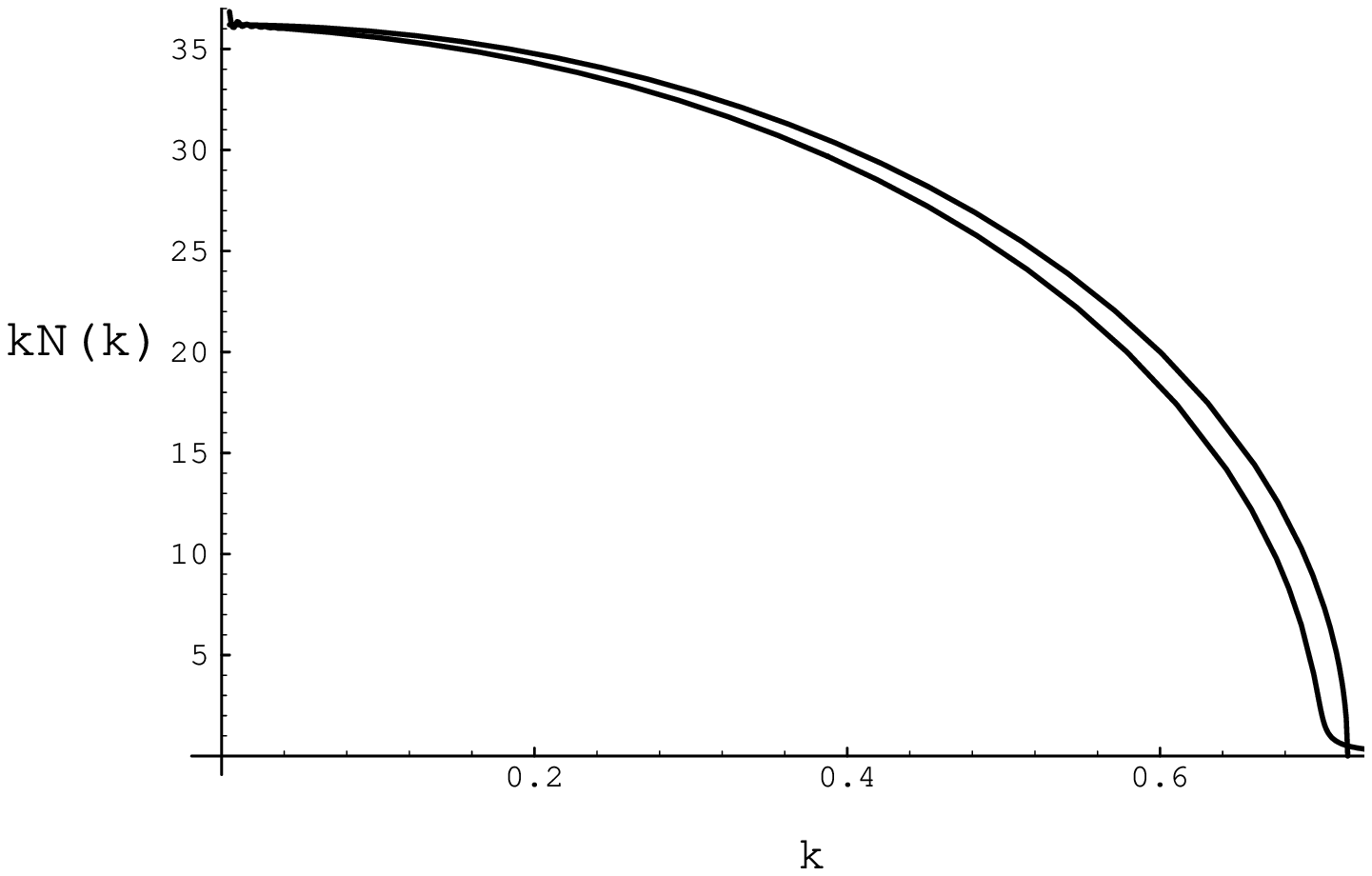}}
\vspace{.35cm}
{FIG. 17. {\small{Comparison of the the analytic 
approximation to the particle number density of Eqn. (\ref{Nfit})
(upper curve) with the adiabatic number density of the numerical
calculation (bottom curve).}}}\\    

Substituting the form (\ref{Nfit}) into (\ref{Dinvlt}) we obtain
\begin{eqnarray} 
-\tilde D^{-1}(s) &=& {1 \over \lambda_R(s^2/4)} + {v^2 \over
s^2}\label{Dfit}\\ 
&& + {2\omega_0^2 \over \pi} \int_0^1 dy \left(1-y^2\right)^{1\over 2}
\left[ {1 \over\omega_0^2 y^2 + s^2}- {1 \over s^2}\right].\nonumber
\end{eqnarray}
Because of the small prefactor of the logarithm in
$1/\lambda_R(s^2/4)$, the running coupling constant is only a few
percent different from the bare $\lambda$ near the zero of $\tilde
D^{-1}(s)$ and this running with $s$ can be neglected in lowest
order. Again we set $ s=\pm i \omega_{pl} - \gamma $ and look for a
zero of $D^{-1}(s)$ near $\omega_0^2 = \omega_{pl}^2$, with 
${\gamma/\omega}\ll 1$. One could perform the integral in (\ref{Dfit})
exactly but we content ourselves with this simple approximation and
obtain
\begin{eqnarray} 
0  &\cong& {1 \over \lambda_{pl}} -{v^2 \over \omega_0
^2}\nonumber\\  
&& + {2 \over \pi} \int_0^1 ~ dy~ [(1-y^2)^ {1\over 2} -
(1-y^2)^{-{1\over 2}}]\,.  
\end{eqnarray}
Performing the integral we find from the real part of this equation,
\begin{equation}
{\omega_{pl}^2 \over v^2} = { 2 \lambda_{pl}\over 2- \lambda_{pl}}  
\end{equation}
In our simulations where $\lambda_{\Lambda}=1\simeq\lambda_{pl}$, the 
output value of the calculated plasma oscillation
frequency $\omega_{pl} = 1.414 v$ is within $0.6\%$ of the measured
value $\omega_{pl} = 1.405 v$ showing the self-consistency of the
approximation and the fit (\ref{Nfit}), to this accuracy. We also
performed the integration in (\ref{Dinvlt}) numerically using the data
of Fig. 18 for $kN(k)$ and reconfirmed $\omega_{pl} = (1.405 \pm 0.001)
v$ from the location of the minimum of the real part of $\tilde
D^{-1}(s)$.

\vspace{1.3cm}
\epsfxsize=7cm
\epsfysize=5cm
\centerline{\epsfbox{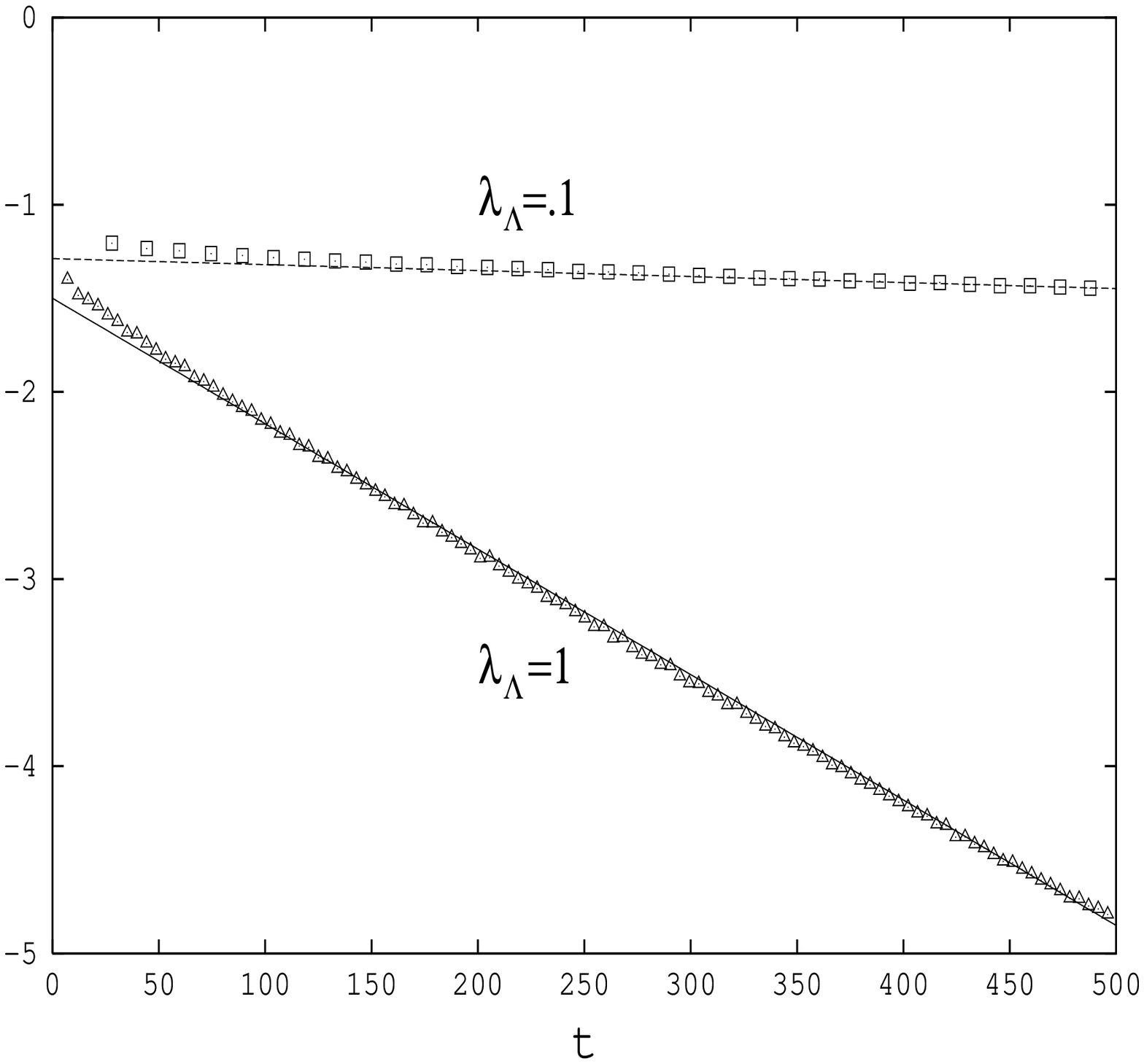}}
{FIG. 18. {\small{The function $y_t$ plotted for $N(k)=0$ initial
conditions at two values of $\lambda_{\Lambda}$. The late time linear
fits to the data are also shown with the intercepts determining the
power law prefator.}}}\\

The envelope of the $\chi$ oscillations obtained numerically is very
well fit by $t^{-\alpha}\exp(-\gamma t)$. This general behavior has
been checked for different values of the bare coupling and we have
verified that it does not depend on which choice of the $\omega_k$
profiles (\ref{initial_modesA}) and (\ref{initial_modesB}) one makes
for the initial conditions. Illustrated in Fig. 19 are the
$\lambda_{\Lambda}=1$ and $\lambda_{\Lambda}=.1$ cases, where
$\alpha=1.5$, $\gamma=.0067$ and $\alpha=1.288$, $\gamma=.00032$
respectively. Figs. 19 and 20 display the complete fit to the damped
oscillation alongside the actual evolution for the case
$\lambda_{\Lambda}=1$. Very good agreement is evident over long times
and large amplitude range. The case of $\lambda_{\Lambda}=.1$ is very
similar.  

\vspace{1.3cm}
\epsfxsize=7cm
\epsfysize=5cm
\centerline{\epsfbox{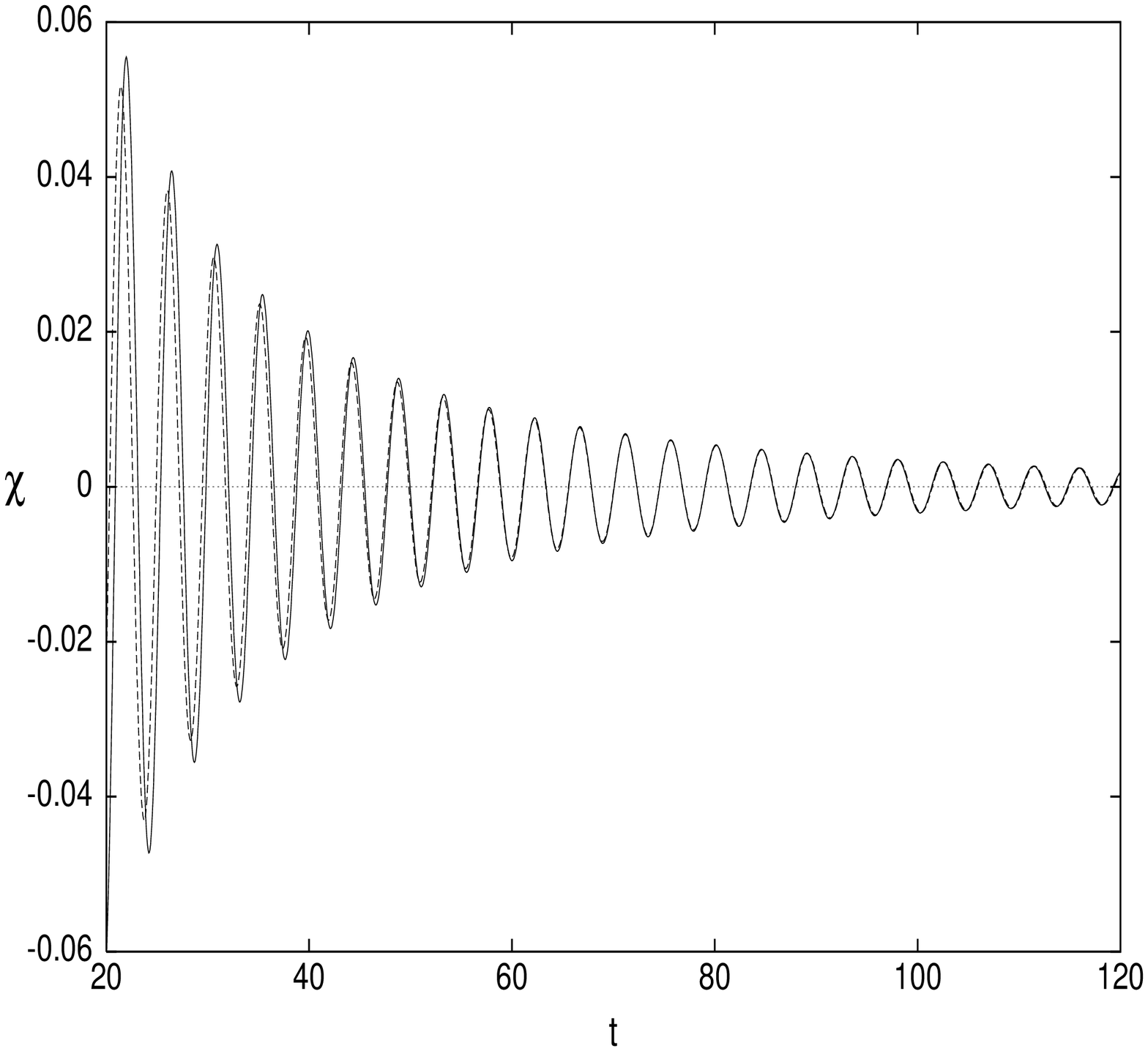}}
{FIG. 19. {\small{The $\chi$ evolution as a function of time (dashed
line) for $\lambda_{\Lambda}=1$. The fit (\ref{envfit}), shown by the
solid curve is $\chi(t)=6.63t^{-1.5}
\exp(-.0067t)\cos(1.405t+.5)$. This figure shows the fit over an early
time range, $t=20,120$.}}}\\   

\vspace{1.3cm}
\epsfxsize=7cm
\epsfysize=5cm
\centerline{\epsfbox{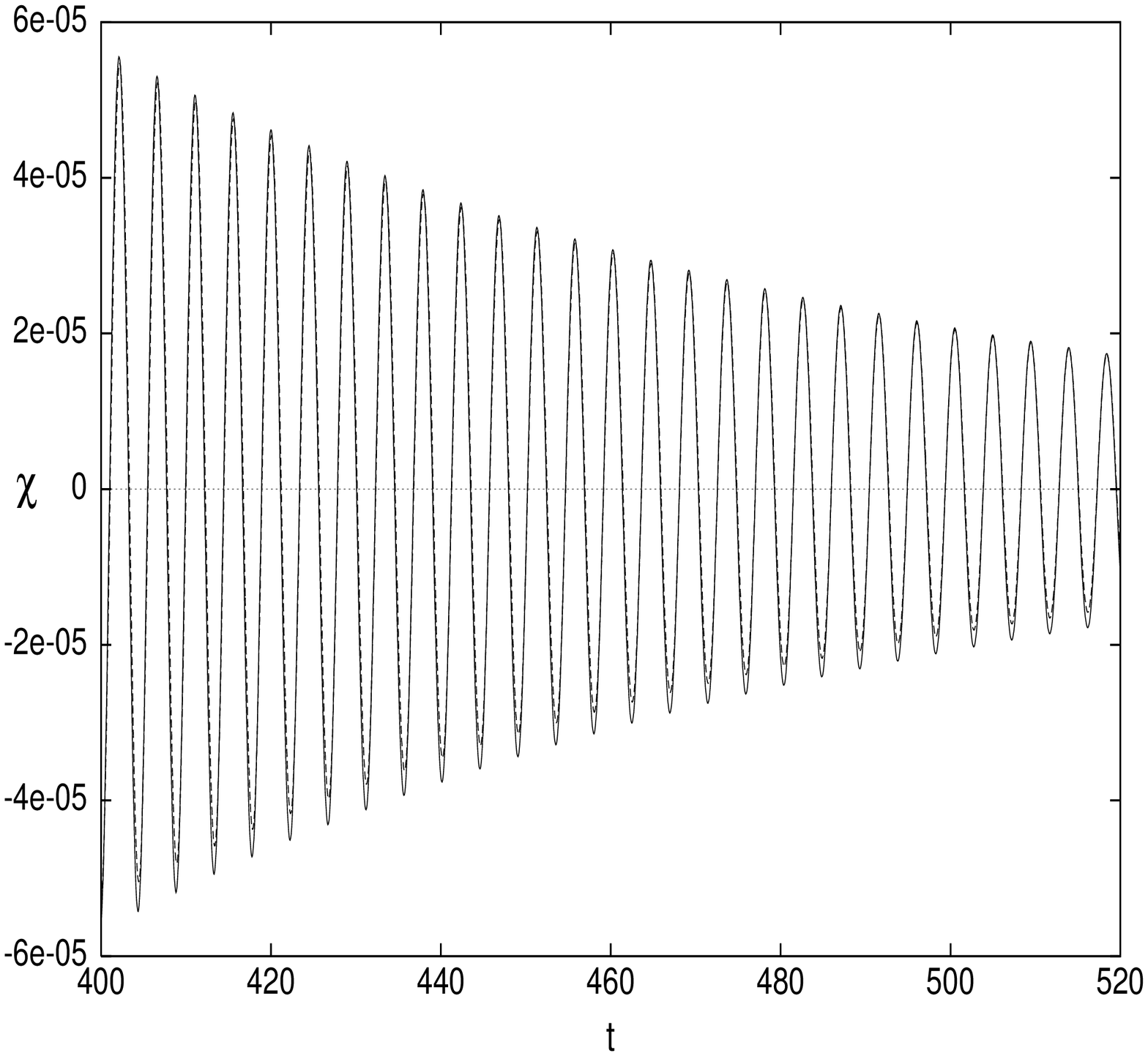}}
{FIG. 20. {\small{The same $\chi$ evolution at late times $t=400,520$
showing the excellence of the simple fit over long times and along with
the results of Fig. 19, very good agreement over three decades of
amplitude.}}}\\

As expected from the time dependence of the envelope of
the oscillations observed in Figs. 18, 19, and 20, the damping rate
cannot be obtained from a simple pole analysis. Indeed the imaginary
part of $\tilde D^{-1}(s)$ evaluated from the numerical data does not
possess a zero for {\em any} $\gamma$ with $\omega_{pl}$ fixed at
$1.405$, although it does decrease monotonically as $\gamma$ is
decreased, thus favoring $\gamma < 0.003\, v$. This value is close to
our momentum resolution $dk$ and hence we could not go to lower values
reliably. In any case, the decay of the envelope at late times is due
to the creation of very low momentum massless bosons which have not
yet dephased efficiently. Hence it is the behavior of the Laplace
transform on the branch cut very close to $s=0$ which determines the
late time damping, and this behavior depends in turn on the the mode
functions $|f_k(t)|^2$ near $k=0$ where neither our replacement,
(\ref{repl}) nor our single pole dominance assumptions are justified.

In earlier studies of damping a $-{3\over 2}$ power law was
encountered and identified as due to the behavior of the relevant
spectral density function near threshold \cite{boydamp}. In the
present case there is no simple argument that we are aware of which
would lead to such a power law at late times.

\section{Summary}
\label{sec:level7} 

In this paper we have presented a study of nonequilibrium evolution
and time dependent behavior of symmetry breaking transitions in $N$
component $\lambda \Phi^4$ field theory. Starting from an effective
action principle for the leading order mean field approximation we
emphasized the correspondence of the equations with the Schr\"odinger
evolution of a Gaussian density matrix according to a certain
effective classical Hamiltonian. This is important for emphasizing
that no fundamental irreversible behavior has been introduced by the
mean field approximation.  Explicit numerical time evolutions from the
unstable spinodal region show effective time irreversibility in the
form of the efficient averaging to zero of the phase information in
the density matrix. The effective von Neumann entropy of the reduced
density matrix measures the information that is lost by discarding
this phase information. Its general increase with particle creation
shows the close connection between dephasing and irreversibility.
Thus our study of effective dissipation and decoherence casts some
light on fundamental issues in quantum statistical mechanics such as
the origin of irreversibility, Boltzmann's H-theorem and the quantum
to classical transition. Here there is much that could be done still
in the context of the framework presented in this paper, most notably
to study quantitatively the Poincar\'{e} recurrence cycle(s) expected
in a Hamiltonian system and their dependence on the various parameters
of the model.

In the specific case at hand, $\Phi^4$ field theory is a prototype of
models of of spontaneous symmetry breaking in a wide variety of
physical systems. It has been common to rely heavily on the
thermodynamic free energy and equilibrium considerations generally in
analyzing these systems.  One broad and generally applicable
conclusion of the present study is that this can be very misleading
where nonequilibrium dynamics is concerned. We have shown that the
distribution of particles created in the mean field time evolution
from an initial configuration in the spinodal region is generally far
from thermal. This leads to a final state in which the mean field need
not be close to minimum of the free energy potential. The simple
observation that there is a sum rule which the nonthermally
distributed particles can saturate is sufficient to resolve this
seemingly paradoxical result.

The linear response analysis of the oscillations about the allowed
stationary configurations is also a new result which is quite
different from what consideration of the free energy function alone
might lead one to expect. We have shown that there is a collective
plasma mode in the radial symmetry breaking direction whose
characteristics depend on the ambient particle distribution. This also
should be generally true of the various systems described by the same
$\Phi^4$ field theory.  While there is excellent agreement in the
thermal case for both the plasmon frequency and damping rate, the
non-thermal situation is more complicated. In this case, due to the
lack of pole dominance, the late time behavior cannot be described by
exponentially damped oscillations. Our numerical results for the
plasmon oscillation envelope are remarkably well-fitted by an
expression of the form $t^{-\alpha}\exp(-\gamma t)$. The plasmon
frequency compares well with an analytical estimate. However, the
precise dependence of the envelope function on the particle density in
the general case remains to be more fully investigated. We have
outlined also how dephasing leads also to effectively classical field
configurations which one can sample in order to extract information on
the creation of topological defects during the phase transition. This
direction is certainly an interesting one to pursue both in the
condensed matter and cosmological applications.

Throughout the paper we have endeavored to bring the theoretical
framework into close contact with practical numerical methods. Indeed
one of the main conclusions of this work is that the real time
dynamics of phase transitions can be studied in a concrete way with
presently available computers. Besides the wide applicability of the
spontaneously broken $\Phi^4$ theory it is interesting for the
existence of massless Goldstone bosons which are created freely during
the phase transition. This essential kinematics is shared by systems
with an exact gauge symmetry such as QCD or general coordinate
invariance such as gravity. The numerical techniques used in this work
and the experience gained in treating the massless case should prove
to be valuable in generalizations to these cases.

\acknowledgments

Numerical computations were carried out on the CM5 at the Advanced
Computing Laboratory, Los Alamos National Laboratory.

\appendix
\section{The Energy-Momentum Tensor and its Renormalization}

The effective action (\ref{Seff}) leads directly to a conserved
energy-momentum tensor at any order of the large $N$ expansion
\cite{largeN}. In this Appendix we give some details about the
energy-momentum tensor and its renormalization.  Because of spatial
homogeneity and isotropy the only nontrivial components of this tensor
are
\begin{eqnarray}
\langle T_{00} \rangle &=& \varepsilon\qquad {\rm and}\nonumber\\
\langle T_{ij} \rangle &=& p\,\delta_{ij} \,, 
\label{app:one}
\end{eqnarray}
where the energy density $\varepsilon$ and isotropic pressure $p$ are
given by
\begin{eqnarray}
\varepsilon &=& {1\over 2} \dot\phi^2 +
{1\over 2}\chi\phi^2 - {\chi\over\lambda} 
\left({\chi\over 2} + \mu^2\right)\nonumber\\
&& + {1\over 2} \int [d{\bf k}]\sigma_k
\left\{|\dot f_k|^2 + (k^2 + \chi)|f_k|^2\right\}\nonumber\\
&=& {1\over 2} \dot\phi^2 + {1\over 2\lambda_{\Lambda}}\chi^2 +
{1\over 4\pi^2}\int_0^{\Lambda} k^2 \, dk\, \sigma_k \left(|\dot
f_k|^2 + k^2|f_k|^2\right)\,, \nonumber\\
\label{enerden}
\end{eqnarray}
and
\begin{eqnarray}
p &=& {1\over 2} \dot\phi^2 - {1\over 2}\chi\phi^2 + 
{\chi\over\lambda} \left({\chi\over 2} + \mu^2\right)\nonumber\\
&& + {1\over 2} \int [d{\bf k}]\sigma_k
\left\{|\dot f_k|^2 - \left({k^2\over 3} + \chi\right)|f_k|^2\right\}
\nonumber\\
&=& {1\over 2} \dot\phi^2 - {1\over 2\lambda_{\Lambda}}\chi^2 +
{1\over 4\pi^2}  \int_0^{\Lambda} k^2 dk \sigma_k \left(|\dot
f_k|^2 - {k^2\over 3}|f_k|^2\right)\,.\nonumber\\ 
\label{press}
\end{eqnarray} 
We have used the gap equation (\ref{gap}) in passing to the latter
expressions in each case. 

The energy-density $\varepsilon$ contains a quartic but constant
cut-off dependence ({\em i.e.} proportional to $\Lambda^4$).  This is
the expected vacuum energy contribution which should be
subtracted. The conservation of the energy density,
\begin{equation}
\dot \varepsilon = 0
\end{equation}
is easily checked from the equations of motion, and is not affected by
the subtraction of the constant quartic divergence. As it will turn
out this single constant subtraction is sufficient to yield a cut-off
independent conserved energy density.

The cut-off dependence of the isotropic pressure $p$ is a bit more
complicated and requires a detailed understanding of how Lorentz
invariance (or more generally, coordinate invariance) is broken by the
spatial momentum cut-off we have introduced in all the mode
integrations. In the first forms of the two expressions
(\ref{enerden}) and (\ref{press}), {\em i.e.} before using the gap
equation, there appear mode integrals whose cut-off dependences
correspond to those of a free theory with the mass $m^2$ replaced by
$\chi(t)$. Divergences in the energy-momentum tensor expectation value
$\langle T_{\mu\nu}\rangle$ have been studied in the literature by
covariant methods such as dimensional regularization or covariant
point splitting \cite{christ} with the result that these divergences
must be proportional either to the metric tensor $g_{\mu\nu} = {\rm
diag}\, (-1, 1, 1, 1)$ in flat Minkowski spacetime or the total
derivative ``improvement term'' of Eqn. (\ref{app:improve})
below. Because we are using a regulator in the form of a cut-off in
the spatial momentum integrations which is {\em not} covariant under
Lorentz or general time dependent coordinate transformations, the
quartic and quadratic cut-off dependence we shall actually obtain will
not have these covariant forms. Hence we shall have to perform the
stress tensor renormalization in a noncovariant manner as well to
correct for the spurious noncovariant $\Lambda^4$ and $\Lambda^2$
cut-off dependence, in order to obtain covariant results in the end.

The actual cut-off dependence in both $\varepsilon$ and $p$ is easily
analyzed by means of an adiabatic expansion to the mode function
equation (\ref{modefn}), the lowest order solution of which is given
by Eqn. (\ref{ftilde}).  By substituting this lowest order adiabatic
approximation to the mode functions into the mode integrals in the
first forms of Eqns. (\ref{enerden}) and (\ref{press}) we can
characterize the most severe dependence on the ultraviolet cut-off
$\Lambda$ in the forms,
\begin{eqnarray}
&&\varepsilon_0 \equiv {\hbar\over 4\pi^2}\int_0^{\Lambda} k^2\, dk\,
\sqrt{k^2 + \chi}  
\nonumber\\
&& ={\hbar\Lambda^4\over 16\pi^2} + {\hbar\Lambda^2 \chi\over 16\pi^2}
- {\hbar\chi^2\over 64\pi^2}\ln \left({4\Lambda^2\over\chi}\right) +
{\hbar\chi^2\over 128\pi^2} + {\cal O}\left( {\chi^3\over
\Lambda^2}\right)\nonumber\\
\label{app:eada}
\end{eqnarray}
and
\begin{eqnarray}  
&&p_0 \equiv {\hbar\over 12\pi^2}\int_0^{\Lambda} {k^4\, dk\over
\sqrt{k^2 + \chi}} \nonumber\\
&& ={\hbar\Lambda^4\over 48\pi^2} - {\hbar\Lambda^2 \chi\over 48\pi^2}
+ {\hbar\chi^2\over 64\pi^2}\ln \left({4\Lambda^2\over\chi}\right) - 
{7\hbar\chi^2\over 384\pi^2} + {\cal O}\left( {\chi^3\over
\Lambda^2}\right)\,. \nonumber\\
\label{app:eadb}
\end{eqnarray}
Clearly the $\Lambda$ dependence is {\em not} such that $\varepsilon_0
= -p_0$, as required by considerations of general covariance. As has
been known for some time the reason for this is that the spatial
momentum cut-off $\Lambda$ acts as a noncovariant point splitting
regulator would, giving terms in the regulated $\langle
T_{\mu\nu}\rangle$ proportional to $\delta_{(\mu}^i\delta_{\nu)}^j$ in
which the spatial directions $i,j = 1,2,3$ are distinguished. Since
such terms do not appear in the $\mu=\nu=0$ time component, the energy
density has the correct $\Lambda$ dependence and requires no
covariantizing correction. However for the pressure we should have
\begin{equation}
p'_0 \equiv -\varepsilon_0 
\label{app:three}
\end{equation}
on grounds of general covariance. This is easy to enforce {\em by
hand} by adding the difference $p'_0 - p_0= -\varepsilon_0 - p_0$ to
$p_0$ above which just corrects for the noncovariant terms induced by
our momentum cut-off.

With this prescription to enforce covariance of the mode integrals in
the pressure, the quartic subtraction required on both $\varepsilon$
and $p$ is the removal of the cosmological vacuum energy
${\hbar\Lambda^4/16\pi^2}$, corresponding to a subtractive
renormalization of the constant cosmological vacuum term.  We are {\em
not} allowed to subtract the subleading quadratic and logarithmic
divergences of $\varepsilon$ appearing in Eqn. (\ref{app:eada}),
since, for one thing, they are multiplied by the time dependent
function $\chi$ and hence would spoil energy conservation, and for
another, such time dependent terms do not correspond to a constant
vacuum energy. Instead we recognize them to be just the correct
cut-off dependent terms needed to combine with the cut-off dependent
terms in the ``classical'' energy density,
\begin{eqnarray}
-{\chi\over\lambda_{\Lambda}} \left({\chi\over 2} +
\mu_{\Lambda}^2\right) &=& -{\chi^2\over 2}\left\{{1\over \lambda_R
(m^2)} - \hbar{1\over 32\pi^2}\ln\left({4\Lambda^2\over
m^2}\right)\right\} \nonumber\\
&& - \chi \left({v^2\over 2} + {\hbar\Lambda^2\over 16\pi^2}\right)
\label{app:four}
\end{eqnarray}
in order to render the total energy density cut-off independent.
Because the analogous terms in $p$ are precisely the negative of those
appearing in $\varepsilon$ after the correction $p'_0 - p_0$ has been
added to its quantum part, we also obtain a (partially) renormalized
pressure by the same manipulations.

The reason that we must still perform one additional subtraction to
obtain a fully renormalized pressure is that at the next order in the
WKB adiabatic expansion of the mode equation (\ref{modefn}), one
obtains for the adiabatic frequency,
\begin{equation}
\omega_k \rightarrow \omega_k - {1\over 4} {\ddot\omega_k\over
\omega_k^2} +{3\over 8} {\dot\omega_k^2\over \omega_k^3} + \cdots~.
\end{equation} 
If the energy density is calculated to this adiabatic order one finds
no additional cut-off dependence in $\varepsilon$, so no additional
subtractions are required for it. Indeed none are permitted consistent
with the principle of general coordinate invariance. However, the
pressure has an additional logarithmic cut-off dependence equal to
\begin{equation}
-{\hbar\over 96\pi^2}\, \ddot\chi\, \ln \left(4\Lambda^2\over
\chi\right)\, . 
\label{app:logdiv}
\end{equation}
Since this divergence appears only in the pressure but not in the
energy it is consistent with the generally covariant and conserved
form of the ``improvement term,'' \cite{ccj}
\begin{equation}
\Delta\langle T_{\mu \nu} \rangle = \xi (g_{\mu\nu}\sq -
\partial_{\mu}\partial_{\nu})\, \langle \Phi^2\rangle
\label{app:timprove} 
\end{equation}
in flat space. This term has no effect on the energy density for
spatially homogeneous mean fields but adds to the pressure the total
derivative,
\begin{equation}
\Delta p = -\xi {d^2\over dt^2} \left\{ \phi^2 + \int [d^3{\bf
k}]\sigma_k |f_k|^2\right\} = -2 {\xi\over \lambda}\ddot\chi  
\label{app:improve} 
\end{equation}
in which $\xi$ is an arbitrary parameter. The fact that a divergence
such as (\ref{app:logdiv}) appears in the pressure means that we
should introduce $\xi$ as a free bare parameter of the theory from the
very beginning, on the same footing as the mass $\mu_{\Lambda}^2$ and
coupling $\lambda_{\Lambda}$, and allow for the possibility that the
bare $\xi_{\Lambda} \ne 0$ will renormalize in general. Indeed this is
known to be the case in $\lambda \Phi^4$ theory \cite{BC} and we have
the renormalization condition,
\begin{equation}
\left(\xi_{\Lambda} - {1\over 6}\right) = Z_{\lambda}^{-1}(\Lambda,
m)\, \left(\xi_R(m^2) - {1\over 6}\right) 
\end{equation}
where $Z_{\lambda}^{-1}$ is the same logarithmic renormalization
constant as that for the coupling constant appearing in
Eqns. (\ref{lren1}) and (\ref{lren2}) of the text. Thus,
\begin{equation}
{\xi_{\Lambda} \over \lambda_{\Lambda}} = {1\over 6\lambda_{\Lambda}}
+ {1\over \lambda_R} \left(\xi_R - {1\over 6}\right)
\end{equation}
and the additional term in the pressure (\ref{app:improve}) becomes
\begin{eqnarray}
\Delta p &=& - {1\over 3\lambda_{\Lambda}}\ddot\chi -
{2\over \lambda_R} \left(\xi_R - {1\over 6}\right)\ddot\chi\nonumber\\
&=& + {1\over 96\pi^2}\ln \left(4\Lambda^2\over m^2\right)\, \ddot\chi
- {2\xi_R\over \lambda_R} \, \ddot\chi\, .
\label{app:renimp}
\end{eqnarray}
Comparing the latter expression with (\ref{app:logdiv}) we observe
that the logarithmic cut-off dependence cancels in the sum $p + \Delta
p$. Hence we {\em must} add the improvement term and renormalize $\xi$
in precisely this way in order to obtain a completely cut-off
independent pressure.

To summarize this discussion of the renormalization of the energy
and pressure we define
\begin{equation}
\varepsilon_R = \varepsilon - {\hbar \Lambda^4 \over 16\pi^2}
\end{equation}
to be the cut-off independent conserved energy density, and
\begin{equation}
p_R = p - p_0 - \varepsilon_0 + {\hbar \Lambda^4 \over 16\pi^2} +
\Delta p 
\end{equation}
to be the renormalized pressure of the $\Phi^4$ theory for the
spatially homogeneous mean fields considered in this paper.

Let us also remark that the value $\xi = {1/6}$ for the improvement
term is also the value chosen for conformal invariance of the scalar
theory in the massless limit. Indeed the trace of the renormalized
energy-momentum tensor is
\begin{eqnarray}
\langle T_{\mu}^{\mu}\rangle_R &=&
 -\varepsilon_R + 3p_R \nonumber\\
&=& v^2\chi + {\hbar\chi^2\over 32\pi^2} + {1\over \lambda_R}
\left(\xi_R - {1\over 6}\right)\ddot \chi
\label{app:tranom}
\end{eqnarray} 
upon using Eqns. (\ref{app:eada}), (\ref{app:eadb}), and 
(\ref{app:renimp}) above. At $\xi = 1/6$ the last term vanishes, 
the first term is the renormalized classical trace of the 
energy-momentum tensor and the second term is the one-loop
quantum trace anomaly,
\begin{equation}
{1\over 2\lambda_R^2}\beta(\lambda_R) \chi^2 = 
-{\chi^2\over 2} \Lambda {d\over d\Lambda}
\left({1\over \lambda_{\Lambda}}\right) = {\hbar\chi^2\over 32\pi^2}
\label{app:anom}
\end{equation}
in terms of the $\beta$ function of the coupling constant $\lambda$.
We note that the trace of the renormalized energy momentum tensor
vanishes (for any value of $\xi$) in the static spontaneously broken
vacuum where $\dot\phi= \chi =0$ and $$f_k ={\sqrt {\hbar\over 2k}}
e^{-ikt}~.$$ This implies that the relativistic equation of state
\begin{equation}
p_R \rightarrow {1\over 3} \varepsilon_R
\end{equation}
holds at late times, independently of the number density distribution.
The numerically obtained approach to this equation of state is shown
in Fig. 20. We emphasize that this equation of state does not imply
relaxation to thermal equilibrium since collisional effects are not
yet taken into account by the leading order large $N$ approximation,
and the nonthermal nature of the late time state is apparent from the
created particle distribution of Fig. 10.

\vspace{.4cm}
\epsfxsize=7.5cm
\epsfysize=5.5cm
\centerline{\epsfbox{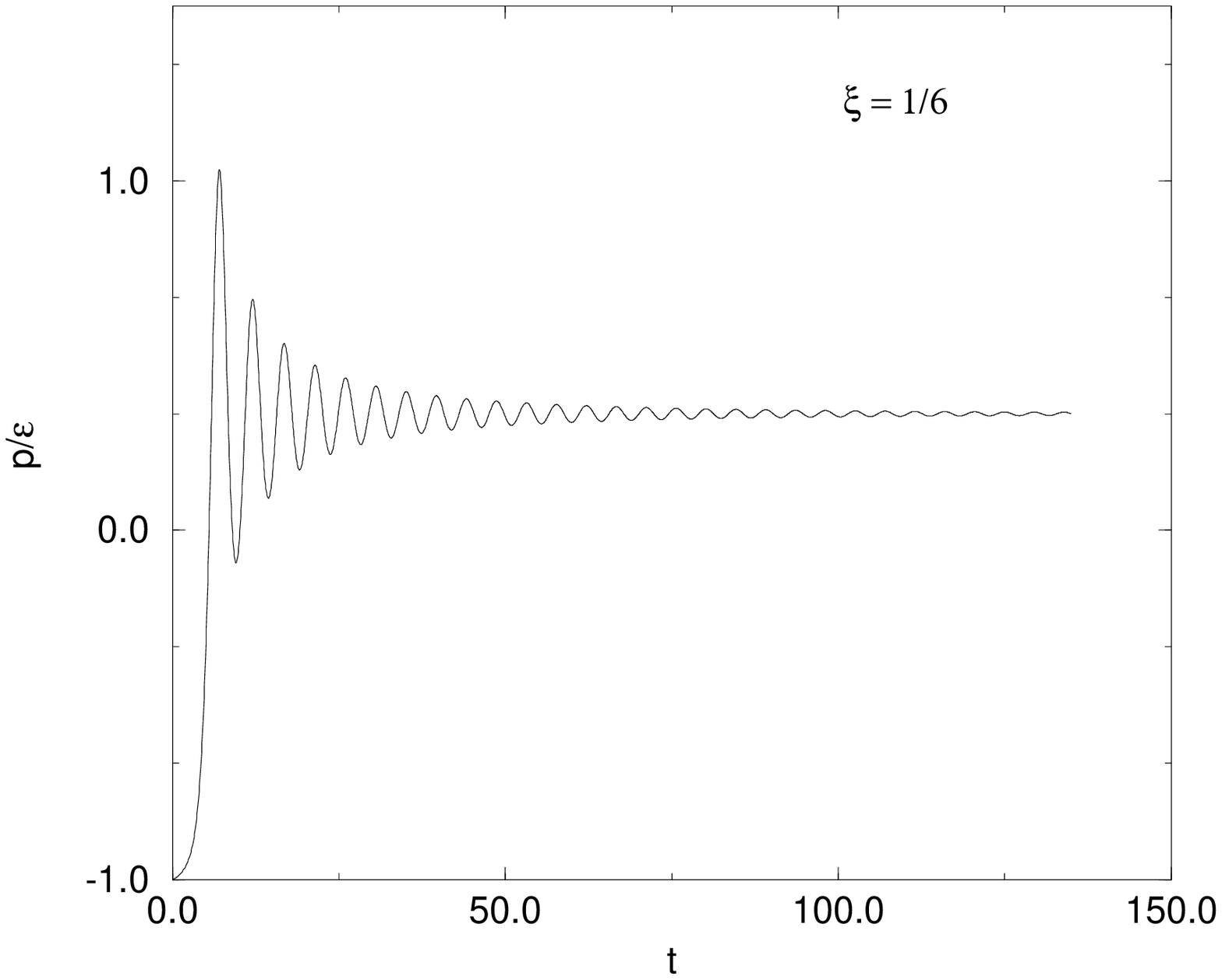}}
\vspace{.35cm}
{FIG. 21. {\small{Evolution of the pressure as a function of time for
$\xi=1/6$. The approach to the equation of state $p=\varepsilon/3$ is
clearly seen.}}}\\

\section{The Gaussian Density Matrix ${\bf \rho}$}

In this second Appendix we examine some of the properties of the
general mixed state Gaussian density matrix (\ref{gauss}) and its time
evolution. The discussion will be restricted mainly to the case of
$d=0$ spatial dimensions to simplify the notation. Generalizations to
$d>0$ are straightforward. First, the time evolution of the density
matrix is unitary:
\begin{equation}
{\bf\rho} (t) = U(t) {\bf\rho} (0) U^{\dagger}(t)\, , \quad U(t) =
\exp \left(-i\int_0^t H_{osc} dt \right)\ , 
\label{app:unitev} 
\end{equation}
where $H_{osc}$ is the time-dependent harmonic oscillator Hamiltonian,
\begin{equation}
H_{osc}({\bf\Phi}, {\bf P}; t) \equiv {1 \over 2}\left({\bf P}^2 
+ \omega^2(t)\,{\bf\Phi}^2\right) 
\label{app:osc}
\end{equation}
that preserves the Gaussian structure of ${\bf\rho}$ under time
evolution, so that the Liouville equation,
\begin{equation}
i\hbar {\partial \over \partial t}{\bf\rho} = [H_{osc},{\bf\rho}]
\label{app:leqn}
\end{equation}
is satisfied without taking the trace.  

It is not difficult to find the explicit form of the unitary operator
$U(t)$ in the coordinate basis,
\begin{eqnarray}
\langle x'|U(t)|x\rangle &=& (2\pi i\hbar
v(t))^{-{1\over2}}\times\nonumber\\  
&&\exp \left\{ {i\over 2\hbar v(t)}\left(u(t)x^2 + \dot v(t)
x'^2 - 2xx'\right)\right\} \nonumber\\
\label{app:unit}
\end{eqnarray}
in terms of the two linearly independent solutions to the classical
evolution equation, 
\begin{equation}
\left({d^2 \over dt^2} + \chi^2 (t)\right) \left( \begin{array}{c}
u\\v 
\end{array}\right) = 0\ ; \qquad \begin{array} {l} u(0) = \dot v (0) =
1\\ \dot u(0) = v(0) = 0~. 
\end{array}
\label{app:evolop}
\end{equation} 
This same $U(t)$ also evolves the quantum operators,
\begin{eqnarray}
{\bf\Phi}(t) &=& U^{\dagger}(t)\,{\bf\Phi}(0)\,U(t) = \phi(t) + a f(t)
+ a^{\dagger} 
f^*(t) \nonumber\\ 
{\bf P}(t) &=& U^{\dagger}(t)\,{\bf P}(0)\,U(t) = p(t) + a \dot f(t) +
a^{\dagger} \dot f^*(t)\ .
\label{app:abas} 
\end{eqnarray}

Mathematically, the three Fock space bilinear operators $aa$,
$a^{\dagger}a^{\dagger}$ and $a^{\dagger}a + aa^{\dagger}$ generate
the Lie algebra of the symplectic group $Sp(2) \cong SU(1,1) \cong
SL(2, R)$ which is the group of homogeneous linear transformations of
phase space $({\bf\Phi},{\bf P})$ which preserves the anti-symmetric
classical Poisson bracket $\{{\bf\Phi},{\bf P}\}=1$, {\it i.e.}
\begin{equation}
\left(\begin{array}{c} {\bf\Phi} \\ {\bf P} \end{array} \right) \rightarrow 
\left(\begin{array}{c} {\bf\Phi}' \\ {\bf P}' \end{array} \right) = 
\left(\begin{array}{cc} a & b \\ c & d \end{array} \right)
\left(\begin{array}{c} {\bf\Phi} \\ {\bf P} \end{array} \right)
\label{app:poisson}
\end{equation}
with
\begin{equation}
ab-cd = 1\ .
\label{app:comm}
\end{equation}
Since this is one condition on four real parameters the group $Sp(2)$
is a three parameter group. If the $a$ and $a^{\dagger}$ Heisenberg
operators are appended to these three, the algebra again closes upon
itself, forming a five parameter group, the inhomogeneous metaplectic
group, $IMp(2)$ \cite{rgl}. The unitary evolution (\ref{app:unitev}),
(\ref{app:unit}) of the Gaussian density matrix (\ref{gauss}) is an
explicit representation of this group's action.

The form of the density matrix in the time-independent Heisenberg
basis is quite easy to obtain from the form of the the transition
matrix element $\langle x\vert n\rangle$ given by Eqn. (\ref{xn}) of
the text. By substituting the integral representation of the Hermite
polynomials,
\begin{equation}
H_n(x) = {n! \over 2\pi i} \oint_{{\cal C}}\ {dz \over z^{n+1}} e^{2xz
-z^2} 
\label{app:hankel}
\end{equation}
where $\cal C$ is a closed contour around the origin of the complex
$z$ plane into the expression 
\begin{equation}
\langle n'\vert \rho\vert n\rangle = \int_{\infty}^{\infty} dx'
\int_{\infty}^{\infty} dx \langle n' \vert x'\rangle 
\langle x'\vert \rho\vert x\rangle \langle x \vert n\rangle
\label{app:rhon}
\end{equation}
and interchanging the orders of integration we can perform the double
Gaussian integral over the shifted vector $(x-\phi, x'-\phi)$
first. Using the standard formula 
\begin{equation}
\int_{\infty}^{\infty} d^2x e^{-x\cdot A\cdot x + B\cdot x} =
{\pi \over \left({\rm det}\, A \right)^{1\over 2}} e^{{1\over
4}B^T\cdot A^{-1} \cdot B}
\label{app:stdform}
\end{equation}
with $A$ the $2\times 2$ matrix
\begin{equation}
A = {\sigma + 1\over 8 \xi^2}\left( \begin{array}{cc} 1+\sigma &
1-\sigma\\ 
1-\sigma & 1 + \sigma \end{array} \right)
\label{app:twotwo}
\end{equation}
and $B$ the column vector
\begin{equation}
B = {\sqrt {2\sigma}\over \xi} \left(\begin{array}{c} z\\z^{\prime
*}\end{array}\right)
\label{app:colvec} 
\end{equation}
the exponent of the resulting expression simplifies
considerably. Letting $z = e^{i\theta}$ and $z'^* = e^{-i\theta'}$ we
are left with 
\begin{eqnarray}
&&\langle n'\vert \rho\vert n\rangle = {2\over \sigma + 1} 
\left( n'! n!\over 2^{n'+n}\right)^{1\over 2}\times\nonumber\\
&&\int_0^{2\pi}{d\theta'\over 2\pi}\, e^{in'\theta'}
\int_0^{2\pi}{d\theta\over 2\pi}\, e^{-in\theta}
\exp\left\{ 2 {\left({\sigma - 1\over\sigma + 1}\right)}e^{i(\theta -
\theta')}\right\}.\nonumber\\ 
\label{app:twofour} 
\end{eqnarray}
Expanding the last exponent in a Taylor series we find that only the
terms with $n'=n$ survive with Eqn. (\ref{diag}) the final result. As 
discussed earlier in Section VI, in this Heisenberg basis the density 
matrix is time-independent and diagonal. This fact allows the writing
of (\ref{app:twofour}) in an operator form,
\begin{equation}
{\bf \bar{\rho}} = {2\over \sigma + 1} \exp\left\{ \ln \left({\sigma
-1\over \sigma + 1}\right) a^{\dagger}a\right\}\,
\label{app:opform} 
\end{equation}
as far as computing matrix elements in this Heisenberg basis is concerned.

By making a different $IMp(2)$ group transformation it is also
possible to diagonalize (\ref{app:osc}) at any given time, bringing the
quadratic Hamiltonian into the standard harmonic oscillator form,
$H_{osc} = {\hbar\omega\over 2}\left(\tilde a \tilde a^{\dagger} +
\tilde a^{\dagger}\tilde a\right)$ with $\tilde a$ time
dependent. This adiabatic particle basis is related to the
time-independent Heisenberg basis by the relations,
\begin{eqnarray}
f &=& \alpha \tilde f + \beta \tilde f^*~,\nonumber\\
a &=& \alpha^* \tilde a - \beta^* \tilde a^{\dagger} + \kappa^*
\label{app:hrels}
\end{eqnarray}
with $\tilde f$ the adiabatic mode function defined by
Eqn. (\ref{ftilde}) of the text, and 
\begin{eqnarray}
\alpha &=& {i\over \hbar} \tilde f^* (\dot f -i\omega f)~,\nonumber\\
\beta &=& -{i\over \hbar} \tilde f (\dot f +i\omega f)~,\nonumber\\
\kappa &=& {i\over \hbar} (\dot\phi f -\phi\dot f)~.
\label{app:twoseven}
\end{eqnarray} 
The Bogoliubov coefficients $\alpha$ and $\beta$ obey
\begin{equation}
\vert \alpha\vert^2 - \vert\beta\vert^2 = 1
\label{app:bogcoff}
\end{equation}
and therefore may be expressed in terms of a real parameter $\gamma$
and two phases,
\begin{eqnarray}
\alpha &=& \cosh \gamma\, e^{i\psi}~,\nonumber\\
\beta &=& - \sinh \gamma\, e^{i(\psi-\theta)}~.
\label{app:phases}
\end{eqnarray}
The density matrix in the adiabatic particle number basis can be
expressed in terms of these time dependent parameters at any given
time $t$. The direct evaluation of the expression analogous to
(\ref{app:rhon}) in the $\tilde n$ number basis is quite tedious, and is
accomplished most rapidly by use of a coherent state basis as
discussed in Ref. \cite{Brown}.  Making the identifications,
\begin{eqnarray}
\alpha &=& {1\over S_{-+}^*} = {S_{+-}\over 1 - \vert
S_{++}\vert^2}~,\nonumber\\ 
\beta &=& {S_{--}\over S_{-+}} = - {S_{+-}S_{++}^* \over 1 - \vert
S_{++}\vert^2} 
\label{app:idents}
\end{eqnarray}
with
\begin{eqnarray}
S_{+-} &=& S_{+-} = {\rm sech}\gamma e^{i\psi}~,\nonumber\\
S_{++} &=& {\rm tanh}\gamma e^{i\theta}~,\nonumber\\
S_{--} &=& - {\rm tanh}\gamma e^{2i\psi -i\theta}
\label{app:threeone}
\end{eqnarray}
and $\chi \rightarrow \gamma$ in the notation of Ref. \cite{Brown},
Eqn. (5.32) of that work yields the desired matrix element in the case
of zero mean field, namely,
\begin{eqnarray}
&&\langle \tilde n'|{\bf\rho}|\tilde n\rangle\big\vert_{\phi =0} = 2
e^{{i\over 2}(\tilde n' -\tilde n)\theta} 
\left({\tilde n'!\over \tilde n!}\right)^{1\over 2}\times\nonumber\\
&&\left[(\sigma^2-1)^2 - 4\sigma^2 \sinh^2 2\gamma\right]^{{1\over 4}
{(\tilde n'+ \tilde n)}}\times\nonumber\\  
&&\left[\sigma^2 + 1 + 2\sigma \cosh 2\gamma\right]^{-{1\over 4}
{(\tilde n' + \tilde n + 1)}}\times\nonumber\\ 
&&P^{{1\over 2}(\tilde n - \tilde n')}_{{1\over 2}(\tilde n 
+ \tilde n')} \left(\left[1 - {4\sigma^2\over (\sigma^2 -1)^2}\sinh^2 
2\gamma\right]^{-{1\over 2}}\right)\,, 
\label{app:rhofd}
\end{eqnarray}
where $P^m_n$ is an associated Legendre polynomial.

The first important feature of this expression for our purposes
is the phase factor when $\tilde n \neq \tilde n'$. If the exact mode
function $f$ is rewritten in the form
\begin{equation}
f (t) = \sqrt{\hbar\over 2\Omega(t)}
\exp\left(-i\int_0^t\,dt'\,\Omega(t')\right)~,
\label{app:modefn}
\end{equation}
this phase
\begin{eqnarray}
\theta &=& {\rm arg}\, \left({\alpha\over -\beta}\right)\nonumber\\
&=& 2\int_0^t\,dt'\, \omega(t') + 
\tan^{-1}\left({{\dot\Omega\omega\over \Omega}\over \Omega^2 -
\omega^2 + {\dot\Omega^2\over 4 \Omega^2}}\right)~.
\label{app:phase}
\end{eqnarray}
Thus, even in the adiabatic limit, where ${\dot\Omega\over
\Omega^2}\ll 1$, the phase angle $\theta$ depends linearly on time and
the off-diagonal elements of the density matrix (\ref{app:rhofd}) are
rapidly varying in time.  On the other hand, the diagonal elements of
(\ref{app:rhofd}) are independent of this phase angle and consequently
much more slowly varying functions of time. Indeed, in the case of
zero mean field $\phi =0$, the adiabatic invariant $W$ of
Eqn. (\ref{adbinv}) is
\begin{equation}
W = \tilde N - N = \sigma\sinh^2\gamma 
\label{app:adinv}
\end{equation}
and $\langle \tilde n|{\bf\rho}|\tilde n\rangle$ depends only upon
$\sigma$ (a constant) and $\gamma$:
\begin{eqnarray}
\langle \tilde n|{\bf\rho}|\tilde n\rangle &=& 
2 \left[(\sigma^2-1)^2 - 4\sigma^2 \sinh^2 2\gamma\right]^{{1\over 2}
{\tilde n}}\nonumber\\
&&\times \left[\sigma^2 + 1 + 2\sigma \cosh 2\gamma\right]^{-{1\over 4}
{(2\tilde n + 1)}} \nonumber\\
&&\times P_{\tilde n} \left(\left[ 1 - {4\sigma^2\over (\sigma^2
-1)^2}\sinh^2 2\gamma\right]^{-{1\over 2}}\right) 
\label{app:rhodab}
\end{eqnarray}
Now if $\sigma > 1$ the Legendre polynomial is not necessarily
positive which means that we cannot interpret the diagonal matrix
elements of the density matrix in the adiabatic particle number basis
as a classical probability distribution in the general mixed state
case. However, if $\sigma =1$ then the argument of the Legendre
polynomial vanishes for any $\gamma \ne 0$. Since
\begin{equation}
P_{2\ell}(0) = (-1)^{\ell} {(2\ell - 1)!!\over 2^{\ell}\ell !}
\label{app:p2l}
\end{equation}
for $\tilde n = 2\ell$ even but $P_{2\ell + 1}(0) = 0$ for $\tilde n =
2\ell + 1$ odd, the diagonal matrix element (\ref{app:rhodab}) simplifies
considerably in the pure state case,
\begin{equation}
\langle \tilde n=2\ell|\rho|\tilde n = 2\ell\rangle
\big\vert_{_{\stackrel{\sigma =1}{\phi =\dot\phi = 0}}} 
= {(2\ell -1)!!\over 2^{\ell}\, \ell !} {\rm sech} \gamma
\,\tanh^{2\ell} \gamma 
\label{app:pstate} 
\end{equation}
with the mean number of created particles,
\begin{equation}
\tilde N = \sinh^2 \gamma  = {\vert \dot f + i \omega f\vert^2 \over 2
\hbar\omega}\,, 
\label{app:parts} 
\end{equation}
which are Eqns. (\ref{rhopair}) and (\ref{bogg}) of the text. 
These results were reported in \cite{ourprl}. 

In the pure state case the even $\tilde n$ diagonal matrix elements of
the density matrix are positive definite and may be interpreted as the
probabilities for observing $\ell$ uncorrelated particle pairs in the
adiabatic particle basis. The odd $\tilde n$ diagonal matrix elements
vanish since particles can only be created in pairs from the
vacuum. Otherwise ({\em i.e.} for $\sigma > 1$), the much more
complicated and non-positive definite expression (\ref{app:rhodab}) shows
that there is no simple classical probability interpretation for the
density matrix. The restriction to zero mean fields $\phi =\dot\phi
=0$ is not serious for spatially homogeneous backgrounds because of
Eqn. (\ref{meanfk}) which shows that all the mean fields vanish except
for the $k = 0$ mode. Hence, at late times all the $k > 0$ modes may
be treated classically if dephasing is effective and typical classical
field configurations in the ensemble can be constructed as in Eqns.
(\ref{sample}) and (\ref{ranph}) of the text, provided the $k=0$ mode
is excluded.

\section{Numerical Methods}

In order to solve the system of equations (\ref{feq}), (\ref{modefn}),
and (\ref{cren}) as an initial value problem, we have to specify the
initial conditions of the mean field $\phi$, its time derivative and
the mode function and its time derivative. Then we have to solve the
gap equation (\ref{cren}) at the initial time $t_0$.  In order to have
a finite set of renormalized equations we have to choose the mode
functions so that the high momentum modes coincide with the zeroth
order adiabatic vacuum described by
\begin{eqnarray}
f_k(t_0)&=&{1\over\sqrt{ 2\omega_k(t_0)}}~,
\label{initial_modes}\\
\dot{f}_k(t_0)&=&
\left[-i\omega_k(t_0)-{\dot{\omega}_k(t_0)\over
2\omega_k(t_0)}\right]f_k(t_0)~, 
\label{initial_modesdot}
\end{eqnarray}
with $\omega_k^2(t_0)=k^2+\chi(t_0)$.  It is easy to verify that for
initial conditions with $\dot{\phi}(t_0)=0$, it is also true that
$\dot{\chi}(t_0)=0$ by inspecting the time derivative of the gap
equation. This simplifies the form of $\dot{f}_k(t_0)$ to
\begin{equation}
\dot{f}_k(t_0)=-i\omega_k(t_0)f_k(t_0)~. 
\end{equation}
For an initial state with positive square effective mass $\chi$ we
could use the same form of the mode functions of (\ref{initial_modes})
also for the low momentum modes.  However, if we wish to investigate
the case of a ``quench'' within the unstable spinodal region by
initial conditions with a negative $\chi(0)$, we have to modify the
initial values of the low momentum modes in order to avoid the
singularity of $f_k(t_0)=1/\sqrt{2\omega_k}$ at
$k^2=-\chi(t_0)$.  We have used two different profiles of the
frequency $\omega_k$ for the initial mode functions:
\begin{eqnarray}
\omega^2_k(t_0)&=&k^2+\chi(t_0)\tanh\left({k^2+\chi(t_0)\over
|\chi(t_0)|}\right)~, \label{initial_modesA}\\ 
\omega^2_k(t_0)&=&k^2+\chi(t_0)\exp(-v^4/k^4)~. 
\label{initial_modesB}
\end{eqnarray}
At large momentum these profiles coincide with the adiabatic vacuum
frequency as required. In all cases, except for thermal initial
conditions, the initial number of quasi-particles, $N(k)=0$.

Numerical simulations were performed on a massively parallel computer
using a momentum grid with $32000$ modes. The upper cut-off was set at
$\Lambda=5v$, implying a grid resolution: $$dk = { \Lambda \over
32000}~. $$  The field $\phi$ was scaled in units of $v$ so that $k$ is
also given in units of $v$, $\chi$ in units of $v^2$, and the time $t$
in units of $v^{-1}$. Most of the results discussed here are for a
bare coupling constant $\lambda_\Lambda=1$ but other values were also
investigated. The corresponding renormalized coupling constants at the
renormalization point $|\chi(t_0)|$ are given by (\ref{lren1}) with
$m^2$ replaced by $|\chi(t_0)|$. With $\lambda_\Lambda=1$, and for
$\phi(t_0)/v=0$, $\lambda_R= 0.990036$, while for $\phi(t_0)/v=0.5$,
$\lambda_R= 0.990248$.  The energy densities in these cases are
$\varepsilon=0.1237$ and $\varepsilon=0.06978$ respectively.

The mode equations were stepped forward in time using a sixth-order
adaptive time-step Runge-Kutta integrator. The time-steps were
controlled by tracking the evolution of $\chi$. Energy conservation to
parts per million was achieved over the temporal range of typical
evolutions. 

\end{document}